\documentclass[pdflatex,sn-mathphys-num]{sn-jnl}

\usepackage{graphicx}%
\usepackage{multirow}%
\usepackage{amsmath,amssymb,amsfonts}%
\usepackage{amsthm}%
\usepackage{mathrsfs}%
\usepackage[title]{appendix}%
\usepackage{xcolor}%
\usepackage{textcomp}%
\usepackage{manyfoot}%
\usepackage{booktabs}%
\usepackage{mhchem}
\usepackage{algorithm}%
\usepackage{algorithmicx}%
\usepackage{algpseudocode}%
\usepackage{listings}%
\usepackage{makecell}
\usepackage{booktabs} %
\usepackage{tabularx} %
\usepackage{array}
\usepackage{bm}

\usepackage[most]{tcolorbox}
\usepackage{listings}
\usepackage{xcolor}

\lstdefinestyle{promptstyle}{
    basicstyle=\ttfamily\footnotesize,
    breaklines=true,
    breakatwhitespace=false,
    columns=fullflexible,
    keepspaces=true,
    showstringspaces=false,
    frame=none
}

\newtcblisting{promptbox}[2][]{
    enhanced,
    breakable,
    colback=gray!3,
    colframe=gray!60,
    boxrule=0.5pt,
    arc=1mm,
    left=1.5mm,
    right=1.5mm,
    top=1mm,
    bottom=1mm,
    title={#2},
    fonttitle=\bfseries,
    listing only,
    listing options={style=promptstyle},
    #1
}

\usepackage{color}
\usepackage{xcolor}
\usepackage{caption}
\usepackage{subfigure}
\usepackage[table]{xcolor}
\usepackage{tcolorbox}
\definecolor{lightblue}{HTML}{F0F8FF}
\definecolor{top1}{rgb}{0.68, 0.85, 0.9}
\definecolor{top2}{rgb}{0.78, 0.93, 0.96}
\definecolor{top3}{rgb}{0.88, 0.97, 0.99}

\theoremstyle{thmstyleone}

\newcommand{\rre}[1]{{\color{black}#1}}

\theoremstyle{thmstyletwo}

\theoremstyle{thmstylethree}

\begin{document}

\title[Article Title]{\rre{MOFFieldBench: A Machine Learning Benchmark for Continuous Electronic Fields in Metal–Organic Frameworks}}

\author[1]{\fnm{\rre{Zhaolin}} \sur{Hu}}
\author[1,3]{\fnm{\rre{Hehe}} \sur{\rre{Fan}}}
\author[2]{\fnm{\rre{Chenhao}} \sur{Rao}}
\author[1]{\fnm{\rre{Wangyihan}} \sur{Guo}}
\author[2]{\fnm{\rre{Meng}} \sur{Xu}}
\author[2]{\fnm{\rre{Qiwei}} \sur{Yang}}
\author[1,3]{\fnm{\rre{Yi}} \sur{Yang}}

\affil[1]{\orgdiv{College of Artificial Intelligence}, \orgname{Zhejiang University }, \orgaddress{\city{Hangzhou}, \country{China}}}
\affil[2]{\orgdiv{College of Chemical and Biological Engineering}, \orgname{Zhejiang University } \orgaddress{\city{Hangzhou}, \country{China}}}
\affil[3]{\orgdiv{Institute of Fundamental and Transdisciplinary Research}, \orgname{Zhejiang University } 
\orgaddress{\city{Hangzhou}, \country{China}}}

\abstract{
Machine-learning models for metal--organic frameworks (MOFs) predominantly rely on atomic structures, whereas electron-density (ED) and electrostatic-potential (ESP) fields remain underexplored as scalable learning representations. These continuous fields encode complementary spatial and electronic information relevant to structure--property relationships, particularly pore geometry and guest--framework interactions, but costly first-principles calculations, storage requirements and the scarcity of systematically aligned datasets limit their broader use. Here we introduce MOFFieldBench, a multimodal benchmark comprising approximately 46,000 MOFs with one-to-one aligned crystallographic structures, ED fields and ESP fields. A unified workflow integrates field generation, error-controlled compression and sparse point-cloud representation, supporting field reconstruction, prediction of 13 material properties, multimodal property prediction and pairwise cross-modal retrieval. ED and ESP provide competitive representations for pore geometry, adsorption and separation properties. Multimodal fusion improves several prediction tasks relative to the best unimodal reference, supporting complementary predictive roles for atomic structures and electronic fields. Learned ED reconstruction closely approximates first-principles reference densities, while predicted-density initialization reduces self-consistent-field calculation time by 13.59\% on average in paired tests. These results support electronic fields as practical, informative learning modalities and establish MOFFieldBench as a resource for studying representation choice, multimodal learning and electronic-field-driven MOF modelling.
}

\maketitle

\section{Introduction}\label{sec: introduction}

Metal--organic frameworks (MOFs) are a chemically and structurally diverse class of crystalline porous materials assembled from inorganic nodes and organic linkers, whose modular construction enables broad control over composition, topology, pore geometry and chemical functionality~\cite{furukawa2013chemistry,wang2013metal}. Their structural and chemical tunability has enabled applications ranging from gas storage and separation to catalysis, energy conversion and molecular recognition~\cite{freund2021current,knebel2022metal}. The rapid expansion of experimental and hypothetical MOF databases has consequently motivated machine-learning approaches for navigating this vast design space, predicting material properties and prioritizing candidate structures~\cite{chung2019advances,wilmer2012large,jablonka2020big}. Most existing models describe MOFs through atomic species, coordinates, connectivity, or geometric and chemical descriptors derived from these quantities~\cite{jablonka2020big,moosavi2020understanding,wang2024comprehensive}. Such atom-centric representations provide compact and effective descriptions of the framework and have enabled accurate learning across diverse materials properties, particularly gas adsorption, diffusion and adsorption-based materials screening~\cite{wang2020accelerating,kang2023multi,sung2025machine}. Crystal-graph models, for example, have been used to accelerate methane-adsorption screening across thousands of experimentally reported MOFs~\cite{wang2020accelerating}, while large-scale representation-learning approaches have extended structure-based prediction to multiple gas adsorption and transport-related tasks~\cite{kang2023multi}. Yet these representations primarily encode the discrete atomic framework and do not explicitly expose the continuous electronic environment generated by a given atomic structure. Spatially distributed interactions relevant to adsorption, separation and guest--framework recognition must therefore be learned indirectly from atomic identities, geometry or derived descriptors. This raises a broader question of representation: what can be gained by describing porous materials through the continuous electronic fields that emerge from their atomic structures?

Among the continuous electronic fields that emerge from an atomic structure, electron density (ED), \(\rho(\mathbf r)\), and electrostatic potential (ESP), \(V(\mathbf r)\), provide two fundamental real-space descriptions of the electronic environment. The ground-state electron density contains, in principle, sufficient information to determine ground-state properties, while its spatial distribution reflects how electronic charge is localized and redistributed across atoms, bonds and coordination environments~\cite{hohenberg1964inhomogeneous}. In MOFs, ED therefore provides information beyond atomic positions alone, capturing metal--ligand bonding and local variations in electronic structure while delineating the chemically occupied framework and pore boundaries~\cite{sarkar2022electron}. ESP describes the potential generated by the underlying nuclear and electronic charge distribution throughout space, including low-density regions within the pores. At positions accessible to adsorbates, it characterizes the electrostatic environment encountered by guest molecules and is therefore directly related to the electrostatic component of guest--framework interactions. Open metal sites, heteroatoms and polar functional groups can consequently create heterogeneous electrostatic environments that strongly affect guest--framework interactions, particularly for polar or quadrupolar adsorbates~\cite{hamad2015atomic}. Such pore electrostatics can directly influence adsorption affinity and molecular selectivity; electrostatic-potential matching between a porous framework and guest molecules has, for example, been shown to enable selective molecular recognition and gas separation~\cite{xie2021electrostatic}. ED and ESP thus encode related but physically distinct aspects of a MOF: ED emphasizes the electronic distribution associated with the framework and its bonding environment, whereas ESP describes the electrostatic landscape experienced by molecules within the pore space. These physical and spatial characteristics make them promising learning representations for capturing structure--property relationships that may not be fully exposed by atom-centric descriptions alone.

The spatial characteristics of MOF pore environments have also been explored through interaction-energy descriptors in representation learning. These descriptors characterize pore environments using simplified, force-field-derived interaction-energy fields rather than first-principles ED or ESP fields. For example, MOFTransformer combines atomic graphs with energy grids~\cite{kang2023multi}, while SpbNet integrates structural features with potential energy surface (PES) basis functions~\cite{zou2026dataefficient}. However, these energy descriptors depend on prescribed potential forms and atom-type parameters and do not explicitly resolve the self-consistent electronic redistribution associated with framework bonding and coordination. Moreover, SpbNet's implemented PES descriptors omit partial charges and Coulombic contributions~\cite{zou2026dataefficient}, leaving the electrostatic component of interactions with polar or quadrupolar adsorbates unrepresented in that input channel. First-principles ED directly describes the framework electron distribution, including the effects of bonding and coordination, whereas ESP maps the electrostatic environment within the pores. These fields therefore provide electronic and electrostatic information that the parameterized interaction-energy descriptors do not explicitly contain, motivating their systematic evaluation as learning representations for MOFs.

Direct investigation of first-principles electronic fields has become increasingly feasible through advances in electronic-structure datasets and machine learning. Large-scale charge-density collections now provide electronic fields for diverse periodic materials~\cite{shen2022charge}, while equivariant neural networks have demonstrated accurate reconstruction of electron density directly from atomic structures~\cite{koker2024charge3net}. Systematic benchmarks have further evaluated charge-density prediction across different levels of electronic-structure accuracy~\cite{chen2025ecd}. Beyond reconstruction, electron density has also been explored directly as a physically informed representation for materials property prediction, for example through three-dimensional convolutional models operating on volumetric charge-density fields~\cite{zhao2020elastic}. Together, these developments establish electron density as both a learnable physical quantity and a potential input modality for materials learning. However, existing studies have focused predominantly on electron density in conventional inorganic crystals and small molecules~\cite{zhang2026unifield}, with electrostatic potential receiving far less attention as an independent learning representation. More fundamentally, it remains unclear how ED and ESP compare with atomistic representations across different classes of MOF properties, and whether combining these modalities improves predictive performance beyond the strongest individual representation.

These open questions are particularly consequential for MOFs, where representing electronic fields at scale is itself nontrivial. Their large and structurally diverse unit cells, together with substantial pore volumes, make dense three-dimensional fields considerably more expensive to generate, store and process than conventional atomistic representations~\cite{jablonka2020big,shen2022charge}. Moreover, the spatial distribution of information differs markedly between the two field modalities: electron density is concentrated primarily around the framework, whereas electrostatic potential can remain strongly structured throughout the pore volume and encode interactions relevant to guest binding~\cite{scatena2019electron}. Efficient compression must therefore preserve numerical fidelity, while compact learning representations must retain the modality-specific spatial information that may be relevant to downstream properties. Establishing ED and ESP as scalable learning representations consequently requires approaches that balance physical fidelity with computational efficiency, together with systematic evaluation of the predictive utility of the resulting representations. The challenge is thus not simply to reduce the size of electronic-field data, but to determine whether their physically meaningful information can be retained in compact representations and translated into useful signals for materials learning.

Here we introduce MOFFieldBench, a multimodal benchmark for learning with electron density and electrostatic potential in metal--organic frameworks. The benchmark comprises 45,909 MOFs with one-to-one aligned crystallographic structures, ED and ESP fields, providing a common basis for comparing atomistic and electronic-field representations. Using task-specific subsets, we first evaluate structure-conditioned reconstruction of dense ED and ESP fields and the use of predicted ED to initialize self-consistent-field (SCF) calculations. We also develop error-controlled compression and sparse point-cloud representations to reduce the storage and computational burden of volumetric electronic data. We then benchmark models based on CIF, ED and ESP across 13 material properties spanning geometric, adsorption and electronic characteristics, and evaluate multimodal combinations to determine whether atomic structures and electronic fields provide complementary predictive information. Finally, we establish cross-modal retrieval across CIF, ED and ESP to evaluate how reliably representations from different physical modalities can be associated with the same underlying material. Together, these tasks provide a systematic framework for assessing how electronic fields can be represented, learned and integrated with conventional atomistic descriptions.

The benchmark identifies complementary predictive strengths of
atomistic and electronic-field representations.
CIF retains a clear advantage for band-gap prediction, while ED and
ESP provide effective inputs for pore geometry and adsorption- and
separation-related properties.
Multimodal fusion reduces MAE across all 13 targets, with at least
one modality combination outperforming the lowest-MAE unimodal
reference for each property, supporting the predictive value of
integrating structural and electronic information.
Pairwise cross-modal retrieval further shows that a query in one
modality can identify the corresponding representation of the same
MOF in another, providing a basis for materials search using
structural or electronic-field inputs.
By providing aligned references and complementary evaluation tasks,
MOFFieldBench establishes a reusable foundation for developing and
comparing models that connect atomic structures, electronic
environments and material properties, supporting the broader
integration of electronic fields into materials learning and discovery.

\begin{figure}[!htbp]
\centering
\includegraphics[width=\linewidth]{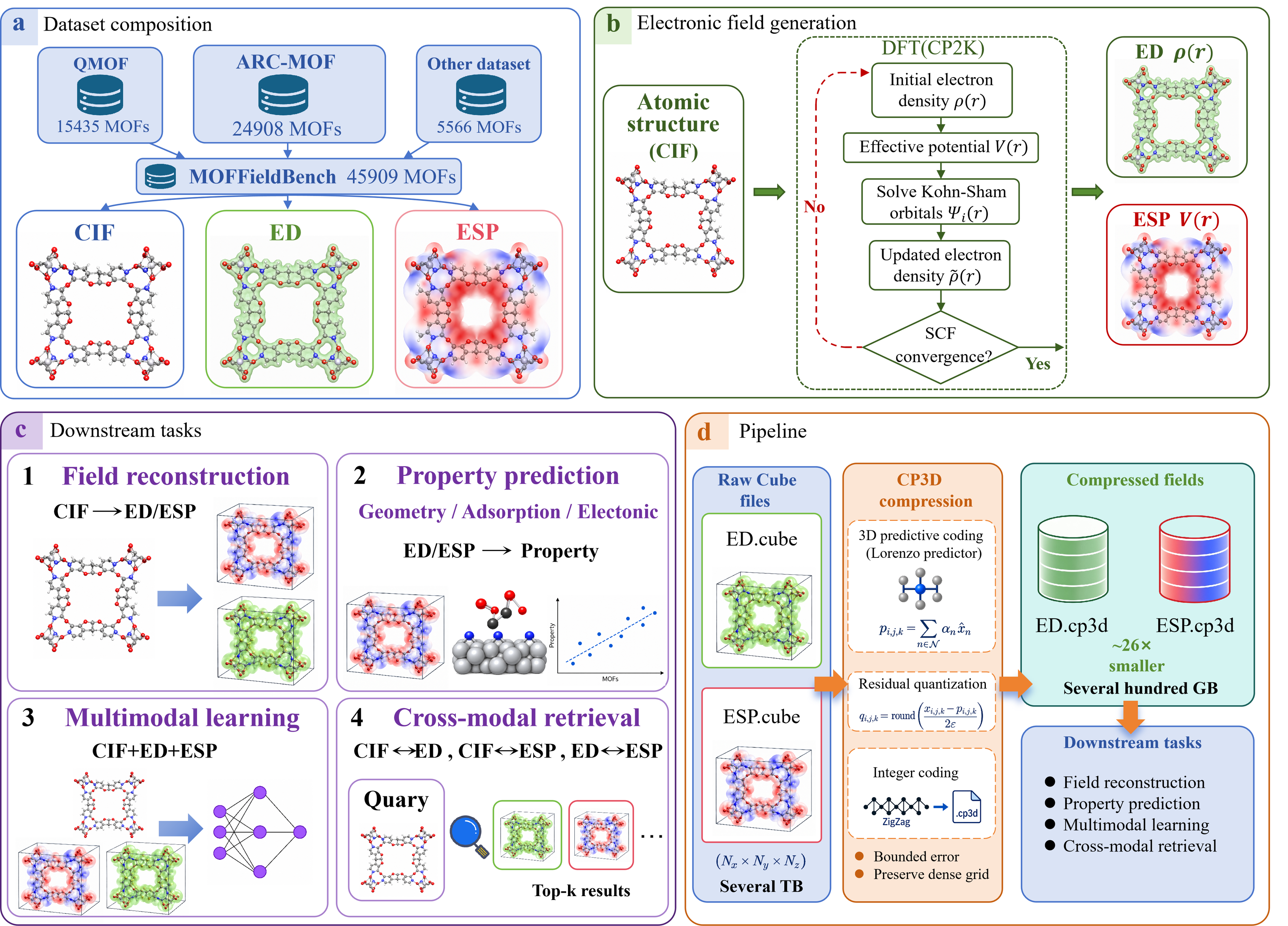}
\caption{
\textbf{Construction and scope of MOFFieldBench.}
\textbf{a,} Structural sources and aligned data modalities.
MOFFieldBench draws primarily on ARC-MOF and QMOF, with additional
structures from CoRE MOF, the Cambridge Structural Database (CSD)
and the Ultrastable MOF database, as well as hypothetical MOFs
constructed using PORMAKE.
Each framework has three aligned representations within the same
periodic unit cell: crystallographic structure (CIF), electron
density (ED) and electrostatic potential (ESP).
\textbf{b,} First-principles field generation.
Self-consistent Kohn--Sham density functional theory calculations
in CP2K generate ED, \(\rho(\mathbf r)\), and ESP, \(V(\mathbf r)\),
on regular three-dimensional grids.
\textbf{c,} Benchmark tasks.
The aligned representations support structure-to-field reconstruction,
prediction of geometric, adsorption and electronic properties,
multimodal property prediction, and cross-modal retrieval.
\textbf{d,} Overall pipeline of MOFFieldBench.
Raw ED and ESP cube files are compressed using CP3D for compact
storage while preserving the original grid geometry with bounded
reconstruction errors.
The fields are decompressed as needed to support structure-to-field
reconstruction, property prediction, multimodal learning and
cross-modal retrieval.
}
\label{fig:moffieldbench_overview}
\end{figure}

\begin{figure*}[!htbp]
\centering
\includegraphics[width=0.98\textwidth]{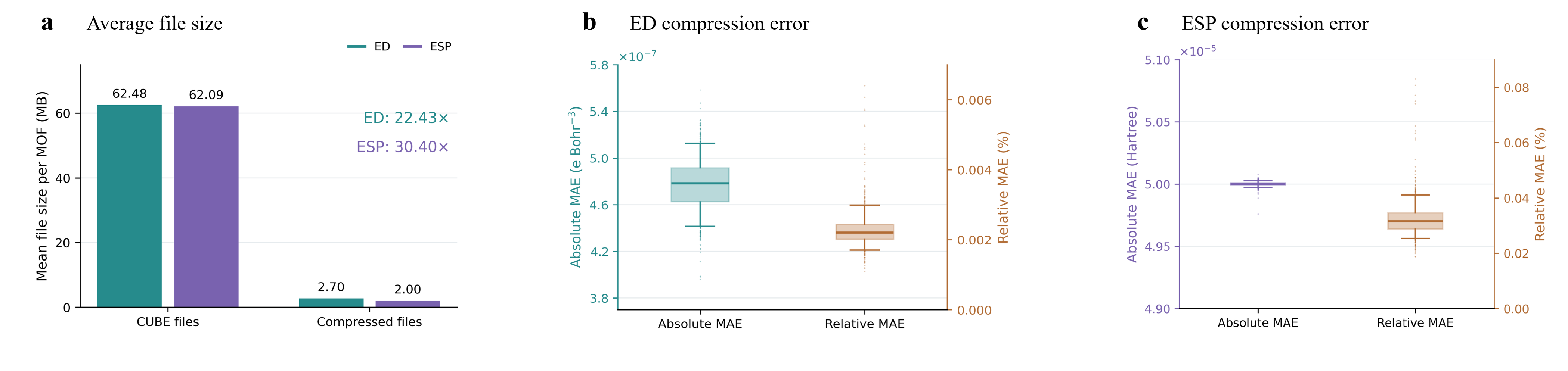}
\caption[Storage efficiency and numerical fidelity of CP3D compression.]{
\textbf{Storage efficiency and numerical fidelity of CP3D compression.}
\textbf{a,} Mean file size per MOF before and after CP3D compression, shown separately for ED and ESP.
Annotated values indicate the compression ratio for each field.
\textbf{b,c,} Distributions of absolute and relative mean absolute errors (MAEs) between the original and decompressed fields for ED (\textbf{b}) and ESP (\textbf{c}).
Absolute MAEs are shown on the left axes, in \(\mathrm{e\,Bohr^{-3}}\) for ED and Hartree for ESP.
Relative MAEs are expressed as percentages and shown on the right axes.
}
\label{fig:compression}
\end{figure*}

\section{Results}\label{sec2}

\subsection{MOFFieldBench framework}


MOFFieldBench integrates metal--organic frameworks from multiple data sources with aligned crystallographic structures, electron-density (ED) fields and electrostatic-potential (ESP) fields (Fig.~\ref{fig:moffieldbench_overview}a). ARC-MOF~\cite{burner2023arc} and QMOF~\cite{rosen2021machine} provide the main structural sources, contributing 24,908 and 15,435 frameworks, respectively. These are supplemented by structures from additional resources, including CoRE MOF~\cite{chung2019advances}, the Cambridge Structural Database (CSD)~\cite{groom2016cambridge} and the Ultrastable MOF database~\cite{nandy2023database}. For each framework, the crystallographic structure and corresponding electronic fields are aligned within the same periodic unit cell. The CIF records the atomic species, coordinates and lattice geometry, while ED, \(\rho(\mathbf r)\), and ESP, \(V(\mathbf r)\), describe the spatial electron distribution and electrostatic potential obtained from first-principles calculations.

The electronic fields were generated using periodic first-principles calculations under a unified workflow (Fig.~\ref{fig:moffieldbench_overview}b). Starting from each crystallographic structure, an initial electron density is used to construct the effective Kohn--Sham potential~\cite{KohnSham1965}, after which the electronic states are solved and the density is updated. This self-consistent cycle is repeated until the electron density satisfies the prescribed convergence criterion, yielding the converged electronic ground state. The resulting electron density is then evaluated over the periodic simulation cell, while the corresponding electrostatic potential is obtained from the converged charge distribution, producing spatially resolved ED and ESP fields on regular three-dimensional grids. Because both fields are derived from the same converged electronic structure and registered to the original crystallographic lattice, their spatial correspondence with the atomic framework is preserved by construction.

Representing ED and ESP as dense three-dimensional grids introduces a substantial storage and data-movement burden at benchmark scale. The original volumetric Cube files collectively occupy several terabytes, which complicates dataset distribution, repeated access and large-scale model training. To address this limitation, we incorporate CP3D(Compress 3D), a three-dimensional predictive compression scheme with configurable
nominal absolute quantization tolerances (Fig.~\ref{fig:moffieldbench_overview}d). CP3D predicts each grid value from previously reconstructed neighbouring points, quantizes the resulting residual under a prescribed absolute error tolerance and subsequently entropy-encodes the quantized residuals. Because the compression operates directly on the regular volumetric grid, the original spatial resolution and grid geometry are retained after reconstruction. Using field-specific error bounds for ED and ESP, CP3D achieves an average compression ratio of \(26.4\times\) while maintaining high numerical fidelity (Fig.~\ref{fig:compression}), reducing the complete electronic-field collection from several terabytes to several hundred gigabytes. This reduction substantially lowers the barrier to distributing and reusing the benchmark, while also decreasing disk I/O, network-transfer and data-loading overhead during repeated machine-learning training. More importantly, 
It provides a scalable storage strategy that reduces storage and data-transfer requirements as electronic-field datasets expand.

The aligned multimodal data support four complementary benchmark tasks (Fig.~\ref{fig:moffieldbench_overview}c). First, structure-to-field reconstruction evaluates whether dense ED and ESP distributions can be recovered from the underlying atomic structure, providing a direct test of the structure--field relationship. Second, property prediction compares CIF-, ED- and ESP-based representations across 13 material properties spanning geometric, adsorption and electronic characteristics, allowing the predictive utility of each modality to be assessed under consistent data splits. Third, multimodal learning combines atomistic and electronic-field representations to test whether their information is complementary for downstream prediction. Finally, cross-modal retrieval evaluates whether representations derived from different physical modalities can be correctly associated with the same underlying MOF, without assuming that the modalities encode identical information. Together, these tasks enable systematic evaluation of how atomic structure, electron density and electrostatic potential capture different aspects of material behaviour and how they can be used individually or jointly for materials learning.

\begin{figure*}[!htbp]
\centering
\includegraphics[width=0.98\textwidth]{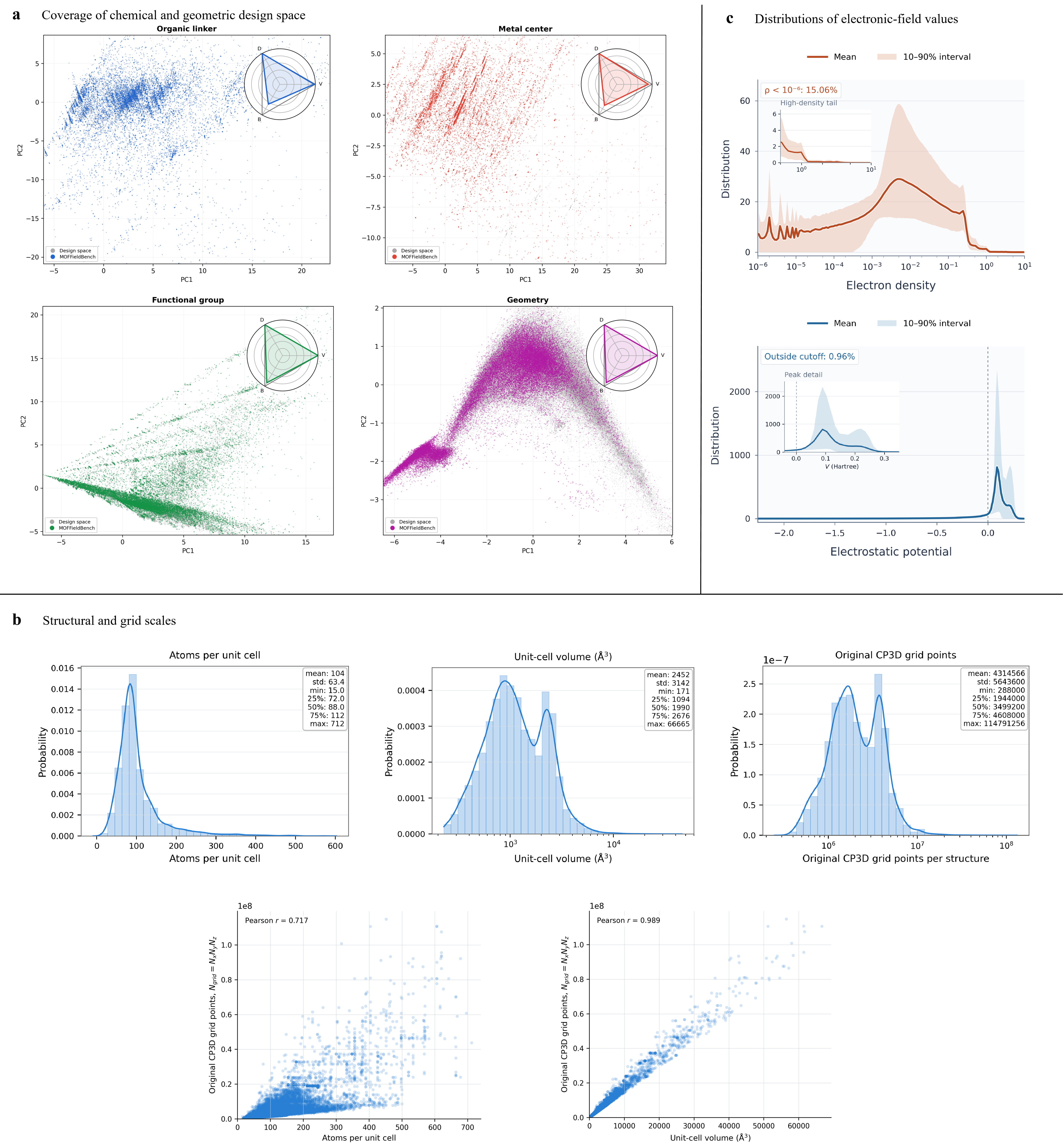}
\caption[Chemical coverage, structural scale and electronic-field statistics of MOFFieldBench.]{
\textbf{Chemical coverage, structural scale and electronic-field statistics of MOFFieldBench.}
\textbf{a,} Principal-component projections of organic-linker,
metal-centre, functional-group and geometric descriptors for
MOFFieldBench (coloured) and the reference MOF population (grey).
Radar plots summarize normalized variety (V), disparity (D) and balance (B).
\textbf{b,} ED (\(\mathrm{e\,Bohr^{-3}}\)) and ESP
(\(\mathrm{Hartree}\)) value distributions.
Solid curves and shaded regions indicate the mean and 10th--90th
percentile intervals across structures, respectively.
ED is shown on a logarithmic horizontal axis.
\textbf{c,} Distributions of atom count, unit-cell volume and grid size,
and relationships of grid size with atom count and unit-cell volume.
}
\label{fig:dataset_characterization}
\end{figure*}

\begin{figure*}[!htbp]
\centering
\includegraphics[width=0.98\textwidth]{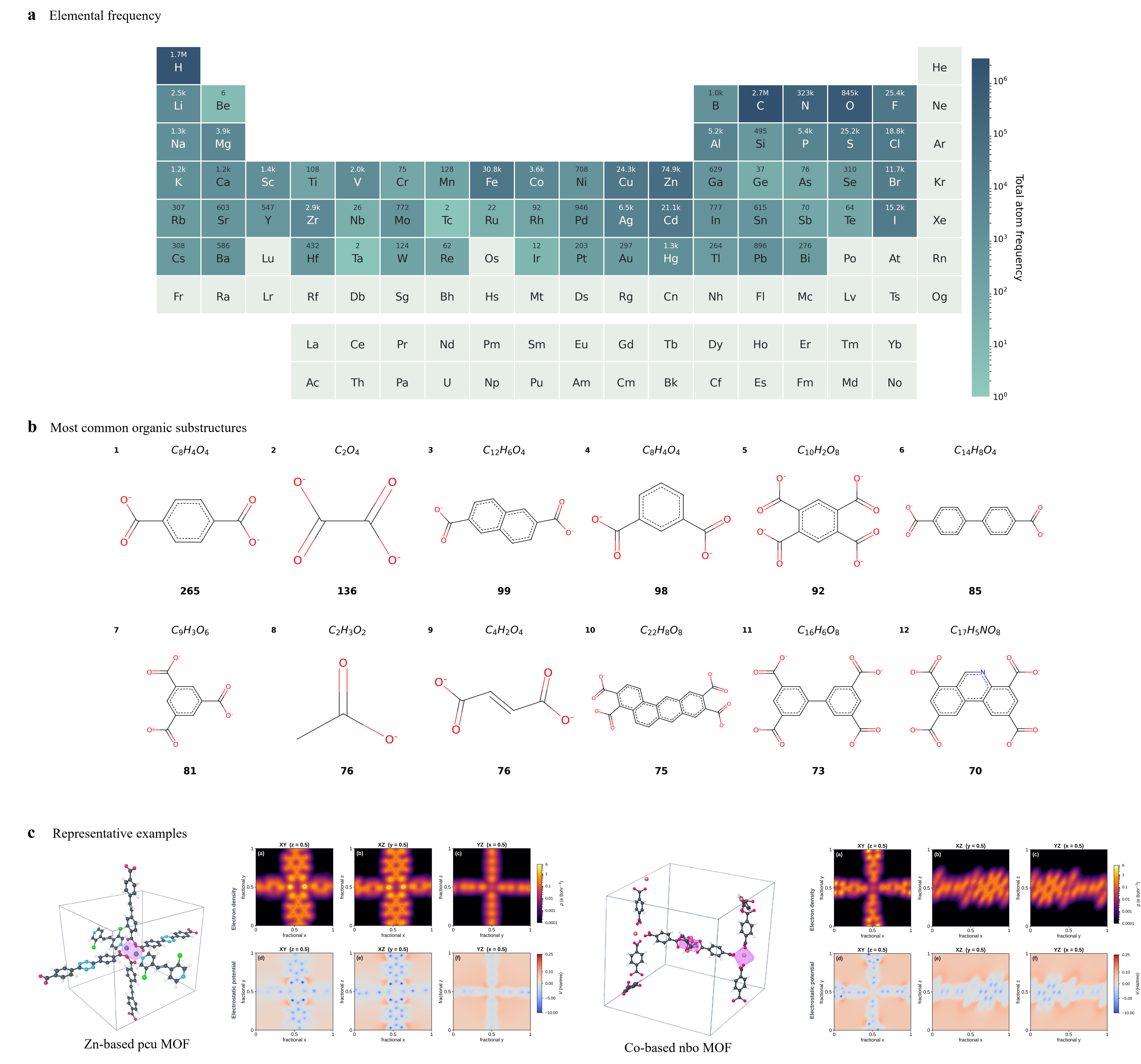}
\caption[Elemental composition, common organic substructures and representative electronic fields in MOFFieldBench.]{
\textbf{Elemental composition, common organic substructures and representative electronic fields in MOFFieldBench.}
\textbf{a,} Elemental composition, showing cumulative atom counts
with labels and a logarithmic colour scale.
\textbf{b,} Common organic substructures identified by clustering
ligand revised autocorrelation (RAC) descriptors, with occurrence
counts shown below.
\textbf{c,} Representative atomic structures and ED (upper rows)
and ESP (lower rows) slices through the unit-cell centre.
ED uses a logarithmic colour scale; units are
\(\mathrm{e\,Bohr^{-3}}\) for ED and Hartree for ESP.
}
\label{fig:substructures_fields}
\end{figure*}

\subsection{Dataset characterization}

MOFFieldBench was characterized in terms of framework chemistry,
structural scale and the statistical and spatial organization of its
electronic fields (Figs.~\ref{fig:dataset_characterization} and \ref{fig:substructures_fields} ).
Its chemical and structural coverage was compared with a broader
reference population in four descriptor spaces: organic linkers,
metal centres, functional groups and framework geometry
(Fig.~\ref{fig:dataset_characterization}a).
The reference population combines experimental and hypothetical
MOFs from the CoRE MOF, CSD-MOF, Wilmer, Boyd--Woo and Majumdar
collections~\cite{wilmer2012large,boyd2016general,chung2019advances,li2021csd,majumdar2021diversifying}.
Linker, metal-centre and functional-group environments are represented
by revised autocorrelation (RAC) descriptors, while framework geometry
is described by pore- and density-related descriptors.
Principal-component projections place MOFFieldBench and the reference
population in the same coordinate spaces, with coloured benchmark
points overlaid on the grey reference background.

The benchmark points span the main populated regions in all four projections, indicating coverage of multiple chemical and structural families. Overlap is particularly extensive in the functional-group and geometric spaces; the linker and metal-centre projections show greater variation in sampling density. The radar plots distinguish three aspects of diversity through normalized variety (V), disparity (D) and balance (B). Variety measures how many regions of the reference space are represented, disparity describes the separation of the occupied regions, and balance quantifies how evenly benchmark structures are distributed among them. High variety and disparity across these spaces support broad coverage of the characterized environments, while variation in balance identifies differences in their relative representation.

Elemental composition and common organic substructures provide a chemical context for the descriptor-space coverage (Fig.~\ref{fig:substructures_fields}a,b). Elemental abundance is reported as cumulative atom counts across crystallographic unit cells. Carbon, hydrogen, oxygen and nitrogen dominate these counts, and Zn, Fe, Cu and Cd are prominent among the metals; Mg, Al, Zr and additional, less frequently represented metallic species are also present. At the linker level, clustering of ligand RAC descriptors across the database identifies the 12 most common organic substructures. Carboxylate-containing motifs predominate, with backbones ranging from compact aliphatic units to single-ring, fused-ring and linked aromatic systems, including a nitrogen-containing aromatic motif. The occurrence counts shown beneath the structures range from 70 to 136 for all but the most frequent displayed motif, which occurs 265 times.

The variation in structural size is accompanied by a wide range of volumetric input sizes (Fig.~\ref{fig:dataset_characterization}c). Atom count and unit-cell volume have medians of 88 atoms per unit cell and approximately \(1.99\times10^{3}\,\text{\AA}^{3}\), respectively, and both distributions have extended upper tails. Measured by point count, the original volumetric grids span more than two orders of magnitude; most contain millions of points, with a long upper tail towards larger grids. Grid size has a strong, nearly linear relationship with unit-cell volume (\(r=0.989\)), consistent with the approximately fixed real-space resolution used for field generation. Its correlation with atom count is weaker (\(r=0.717\)): structures with comparable atom counts can require markedly different numbers of grid points. Because the spatial extent of the cell primarily determines volumetric input size, frameworks with moderate atom counts can still require large grids and substantial storage and computational resources.

ED and ESP also differ in the distributions of values over their grids
(Fig.~\ref{fig:dataset_characterization}b). The ED distribution is broad
and strongly non-uniform, with \(15.06\%\) of grid points below
\(10^{-6}\,\mathrm{e\,Bohr^{-3}}\). It combines extensive low-density
pore regions with high-density features localized near framework atoms
and along bonds, reflecting the rapid decrease in density away from
the framework. The inset resolves the high-density tail.
ESP is generated by the collective nuclear and electronic charge
distribution and extends into pore regions beyond the immediate
vicinity of the framework. Most ESP values lie within a relatively
narrow central range, with a pronounced positive peak near
\(0.1\,\mathrm{Hartree}\) and positive and negative tails corresponding
to larger deviations in potential. Only \(0.96\%\) of ESP values fall
outside the plotted range. The solid curves summarize the mean
field-value distributions across structures, while the shaded regions
indicate the corresponding 10th--90th percentile intervals.
These statistics characterize the main distributional features and
their variability across the collection.

Real-space slices from two structurally different, representative MOFs show how these statistical differences appear within individual unit cells (Fig.~\ref{fig:substructures_fields}c). In each framework, regions of high ED follow the atomic skeleton and bonding environments, whereas the pore interiors have substantially lower density. ESP varies continuously through the framework and surrounding void space, forming positive and negative regions that extend into the pores. Areas that are nearly featureless in ED can therefore exhibit pronounced electrostatic variation in ESP. This contrast appears across all three slice orientations. The \(xy\), \(xz\) and \(yz\) sections pass through fractional coordinates \(z=0.5\), \(y=0.5\) and \(x=0.5\), respectively. The maps give a spatial interpretation of the field-value distributions: ED emphasizes local electronic structure around the framework, whereas ESP retains longer-range electrostatic information within the pores.

These observations identify two challenges for learning from MOF electronic fields. First, differences in unit-cell dimensions produce large variations in volumetric input size, leading to substantial heterogeneity in memory and computational demands. Second, the two fields distribute information differently: ED features are concentrated near the framework, whereas ESP variations extend throughout the pore volume. Compact representations must therefore lower the cost of dense grids while preserving both local framework features and longer-range information within the pores. These requirements guide the sampling and point-cloud representations introduced in Section~\ref{sec:pointcloud}.

\begin{figure*}[!htbp]
\centering
\includegraphics[width=0.98\textwidth]{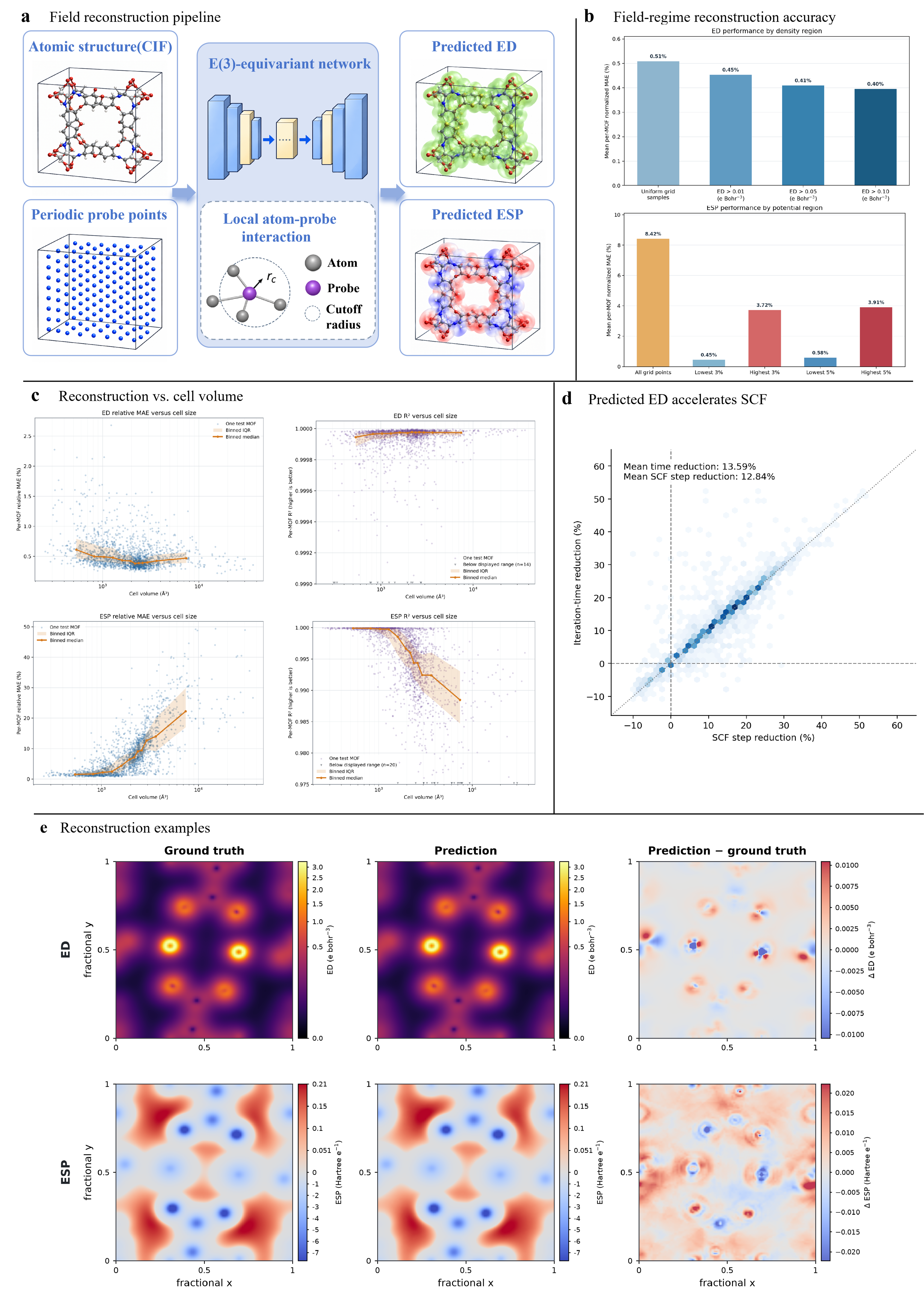}
\caption{
\textbf{Structure-to-field reconstruction and predicted-ED initialization.}
\textbf{a,} E(3)-equivariant reconstruction of ED and ESP.
\textbf{b,} Mean per-MOF normalized MAE for complete grids and selected field subsets, with normalization performed within each regime.
\textbf{c,} Normalized MAE and $R^2$ versus unit-cell volume for ED (\textbf{c}) and ESP (\textbf{d}).
Points represent test MOFs; curves and bands show binned medians and interquartile ranges.
\textbf{d,} Percentage reductions in SCF iterations and subsequent DFT time with predicted-ED initialization relative to standard initialization; timings exclude ED reconstruction.
\textbf{e,} Reference, predicted and residual (prediction minus reference) ED/ESP slices in fractional coordinates.
ED and ESP units are $\mathrm{e\,Bohr^{-3}}$ and $\mathrm{Hartree\,e^{-1}}$, respectively.
}
\label{fig3}
\end{figure*}

\subsection{Structure-to-field reconstruction}

A central objective of MOFFieldBench is to determine whether dense electronic fields can be recovered directly from periodic atomic structures (Fig.~\ref{fig3}a). We formulate this task as structure-to-field reconstruction: given the atomic species, coordinates and lattice geometry of a MOF, spatial probe points are introduced throughout the unit cell and connected to nearby framework atoms within a local neighbourhood. Atomic information is then encoded using an E(3)-equivariant graph neural network~\cite{koker2024charge3net}, which propagates geometric and chemical features from the framework atoms to the probe locations while respecting rotational and translational symmetries. Each probe is finally mapped to a scalar field value, corresponding to either the electron density, $\rho(\mathbf r)$, or the electrostatic potential, $V(\mathbf r)$. Evaluating the probes at the original grid locations reconstructs the dense three-dimensional field and enables direct comparison with the corresponding first-principles reference. This formulation provides a unified setting for evaluating how faithfully different electronic fields can be inferred from atomic structure and whether the resulting predictions are sufficiently accurate for downstream physical use.

To assess reconstruction across the non-uniform ED and ESP distributions (Fig.~\ref{fig:dataset_characterization}b), we evaluated uniformly sampled points across
  the entire unit cell and points within selected reference-field regimes (Fig.~\ref{fig3}b). Normalized MAE was calculated separately for each MOF within each regime and then averaged across structures. For ED, most grid points lie in low-density pore space, whereas appreciable density is concentrated near atoms and bonds. We therefore additionally examined higher-density subsets defined by $\rho>0.01$, $\rho>0.05$ and $\rho>0.10\,\mathrm{e\,Bohr^{-3}}$. The mean per-MOF normalized MAE is 0.51\% over the complete grid and 0.45\%, 0.41\% and 0.40\% within these subsets, respectively, demonstrating low relative errors in the electronically concentrated regions around the framework.

For ESP, most values lie near the centre of the distribution, while strongly positive and negative potentials occupy smaller spatial regions (Figs.~\ref{fig:dataset_characterization}b and \ref{fig:substructures_fields}c). We therefore separately evaluated the highest and lowest 3\% and 5\% of potential values. The mean per-MOF normalized MAE is 8.42\% over the complete grid. The lowest 3\% and 5\% subsets yield normalized errors of 0.45\% and 0.58\%, respectively, compared with 3.72\% and 3.91\% for the corresponding highest-potential subsets. These results reveal an asymmetry between the two tails and substantial regional variation in normalized reconstruction performance. For both fields, normalization is performed within each regime, so comparisons reflect differences in both residual magnitudes and normalization scales rather than differences in absolute error alone.

Reconstruction performance shows contrasting dependencies on unit-cell volume (Fig.~\ref{fig3}c). ED reconstruction remains stable across the evaluated volume range, with only weak variation in normalized MAE and $R^2$ values close to unity. ESP exhibits a stronger size dependence: reconstruction error increases and $R^2$ decreases with cell volume, while the spread of errors broadens for larger structures.

Real-space reconstructions reveal the corresponding differences in residual patterns (Fig.~\ref{fig3}e). The predicted ED closely reproduces the locations and spatial profiles of high-density features around the framework. Its largest visible residuals occur near these sharply varying features, while errors remain small across the low-density pore space. The predicted ESP recovers the main positive and negative regions but exhibits more spatially extended discrepancies, particularly in high-potential regions farther from framework atoms. The substantially stronger performance for ED may partly reflect its spatial localization: high density is concentrated near atoms and bonds, with values approaching zero through much of the pore space, whereas ESP retains substantial variation away from the framework and depends on the collective nuclear and electronic charge distribution. The current method therefore reconstructs ED reliably, while ESP prediction requires further improvement. These findings motivate new paradigms for ESP prediction that explicitly incorporate longer-range electrostatic contributions and global structural context beyond predominantly local atom--probe message passing.

Despite the remaining challenge of ESP reconstruction, the high fidelity achieved for ED already provides direct practical value for first-principles calculations. To examine whether the reconstructed field is physically useful beyond numerical agreement with the reference density, we used the predicted ED to initialize subsequent self-consistent DFT calculations (Fig.~\ref{fig3}d). Across the evaluated structures, this initialization reduced the average number of SCF iterations by 12.84\% and the average wall-clock time by 13.59\% relative to the standard initialization. The paired scatter plot compares these two percentage reductions for each structure, showing that reductions in iteration count are generally accompanied by reductions in computational time, although the magnitude of the benefit varies across MOFs. The closely matched average reductions are consistent with a decrease in the iterative refinement required to reach self-consistency. This result demonstrates that structure-to-field reconstruction can serve not only as a predictive task, but also as a practical component of electronic-structure workflows by accelerating the convergence of subsequent DFT calculations.

\subsection{Electronic-field representations and property prediction}
\label{sec:pointcloud}
\begin{figure*}[!htbp]
\centering
\includegraphics[width=0.98\textwidth]{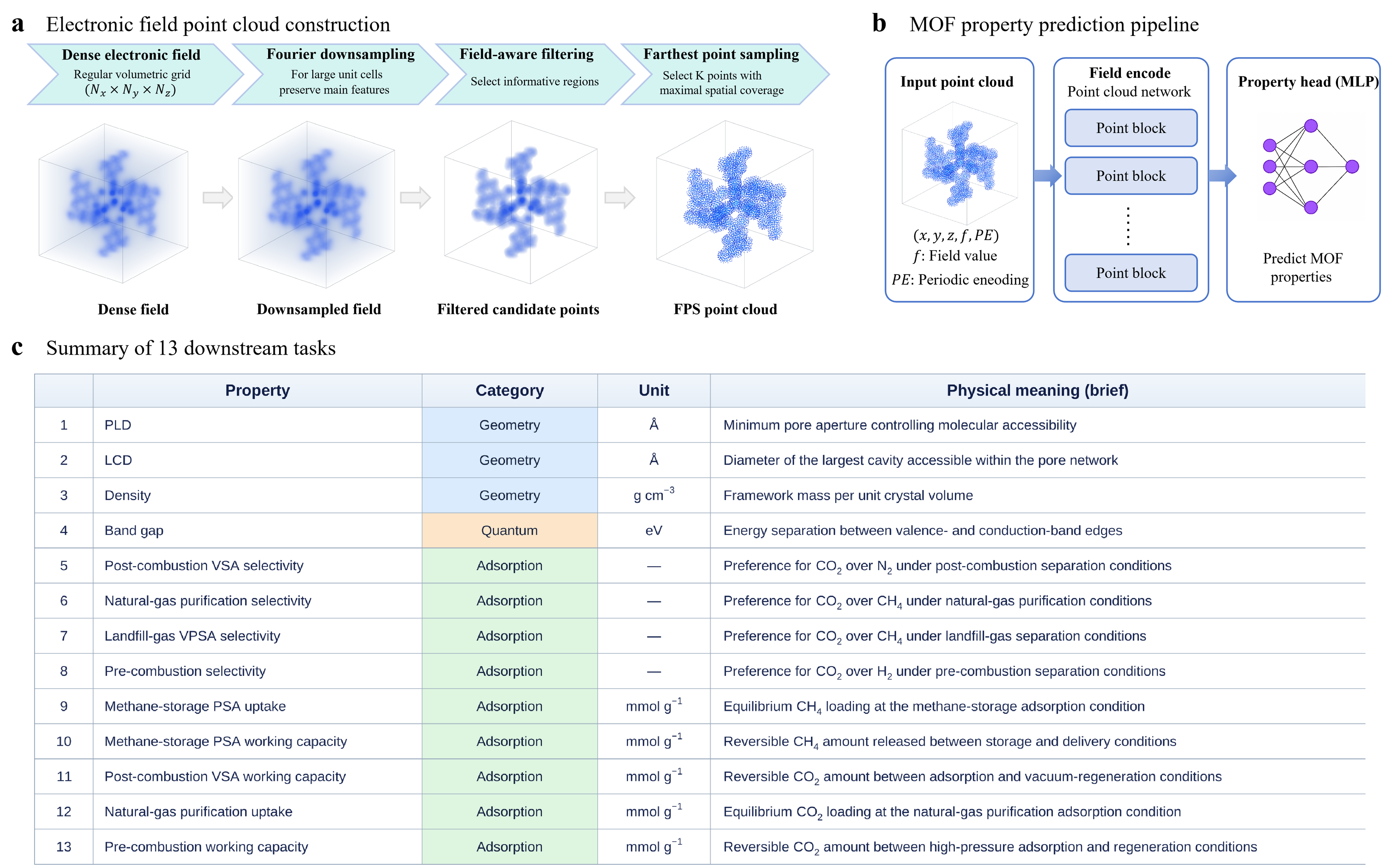}
\caption[Electronic-field point-cloud construction and MOF property prediction.]{
\textbf{Electronic-field point-cloud construction and MOF property prediction.}
\textbf{a,} Construction of electronic-field point clouds.
Large volumetric grids are first downsampled in the Fourier domain, followed by modality-dependent filtering.
FPS is applied to the retained candidates to obtain a fixed-size point cloud.
\textbf{b,} Property-prediction pipeline.
Sampled points combine electronic-field values with spatial features, including periodic sine and cosine encodings of fractional coordinates.
The resulting point cloud is processed by a point-cloud encoder and a regression head.
\textbf{c,} Summary of the 13 prediction tasks, comprising three geometric properties, one electronic property and nine adsorption or separation properties.
Units and brief physical interpretations are listed for each target.
}
\label{fig4}
\end{figure*}

Dense ED and ESP fields provide spatially resolved descriptions of the
electronic environment, but their volumetric form introduces substantial
variation in input size and computational requirements.
As shown in Fig.~\ref{fig:dataset_characterization}, grid size increases
strongly with unit-cell volume, while field values are distributed
non-uniformly across space.
Variations in lattice dimensions and cell geometry also complicate the
construction of a common voxel representation without extensive
resampling or padding.
We therefore convert each dense field into a fixed-size point cloud
before downstream learning (Fig.~\ref{fig4}a).
Each point retains its spatial position and local field value, providing
a common input format across frameworks with different unit-cell sizes
and geometries.

Point-cloud construction combines grid reduction, modality-dependent filtering and spatial sampling. For particularly large grids, Fourier downsampling first reduces the candidate point pool. Filtering is guided by the contrasting field-value distributions (Fig.~\ref{fig:dataset_characterization}b). ED is concentrated around atoms and bonds, while much of the pore space has near-zero density. Removing points below a prescribed threshold therefore directs the sampling budget towards higher-density regions associated with the framework. ESP values cluster within a relatively narrow central range, with less populated high- and low-potential tails. Selecting candidates from both tails prioritizes regions with pronounced potential deviations, allowing contrasting electrostatic environments to be represented within a limited point budget. Farthest point sampling (FPS) is then applied to the retained candidates to obtain a fixed number of spatially dispersed points, controlling input size while providing spatial coverage of the selected field regions.

To encode crystal periodicity, sampled positions are expressed in
fractional coordinates and supplemented with periodic sine and cosine
features (Fig.~\ref{fig4}b).
Periodically equivalent coordinates receive identical periodic
encodings, providing positional information that is continuous across
unit-cell boundaries.
The resulting point features combine spatial information with the
corresponding ED or ESP value and are processed by a point-cloud encoder
and regression head for property prediction.

The representations were evaluated on 13 properties comprising three
geometric targets, one electronic target and nine adsorption or
separation targets (Fig.~\ref{fig4}c).
The geometric tasks include pore limiting diameter (PLD), largest cavity
diameter (LCD) and crystal density, while band gap probes electronic
properties.
The remaining tasks cover gas selectivity, uptake and working capacity
under different storage and separation conditions.
These targets provide a common benchmark for comparing CIF, ED and ESP
across properties associated with pore geometry, electronic structure
and guest--framework environments.

\begin{figure*}[!htbp]
\centering
\includegraphics[width=0.98\textwidth]{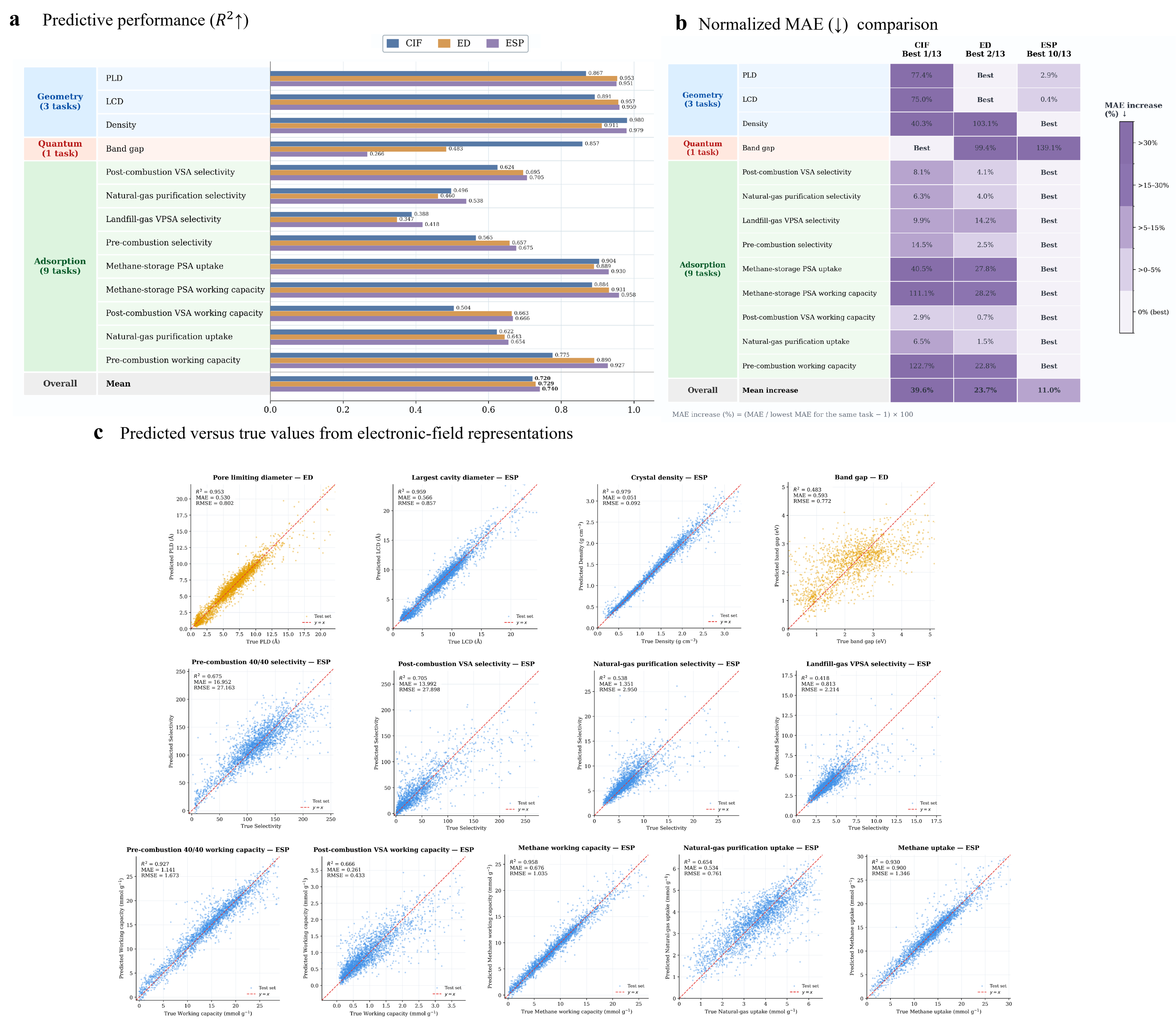}
\caption[Task-dependent predictive performance of atomic and electronic-field representations.]{
\textbf{Task-dependent predictive performance of atomic and electronic-field representations.}
\textbf{a,} Test-set \(R^2\) for CIF, ED and ESP across 13 properties.
The bottom row shows the  mean across tasks.
\textbf{b,} Percentage MAE increases relative to the lowest-MAE modality
for each task. Lower values indicate better accuracy.
The lowest-MAE modality is labelled ``Best''.
Column headings show task wins, and the bottom row shows the
 mean increase
\textbf{c,} Test-set parity plots for the electronic-field modality
indicated in each panel. Dashed lines denote \(y=x\).
}
\label{fig5}
\end{figure*}

Predictive performance varies across target properties and evaluation metrics (Fig.~\ref{fig5}a,b). ESP achieves the lowest MAE for 10 of the 13 tasks, including density and all nine adsorption or separation properties. ED achieves the lowest MAE for PLD and LCD, while CIF performs best for band gap. The unweighted mean \(R^2\) across tasks is 0.720 for CIF, 0.729 for ED and 0.740 for ESP. To compare errors across targets with different units, MAE is expressed as its percentage increase over the lowest MAE for each task, with the best-performing modality assigned \(0\%\) (Fig.~\ref{fig5}b). The mean relative MAE increase across the 13 tasks is \(39.6\%\) for CIF, \(23.7\%\) for ED and \(11.0\%\) for ESP. ESP achieves the strongest overall predictive performance across the evaluated tasks. 

Electronic-field point clouds perform particularly well for pore dimensions. The highest field-model \(R^2\) values are 0.953 for PLD and 0.959 for LCD, exceeding the corresponding CIF values of 0.867 and 0.891 (Fig.~\ref{fig5}a). For density, ESP achieves the lowest MAE, approximately \(0.051\,\mathrm{g\,cm^{-3}}\), while CIF and ESP attain similar \(R^2\) values of 0.980 and 0.979, respectively. Thus, even for a single target, the lowest MAE need not coincide with the highest \(R^2\).

Band-gap prediction instead favours CIF, which achieves \(R^2=0.857\), compared with 0.483 for ED and 0.266 for ESP. The ED and ESP MAEs are \(99.4\%\) and \(139.1\%\) higher than that of CIF, respectively (Fig.~\ref{fig5}b). One possible explanation is the absence of explicit elemental identities in the current ED and ESP point-cloud inputs. Elemental composition and the associated orbital energies and hybridization influence the band gap, and the CIF model can use atom-type features together with their structural environments as direct chemical cues. The ED and ESP models must instead infer these distinctions indirectly from sampled field values and positional features, which may make the relevant chemical information less accessible under the present sampling and encoding scheme.

Across all nine adsorption and separation tasks, ESP achieves both the highest \(R^2\) and the lowest MAE (Fig.~\ref{fig5}a,b). For methane-storage PSA working capacity, \(R^2\) increases from 0.884 with CIF to 0.958 with ESP; for pre-combustion working capacity, the corresponding values are 0.775 and 0.927. The CIF MAEs for these tasks are \(111.1\%\) and \(122.7\%\) higher than the ESP MAEs, respectively. ED also improves upon CIF for several targets, although its performance is less consistent across the adsorption tasks. These results support the utility of spatially resolved electrostatic features for the gas properties considered here, with the magnitude of the benefit varying across targets and metrics.


The contrasting performance of structural and electronic-field models motivates the multimodal experiments in Section~\ref{sec:multimodal}, which assess whether combining these features improves prediction beyond the strongest unimodal model.

\subsection{Multimodal learning and cross-modal retrieval}
\label{sec:multimodal}

\begin{figure*}[!htbp]
\centering
\includegraphics[width=0.98\textwidth]{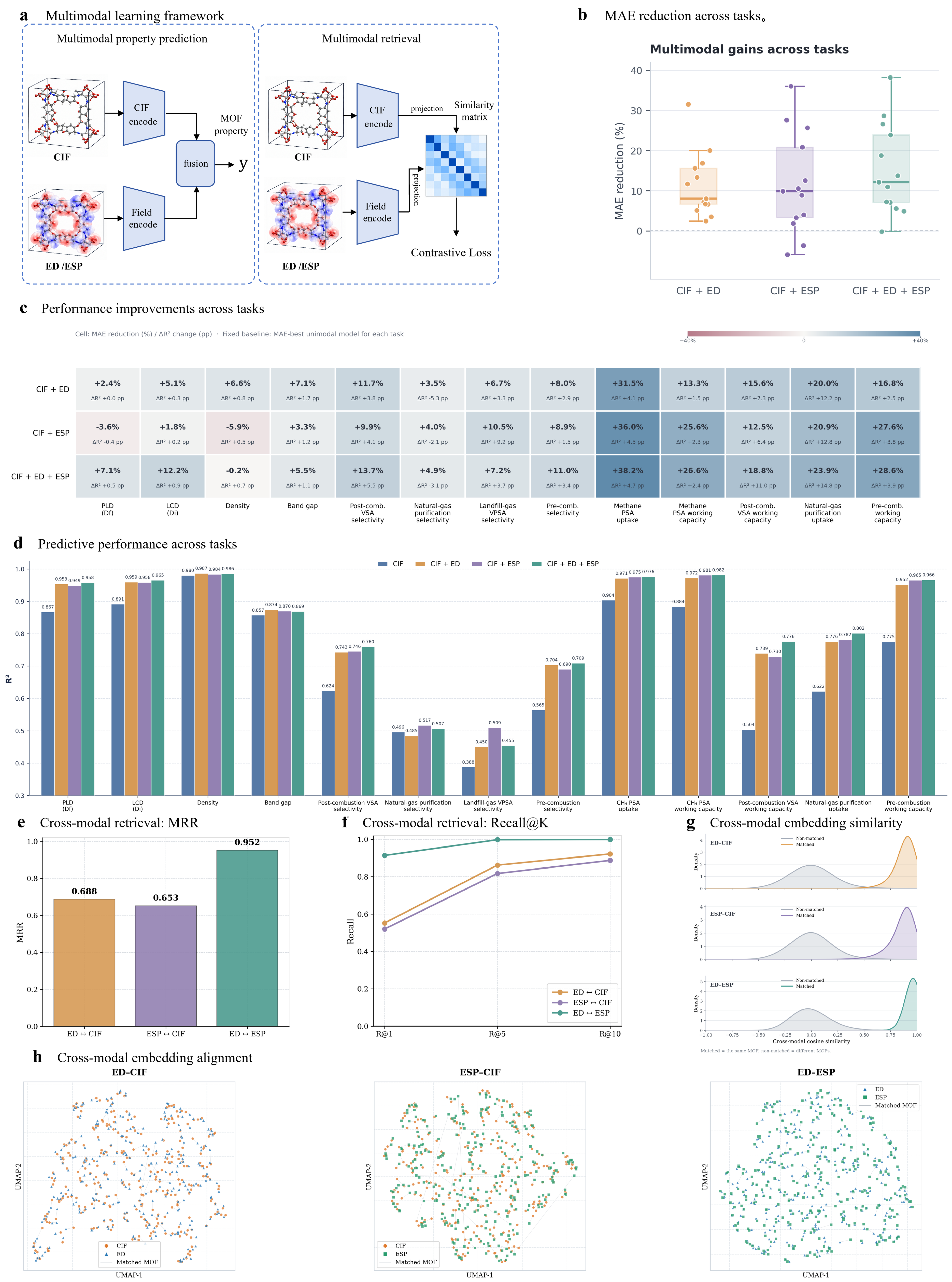}
\caption{
\textbf{Multimodal property prediction and cross-modal retrieval.}
\textbf{a,} Feature fusion for property prediction (left) and pairwise contrastive retrieval (right).
\textbf{b,} Relative MAE reductions for CIF+ED, CIF+ESP and CIF+ED+ESP across 13 tasks.
Points denote tasks; boxes, central lines and whiskers indicate interquartile ranges, medians and observed ranges.
\textbf{c,} Task-wise MAE reductions (colours and large labels) and changes in \(R^2\) (small labels).
Both metrics in \textbf{b,c} use the lowest-MAE unimodal model for each task as the reference.
\textbf{d,} \(R^2\) comparisons between CIF and fusion models.
\textbf{e,f,} MRR (\textbf{e}) and Recall@\(K\) (\textbf{f}; \(K=1,5,10\)) for ED\(\leftrightarrow\)CIF, ESP\(\leftrightarrow\)CIF and ED\(\leftrightarrow\)ESP retrieval.
\textbf{g,} Cosine similarities for matched (same MOF, coloured) and unmatched (different MOFs, grey) pairs.
\textbf{h,} UMAP projections: CIF, orange circles; ED, blue triangles; ESP, green squares.
Grey lines connect matched embeddings.
}
\label{fig6}
\end{figure*}

Multimodal learning combines atomic and electronic-field representations
for property prediction and evaluates whether the same MOF can be
identified across modalities (Fig.~\ref{fig6}a). Property prediction
considers CIF+ED, CIF+ESP and CIF+ED+ESP, while cross-modal retrieval
comprises three independently trained pairwise tasks:
ED\(\leftrightarrow\)CIF, ESP\(\leftrightarrow\)CIF and
ED\(\leftrightarrow\)ESP.

For property prediction, modality-specific encoders extract features
from CIF, ED and ESP, which are fused and passed to a regression head.
For each task and modality combination, the lowest-MAE configuration
among the three evaluated fusion strategies is reported together
with the \(R^2\) of that same configuration. MAE reductions and
\(\Delta R^2\) are both calculated relative to the unimodal model
with the lowest MAE for the corresponding task.

Multimodal fusion improves prediction across the 13 benchmark
properties, which span pore geometry, density, band gap, gas
selectivity, uptake and working capacity (Fig.~\ref{fig6}b--d).
CIF+ED reduces MAE for all 13 targets, while CIF+ESP and
CIF+ED+ESP improve 11 and 12 targets, respectively. Three-modal
fusion achieves the lowest MAE among the three modality
combinations for 10 properties (Fig.~\ref{fig6}c).
The largest reductions occur for adsorption-related targets.
For methane PSA uptake, MAE reductions range from \(31.5\%\)
for CIF+ED to \(38.2\%\) for CIF+ED+ESP. Three-modal fusion also
reduces MAE by \(26.6\%\) for methane PSA working capacity,
\(23.9\%\) for natural-gas purification uptake and \(28.6\%\)
for pre-combustion working capacity.

Improvements also extend to geometric and electronic properties.
CIF+ED reduces the MAE for density by \(6.7\%\) and for band gap
by \(7.1\%\), with the latter exceeding the \(5.6\%\) reduction
obtained with CIF+ED+ESP. For landfill-gas VPSA selectivity,
CIF+ESP yields a \(10.5\%\) MAE reduction, compared with
\(7.2\%\) for CIF+ED+ESP.

The \(R^2\) comparisons against CIF alone provide a complementary
view of these gains (Fig.~\ref{fig6}d). Three-modal fusion increases
\(R^2\) from 0.504 to 0.776 for post-combustion VSA working capacity
and from 0.622 to 0.802 for natural-gas purification uptake.
For density, all three multimodal combinations achieve higher
\(R^2\) than CIF alone, although CIF+ESP and CIF+ED+ESP do not
reduce MAE relative to the best unimodal baseline. MAE and
\(R^2\) improvements are not always concurrent: for natural-gas
purification selectivity, all three combinations reduce MAE
while yielding lower \(R^2\) than the MAE-selected unimodal
reference (Fig.~\ref{fig6}c).

ED and ESP supplement the CIF representation with explicit
information about electron distributions and electrostatic
environments around the framework and within the pores.
Combining these field features with atomic identities and
structural geometry provides complementary inputs for learning
MOF structure--property relationships. The broad improvements
across the benchmark support electronic-field representations
as useful additions to structure-based property prediction,
with particularly substantial gains for adsorption-related
properties.

Cross-modal retrieval examines whether a model can use one representation
of a MOF to identify its corresponding representation in another modality
(Fig.~\ref{fig6}e,f).
Three independently trained pairwise tasks,
ED\(\leftrightarrow\)CIF, ESP\(\leftrightarrow\)CIF and
ED\(\leftrightarrow\)ESP, evaluate this capability through bidirectional
matching in learned embedding spaces.
These tasks assess whether models can associate atomic structures with
their electronic environments and recognize the same material across
different representations.

ED\(\leftrightarrow\)ESP exhibits the strongest matching performance,
achieving an MRR of 0.952, with Recall@5 and Recall@10 approaching unity.
Matching CIF with either field is more challenging, indicating that
the correspondence between the two electronic-field representations
is more readily learned under the evaluated models.
The cosine-similarity distributions support this pattern:
matched pairs have higher similarities than non-matched pairs,
and matched ED--ESP pairs concentrate most tightly near unity
(Fig.~\ref{fig6}g).
UMAP projections provide a complementary qualitative view,
with representations of the same MOF frequently appearing nearby
across modalities, particularly for ED--ESP
(Fig.~\ref{fig6}h).

The two retrieval directions offer complementary routes to accessing
material information.
CIF-to-field retrieval uses an atomic structure to locate its
corresponding ED or ESP record, providing access to precomputed
electronic information for further analysis.
Conversely, field-to-CIF retrieval identifies framework structures
from electronic-field queries, providing a basis for searching
MOF collections through their electronic environments.
ED--ESP retrieval further connects electron distributions with
their corresponding electrostatic environments, allowing either
field to serve as a query for the other.

Beyond matching paired records, these capabilities provide a foundation
for electronic-information-guided materials search.
Future extensions could examine whether queries specifying desired
electron-density or electrostatic environments can retrieve suitable
framework candidates, supporting virtual screening and retrieval-assisted
inverse design.
MOFFieldBench makes cross-modal recognition a measurable benchmark
objective alongside reconstruction and property prediction, enabling
the development of models that connect structural and electronic
information for materials discovery.

\section{Discussion}
\label{sec:discussion}

MOFFieldBench establishes an aligned data and evaluation setting for
learning from periodic structures and continuous electronic fields in
MOFs. Building on advances in equivariant charge-density
reconstruction~\cite{koker2024charge3net}, it connects field
reconstruction, property prediction and cross-modal matching within a
common material collection. The paired CIF, ED and ESP references
support comparisons across different scientific objectives while
preserving material correspondence. These reference pairs and task
definitions provide a basis for reproducible comparisons as
reconstruction methods, field encoders and multimodal models evolve.

ED and ESP differ substantially in 
reconstruction difficulty under the evaluated 
architecture. 
The local equivariant model recovers ED accurately across the evaluated
density regimes and unit-cell sizes, whereas ESP reconstruction
deteriorates for larger cells. This pattern is consistent with high ED
being concentrated near atoms and bonds and near-zero ED across much
of the pore space, while electrostatic potential reflects more
spatially extended contributions. These observations motivate new ESP
reconstruction strategies that combine local chemical environments
with broader structural context or explicit periodic electrostatic
constraints. MOFFieldBench provides common reference fields for
comparing these approaches across cell sizes and field regimes and
for testing how reconstruction quality affects downstream applications.

Predicted-ED initialization provides a concrete example of
computational reuse. Mean reductions of \(12.84\%\) in SCF iterations
and \(13.59\%\) in subsequent DFT wall-clock time show that learned
fields can assist iterative quantum calculations while retaining the
numerical solver for refinement. This experiment gives reconstruction
a practical role beyond reproducing reference grids. Future studies
can test predicted-ED initialization across additional MOF chemistries
and electronic-structure settings, extending the benchmark's evaluation
of learned fields within first-principles workflows.

The property-prediction results reveal distinct strengths of the three
representations. ESP achieves the lowest MAE for 10 of 13 properties,
ED achieves the lowest MAE for the two pore-diameter targets, and CIF
retains a clear advantage for band gap. The adsorption results suggest
that spatially resolved electrostatic features are useful for predicting
guest--framework behaviour. The CIF advantage for band gap may partly
reflect its explicit atom identities and structural environments,
whereas the current field models infer chemical distinctions from
sampled values and positions. Future field models could incorporate
explicit elemental context, while extending the task set would test
whether the observed strengths generalize to other electronic and
adsorption properties.

Multimodal fusion highlights the benefit of integrating structural
and electronic information for MOF property prediction.
ED and ESP provide spatial descriptions of electron distributions
and electrostatic environments that complement the atomic identities
and geometry encoded by CIF.
Multimodal fusion reduces MAE across all 13 benchmark targets, with
at least one modality combination outperforming the lowest-MAE
unimodal reference for each property.
The pronounced gains for adsorption-related targets suggest that
jointly representing framework structure and electronic environments
is useful for learning properties associated with guest--framework
interactions.
Future work could develop fusion methods that learn interactions
between atomic features and local field regions, adaptively weight
electronic information and integrate features across spatial scales.
MOFFieldBench provides aligned structural and electronic-field
references for evaluating these developments against the strongest
unimodal baselines and extending multimodal learning to additional
material properties.

Pairwise cross-modal retrieval shows that a query in one modality can
identify the corresponding representation of the same MOF in another,
with particularly strong ED--ESP matching.
The two retrieval directions offer complementary uses:
structural queries provide access to precomputed ED or ESP records
for electronic-field analysis, while electronic-field queries identify
their associated framework structures.
This capability connects structural and electronic information within
searchable materials collections.
Future evaluations could extend retrieval to larger candidate
collections and previously unseen chemical or topological families.
Moving beyond exact identity matching to finding MOFs with similar
electronic environments could support virtual screening guided by
electron-density or electrostatic features, providing a foundation
for retrieval-assisted inverse design.
MOFFieldBench supplies aligned references for developing and evaluating
these approaches to materials search.

Error-bounded compression reduces the storage required for ED and ESP
data by approximately \(26\times\), making the reference fields easier
to store and reuse. Field-aware filtering and spatial sampling convert
these fields into fixed-size point clouds for downstream learning.
The same electronic-field data can support different sampling
strategies, point counts and point-cloud encoders without repeating
the underlying DFT calculations. This flexibility allows MOFFieldBench
to support future developments in electronic-field point-cloud
representations and learning methods.

\subsection{Limitations}
\label{sec:limitations}

While MOFFieldBench provides an aligned setting for evaluating
electronic-field learning, several limitations remain. First, the
benchmark currently focuses on MOFs, and its applicability to other
porous materials, such as covalent organic frameworks and zeolites,
remains to be established. Second, the reference fields are specific
to the adopted computational protocol, including pseudopotential
valence densities for ED, and transfer across electronic-structure
settings requires further evaluation. Third, fixed point counts and
modality-dependent filtering retain only selected field regions.
Low-density ED regions and ESP values near the centre of the
distribution that are excluded during filtering may still be relevant
to certain targets, motivating adaptive sampling strategies that can
be evaluated using the same electronic-field references. Finally, the
property benchmark covers 13 targets centred on geometry, band gap
and adsorption. Extending the task set to additional electronic and
functional properties, together with downstream tests using
reconstructed fields, would provide a broader assessment of
electronic-field representations.

\section{Methods}

\subsection{Dataset construction and electronic-field generation}

MOFFieldBench comprises 45,909 metal--organic framework structures
collected from six sources: 24,908 from
ARC-MOF, 15,435 from
QMOF, 688 from
CoRE MOF, 1,955 from the Cambridge Structural
Database (CSD), 1,736 from the database of hypothetical
MOFs constructed using PORMAKE, and 1,187 from the Ultrastable
MOF database.
Each retained structure is associated with a crystallographic
structure in CIF format and corresponding electron-density (ED)
and electrostatic-potential (ESP) fields within the same periodic
unit cell.The electronic-structure calculations required
approximately \(10^{6}\) CPU core-hours.

Periodic single-point density-functional theory calculations were
performed on the supplied structures without additional geometry
optimization, using the Quickstep module of
CP2K~\cite{wilhelm2026cp2k} within the Gaussian and plane-wave
(GPW) framework.
Calculations employed \texttt{RUN\_TYPE ENERGY} and the
Perdew--Burke--Ernzerhof (PBE) exchange--correlation
functional~\cite{perdew1996generalized}.
Element-specific \texttt{DZVP-MOLOPT-SR-GTH} basis sets from
the \texttt{BASIS\_MOLOPT} library~\cite{vandevondele2007}
were used together with \texttt{GTH-PBE}
pseudopotentials~\cite{Goedecker1996}.
The auxiliary plane-wave cutoff and relative cutoff were set
to 350 and 50 Ry, respectively.
Three-dimensional periodic boundary conditions and
\(\Gamma\)-point sampling were used throughout.
No empirical dispersion correction was included.

Self-consistent-field (SCF) calculations used a
diagonalization-based procedure with a maximum of 128 iterations,
\texttt{EPS\_SCF} \(=5\times10^{-6}\) and
\texttt{EPS\_DEFAULT} \(=1\times10^{-11}\).
All systems were treated as charge neutral.
Structure-specific spin multiplicities were assigned, with
unrestricted Kohn--Sham calculations used for open-shell systems.
Electronic smearing and orbital transformation were not used
in the standard workflow.
Complete calculation templates, element-specific basis and
pseudopotential assignments, and structure-specific spin settings
are provided in the Supplementary Methods.

After SCF convergence, ED was exported using
\texttt{E\_DENSITY\_CUBE}.
Under the GPW framework with GTH pseudopotentials, this output
represents the valence electron density.
ESP fields were derived from the \texttt{V\_HARTREE\_CUBE}
output associated with the electronic and ionic charge
distributions.
Both outputs used \texttt{STRIDE 1}, retaining the native
CP2K volumetric grids over the complete periodic unit cell.

Grid dimensions, origins and grid vectors were parsed from
the Cube-file headers and checked against the corresponding
crystallographic cells.
Each CIF and its associated ED and ESP fields were linked
through a common material identifier.
Only records with converged electronic-structure calculations,
valid paired field outputs and consistent periodic-cell
definitions were retained.
Records with failed SCF convergence, missing field outputs
or invalid grid geometry were excluded.

\subsection{Electronic-field compression and point-cloud construction}
Electronic fields were processed separately for storage and model input.
Storage compression operated directly on the native ED and ESP grids
using prescribed nominal absolute quantization tolerances.
For property prediction and cross-modal retrieval, the decompressed
fields were converted into fixed-size point clouds through Fourier
downsampling, modality-specific filtering and farthest point sampling
(FPS).

For storage, CP3D (Compress 3D) was used as a predictive compressor
for three-dimensional fields with configurable nominal absolute
quantization tolerances.
Let \(f_{i,j,k}\) denote the original field value at grid indices
\((i,j,k)\), and \(p_{i,j,k}\) its predicted value.
The prediction was calculated from previously reconstructed
neighbouring values using the three-dimensional Lorenzo
predictor~\cite{https://doi.org/10.1111/1467-8659.00681}:
\begin{equation}
\begin{aligned}
p_{i,j,k}={}&
\hat f_{i-1,j,k}
+\hat f_{i,j-1,k}
+\hat f_{i,j,k-1}\\
&{}-\hat f_{i-1,j-1,k}
-\hat f_{i-1,j,k-1}
-\hat f_{i,j-1,k-1}
+\hat f_{i-1,j-1,k-1}.
\end{aligned}
\end{equation}
Here, \(\hat f\) denotes a reconstructed field value.
Using the same reconstructed neighbours during compression and
decompression ensured identical predictions at each grid point.

For a nominal absolute quantization tolerance $\varepsilon$,
residual quantization and reconstruction were defined as
\begin{equation}
\begin{aligned}
q_{i,j,k}
&=
\operatorname{round}\!\left(
\frac{f_{i,j,k}-p_{i,j,k}}{2\varepsilon}
\right),\\[3pt]
\hat f_{i,j,k}
&=
p_{i,j,k}+2\varepsilon q_{i,j,k}.
\end{aligned}
\end{equation}
In exact arithmetic, the quantization rule above yields
$|f_{i,j,k}-\hat f_{i,j,k}|\leq\varepsilon$.
The quantized residuals were encoded losslessly, while values
outside the quantization range were stored explicitly.
The implementation used nominal absolute quantization tolerances of
$\varepsilon_{\mathrm{ED}}=
10^{-6}\,\mathrm{e\,Bohr^{-3}}$ and
$\varepsilon_{\mathrm{ESP}}=
10^{-4}\,\mathrm{Hartree}$.
Measured maximum reconstruction errors are reported in
Supplementary Table~S4.
Decompression retained the original grid dimensions and spatial
registration without resampling.
Compression errors were evaluated on the decompressed native grids
before Fourier downsampling or point-cloud sampling.

For point-cloud construction, Fourier downsampling was applied
to fields containing more than \(13\times10^{6}\) grid points.
Let \(X\in\mathbb{R}^{N_x\times N_y\times N_z}\) denote a
decompressed field and \((M_x,M_y,M_z)\) the target grid dimensions.
The resampled field was obtained as
\begin{equation}
\widetilde X =
\frac{M_xM_yM_z}{N_xN_yN_z}
\mathcal{F}^{-1}_{M_x,M_y,M_z}
\left[
\mathcal{C}_{M_x,M_y,M_z}
\left(\mathcal{F}_{N_x,N_y,N_z}[X]\right)
\right],
\end{equation}
where \(\mathcal{F}\) and \(\mathcal{F}^{-1}\) denote the
three-dimensional discrete Fourier transform and inverse transform,
and \(\mathcal{C}\) retains the low-frequency coefficients
compatible with the target grid.
The prefactor preserves the mean field value under the change
in grid dimensions.
The reduction factor was selected according to the native grid
size, with stronger downsampling applied to larger fields.
Fields below the grid-size threshold were retained at their
native resolution.

Modality-specific filtering was subsequently applied to construct
the sampling candidates.
For ED, points with density above a threshold \(\tau_\rho\)
were retained.
We evaluated
\(\tau_\rho\in\{0.02,0.03,0.05,0.08,0.10\}\,
\mathrm{e\,Bohr^{-3}}\).
For ESP, points were retained from both the lower and upper tails
of the potential distribution.
The fraction retained from each tail was varied over
\(3\%\), \(5\%\) and \(7\%\), corresponding to
\(\alpha\in\{0.03,0.05,0.07\}\).
The candidate sets were defined as
\begin{equation}
\begin{aligned}
\mathcal{C}_{\mathrm{ED}}
&=
\left\{
\mathbf r_i:
\widetilde\rho(\mathbf r_i)>\tau_\rho
\right\},\\
\mathcal{C}_{\mathrm{ESP}}
&=
\left\{
\mathbf r_i:
\widetilde V(\mathbf r_i)\leq Q_\alpha
\ \text{or}\
\widetilde V(\mathbf r_i)\geq Q_{1-\alpha}
\right\}.
\end{aligned}
\end{equation}
Here, \(\widetilde\rho\) and \(\widetilde V\) denote the fields
after optional Fourier resampling, and \(Q_\alpha\) and
\(Q_{1-\alpha}\) are the corresponding quantiles of the ESP
distribution for each individual MOF.
ED filtering therefore removes low-density regions, whereas ESP
filtering retains both the highest- and lowest-potential environments.

FPS~\cite{gonzalez1985clustering,eldar1994farthest} was applied
to each candidate set to select a prescribed number of points \(N\).
We evaluated \(N\in\{4096,6144,8192,10240\}\).
Let \(\mathcal C\) denote the candidate set and \(S_t\) the
points selected after \(t\) iterations.
The next point was chosen as
\begin{equation}
\mathbf r_{t+1}
=
\operatorname*{arg\,max}_{\mathbf r\in\mathcal C\setminus S_t}
\min_{\mathbf q\in S_t}
\|\mathbf r-\mathbf q\|_2,
\end{equation}
where \(\|\mathbf r-\mathbf q\|_2\) is the Euclidean distance
between Cartesian coordinates.
Sampling continued until \(N\) points were selected.
Each point retained its spatial position and corresponding field
value, yielding a fixed-size input across MOFs with different
lattice dimensions and grid sizes.
ED and ESP point clouds were constructed separately using their
respective filtering criteria.
The grid-reduction factors and task-specific choices of filtering
thresholds and point counts are specified in the Supplementary Methods.

\subsection{Electronic-field reconstruction}

Electronic-field reconstruction was formulated as a probe-based
scalar regression task.
Given a periodic MOF structure
\(\mathcal S=\{(Z_i,\mathbf r_i)\}_{i=1}^{N_{\mathrm{atom}}}\),
where \(Z_i\) and \(\mathbf r_i\) denote atomic species and
Cartesian coordinates, together with the lattice matrix \(\mathbf A\),
the model predicts the field value at a spatial probe position
\(\mathbf r_p\):
\begin{equation}
\hat f_\theta(\mathbf r_p)
=
F_\theta(\mathcal S,\mathbf A,\mathbf r_p).
\end{equation}
Here, \(f\) denotes either ED, \(\rho\), or ESP, \(V\), and
\(\theta\) represents the model parameters.
Separate models were trained for ED and ESP using the same
general architecture.
Evaluating the model at spatial query points allows reconstruction
of fields with different grid dimensions and unit-cell geometries.

The reconstruction network was based on the probe-based
E(3)-equivariant architecture of
ChargE3Net~\cite{koker2024charge3net}.
Periodic atomic environments were represented using an atom--atom
graph and an atom--probe graph, with lattice-translated atomic
images included in both graphs.
For a probe at \(\mathbf r_p\), the neighbouring atoms were
defined as
\begin{equation}
\mathcal N(p)
=
\left\{
(i,\mathbf n):
\left\|\mathbf r_i+\mathbf A\mathbf n-\mathbf r_p\right\|_2<r_c,
\quad \mathbf n\in\mathbb Z^3
\right\},
\end{equation}
where \(r_c\) is the radial cutoff and \(\mathbf n\) indexes
periodic lattice translations.
This construction includes neighbouring atoms across unit-cell
boundaries according to their periodic separation.

Atomic species were first embedded and updated through
equivariant message passing on the atom--atom graph.
The resulting atomic features were transferred to probe nodes
through atom--probe interactions.
Probe nodes served as spatial query locations and carried no
elemental identity.
An invariant scalar readout mapped each probe representation to
the corresponding ED or ESP value.
Architectural parameters, including interaction layers, radial
basis functions, irreducible representations and cutoff settings,
are provided in the Supplementary Methods.

Training used reference field values at probe positions sampled
from the volumetric grids.
The reconstruction objective was the mean absolute error over
the sampled probes:
\begin{equation}
\mathcal L_{\mathrm{rec}}
=
\frac{1}{N_{\mathrm{probe}}}
\sum_{p=1}^{N_{\mathrm{probe}}}
\left|
\hat f_\theta(\mathbf r_p)-f(\mathbf r_p)
\right|,
\end{equation}
where \(N_{\mathrm{probe}}\) is the number of probe values
included in the loss.
At inference, atomic representations were computed once for each
structure and reused while evaluating probes in batches at the
native CP2K grid locations.
The predictions were assembled into volumetric fields with the
same grid dimensions and spatial registration as the references.

Reconstruction accuracy was evaluated on held-out structures
using MAE, normalized MAE and \(R^2\).
In addition to full-grid evaluation, errors were examined within
reference ED regions satisfying
\(\rho>0.01\), \(0.05\) and
\(0.10\,\mathrm{e\,Bohr^{-3}}\), and within the lowest and
highest \(3\%\) and \(5\%\) of reference ESP values.
For the regime-resolved comparisons, normalized errors were
calculated separately for each MOF within the corresponding
field subset and subsequently averaged across the evaluated
structures.
Errors were also analysed as a function of unit-cell volume
and distance to the nearest framework atom.

To evaluate the use of reconstructed ED for SCF initialization,
the predicted density was supplied in place of the default
initial density input.
For each evaluated MOF, the model was queried at the required
grid positions to generate the initial density,
\(\rho^{(0)}(\mathbf r)=\hat\rho_\theta(\mathbf r)\).
The SCF procedure then iteratively refined this density to
convergence.
A calculation using predicted-ED initialization was compared
with a calculation using the standard initialization for the
same structure.
The paired calculations used the same electronic-structure
settings and convergence criteria, with the initial density
as the changed input.

SCF acceleration was quantified through paired reductions in
iteration count and wall-clock time.
For MOF \(m\), these reductions were defined as
\begin{equation}
\begin{aligned}
\Delta N^{(m)}
&=
100\left(
1-\frac{N_{\mathrm{pred}}^{(m)}}
        {N_{\mathrm{std}}^{(m)}}
\right),\\
\Delta T^{(m)}
&=
100\left(
1-\frac{T_{\mathrm{pred}}^{(m)}}
        {T_{\mathrm{std}}^{(m)}}
\right),
\end{aligned}
\end{equation}
where \(N\) and \(T\) denote the SCF iteration count and
wall-clock time, respectively, and the subscripts
\(\mathrm{pred}\) and \(\mathrm{std}\) indicate predicted-ED
and standard initialization.
Positive values indicate reductions relative to the standard
initialization.
The reported mean reductions were obtained by averaging the
per-MOF percentage reductions across the evaluated structures.

\subsection{Property prediction and evaluation}

Property prediction was formulated as supervised regression from
a MOF representation to a scalar material property.
We evaluated 13 targets comprising three geometric properties,
one electronic property and nine adsorption or separation
properties.
The geometric targets were pore limiting diameter, largest cavity
diameter and crystal density, while the electronic target was
band gap.
The remaining targets described gas uptake, working capacity
and selectivity under the conditions specified for each task.
Three unimodal inputs were compared: crystallographic structures
(CIF), electron density (ED) and electrostatic potential (ESP).

For CIF, a periodic crystal graph containing atomic identities,
coordinates and lattice information was processed by a structure
encoder to obtain a material-level embedding
\(\mathbf z_{\mathrm{CIF}}\).
For ED and ESP, the point clouds constructed using the procedure
described above were processed by point-cloud encoders.
A regression head mapped each material-level embedding to the
target property.
ED and ESP used the same point-cloud backbone architecture
but were trained as separate models.

Electronic-field inputs were augmented with periodic positional
features.
Let \(\mathbf s_i=(u_i,v_i,w_i)\in[0,1)^3\) denote the fractional
coordinates of sampled point \(i\), and let \(\mathbf A\) be
the lattice matrix.
Its Cartesian position is \(\mathbf r_i=\mathbf A\mathbf s_i\).
A \(k\)-order Fourier encoding was constructed as
\begin{equation}
\begin{aligned}
\phi_k(t)
&=
\left[
\sin(2\pi\ell t),\cos(2\pi\ell t)
\right]_{\ell=1}^{k},\\
\Phi_k(\mathbf s_i)
&=
\left[
\phi_k(u_i),\phi_k(v_i),\phi_k(w_i)
\right]
\in\mathbb R^{6k}.
\end{aligned}
\end{equation}
The point feature was
\(\mathbf h_i=[f_i,\Phi_k(\mathbf s_i)]\in\mathbb R^{6k+1}\),
where \(f_i\) is the sampled ED or ESP value.
Cartesian coordinates were retained separately to describe
the physical geometry of the point cloud.
The positional encoding is periodic under integer translations
of the fractional coordinates.
The encoder received the point set
\(\{(\mathbf r_i,\mathbf h_i)\}_{i=1}^{N}\), where \(N\)
is the prescribed point-cloud size.

To assess the influence of periodic positional encoding, we evaluated
\(k\in\{0,1,2,3\}\).
For \(k=0\), the Fourier features were omitted and
\(\mathbf h_i=[f_i]\), while Cartesian coordinates were retained
as the geometric input.
The nonzero orders used the corresponding number of sine--cosine
frequency pairs for each fractional-coordinate component.

Multimodal property prediction combined the material-level
embeddings from modality-specific encoders through feature
concatenation, a fusion network and a regression head.
We evaluated CIF+ED, CIF+ESP and CIF+ED+ESP across the
same 13 targets.
For a selected modality configuration \(\mathcal M\),
the prediction was defined as
\begin{equation}
\begin{aligned}
\mathbf z_{\mathrm{fusion}}
&=
f_{\mathrm{fusion}}
\left([\mathbf z_m]_{m\in\mathcal M}\right),\\
\hat y_{\mathrm{multi}}
&=
g_{\mathrm{multi}}(\mathbf z_{\mathrm{fusion}}),
\end{aligned}
\end{equation}
where \([\mathbf z_m]_{m\in\mathcal M}\) denotes concatenation
of the encoder embeddings for the selected modalities.

Both unimodal and multimodal property-prediction models were
trained using mean squared error (MSE).
For a minibatch containing \(B\) materials, the regression
objective was
\begin{equation}
\mathcal L_{\mathrm{reg}}
=
\frac{1}{B}\sum_{i=1}^{B}(\hat y_i-y_i)^2,
\end{equation}
where \(y_i\) and \(\hat y_i\) denote the target and predicted
values, respectively.
For multimodal models, this loss was applied to the prediction
obtained from the fused representation.

For each target, identical training, validation and test partitions
were used across the unimodal and multimodal models.
Representations of the same MOF were assigned to the same
partition, ensuring that comparisons used identical materials.
Prediction performance was evaluated on the held-out test set.
Detailed task definitions, units, sample counts, label distributions,
point-cloud sizes, Fourier-encoding orders, encoder architectures,
fusion-network settings and optimization parameters are provided
in the Supplementary Methods and Supplementary Tables.

Prediction accuracy was quantified using mean absolute error
(MAE), root mean squared error (RMSE) and the coefficient
of determination \(R^2\):
\begin{equation}
\begin{aligned}
\mathrm{MAE}
&=
\frac{1}{N}\sum_{i=1}^{N}|\hat y_i-y_i|,\\
\mathrm{RMSE}
&=
\sqrt{\frac{1}{N}\sum_{i=1}^{N}(\hat y_i-y_i)^2},\\
R^2
&=
1-
\frac{\sum_{i=1}^{N}(\hat y_i-y_i)^2}
     {\sum_{i=1}^{N}(y_i-\bar y)^2}.
\end{aligned}
\end{equation}
Here, \(N\) is the number of test materials, and \(\bar y\)
is the mean reference value within the test set.
MAE and RMSE were reported in the original target units.

Multimodal models were evaluated using three fusion strategies.
Residual-gated fusion~\cite{he2016deep} combines multiplicative gating of modality
features with a learned residual
correction.
Symmetric gated fusion~\cite{ilse2018attention} applies a shared scoring function to
modality embeddings and combines them through a softmax-weighted
sum.
Attention-gated fusion~\cite{lee2019set} uses a learned query to pool modality
embeddings through multi-head attention before
regression.

For each task and modality combination, the lowest-MAE
configuration across the three strategies was reported together
with its corresponding \(R^2\). The unimodal model with the
lowest MAE among CIF, ED and ESP provided the reference
MAE and \(R^2\) for each task.

To compare errors across targets with different units and scales,
unimodal MAEs were expressed as percentage increases over the
reference, and multimodal MAEs as percentage reductions.
Changes in \(R^2\) were calculated against the same reference:
\begin{equation}
\begin{aligned}
I_m(\%)
&=
100\left(
\frac{\mathrm{MAE}_m}{\mathrm{MAE}_{\mathrm{ref}}}-1
\right),\\
G_{\mathrm{multi}}(\%)
&=
100\left(
1-\frac{\mathrm{MAE}_{\mathrm{multi}}}
        {\mathrm{MAE}_{\mathrm{ref}}}
\right),\\
\Delta R^2
&=
R^2_{\mathrm{multi}}-R^2_{\mathrm{ref}}.
\end{aligned}
\end{equation}
The reference model has \(I_m=0\).
Positive \(G_{\mathrm{multi}}\) and \(\Delta R^2\) indicate
lower MAE and higher \(R^2\), respectively.
The task-averaged \(R^2\) and relative unimodal MAE increase
were calculated as unweighted arithmetic means across
the 13 targets.

\subsection{Cross-modal retrieval and evaluation}

Cross-modal retrieval was formulated as matching different physical
representations of the same MOF.
We evaluated three pairwise tasks: CIF$\leftrightarrow$ED,
CIF$\leftrightarrow$ESP and ED$\leftrightarrow$ESP.
Each modality pair was trained as an independent retrieval model,
yielding a separate pairwise embedding space.
Within each task, the paired modalities used identical training,
validation and test partitions, so that all representations of
a given MOF belonged to the same split.

Modality-specific encoders and projection heads mapped the paired
inputs to embeddings of a common dimension.
For an input \(x_m^{(i)}\) from modality \(m\), the projected
embedding was normalized to unit length, and cross-modal
similarity was calculated as
\begin{equation}
\begin{aligned}
\widetilde{\mathbf z}_m^{(i)}
&=
p_m\left(f_m(x_m^{(i)})\right),
\qquad m\in\{A,B\},\\
\mathbf z_m^{(i)}
&=
\frac{\widetilde{\mathbf z}_m^{(i)}}
     {\|\widetilde{\mathbf z}_m^{(i)}\|_2},\\
s_{ij}
&=
\left(\mathbf z_A^{(i)}\right)^{\top}
\mathbf z_B^{(j)},
\end{aligned}
\end{equation}
where \(f_m\) and \(p_m\) denote the encoder and projection
head, respectively.
The dot product of the normalized embeddings therefore
corresponds to cosine similarity.

Models were trained using a symmetric contrastive objective
with in-batch negatives~\cite{radford2021learning}.
Each minibatch contained \(B\) matched MOF pairs.
For each query, its paired representation in the other modality
was treated as the positive example, while the remaining
\(B-1\) representations in that modality were treated as negatives.
The directional losses and their symmetric average were
defined as
\begin{equation}
\begin{aligned}
\mathcal L_{A\rightarrow B}
&=
-\frac{1}{B}\sum_{i=1}^{B}
\log
\frac{\exp(s_{ii}/\tau)}
     {\sum_{j=1}^{B}\exp(s_{ij}/\tau)},\\
\mathcal L_{B\rightarrow A}
&=
-\frac{1}{B}\sum_{i=1}^{B}
\log
\frac{\exp(s_{ii}/\tau)}
     {\sum_{j=1}^{B}\exp(s_{ji}/\tau)},\\
\mathcal L_{\mathrm{ret}}
&=
\frac{1}{2}
\left(
\mathcal L_{A\rightarrow B}
+
\mathcal L_{B\rightarrow A}
\right),
\end{aligned}
\end{equation}
where \(\tau\) is the temperature parameter.
This objective encourages matched representations to have
higher similarity than unmatched representations within
the batch.

At inference, each test query was compared with the complete
candidate gallery from the paired target modality in the
test set.
Candidates were ranked in descending order of cosine
similarity, and retrieval was evaluated separately in both
directions for each modality pair.
Let \(N_{\mathrm q}\) denote the number of test queries and
\(r_i\) the rank of the correct paired representation for
query \(i\).
Recall@$K$ and mean reciprocal rank (MRR) were calculated as
\begin{equation}
\begin{aligned}
\mathrm{Recall}@K
&=
\frac{1}{N_{\mathrm q}}
\sum_{i=1}^{N_{\mathrm q}}
\mathbb I(r_i\leq K),\\
\mathrm{MRR}
&=
\frac{1}{N_{\mathrm q}}
\sum_{i=1}^{N_{\mathrm q}}
\frac{1}{r_i},
\end{aligned}
\end{equation}
where \(\mathbb I(\cdot)\) is the indicator function.
We report Recall@1, Recall@5 and Recall@10.
For each metric, the main-text results show the arithmetic
mean of the two retrieval directions within each modality pair.
Complete directional results are provided in the
Supplementary Information.

The learned embeddings were additionally examined using
cosine-similarity distributions for matched and non-matched
MOF pairs and UMAP projections~\cite{mcinnes2018umap}
of the pairwise embedding spaces.
Matched pairs corresponded to the same MOF across modalities,
whereas non-matched pairs represented different MOFs.
These analyses used embeddings from the independently trained
model for each modality pair.
Retrieval metrics were computed in the original learned
embedding spaces.
Detailed data partitions, encoder and projection-head
configurations, embedding dimensions, contrastive temperatures,
batch sizes, optimization settings and visualization procedures
are provided in the Supplementary Methods.

\subsection{Model architectures and training}
\label{sec:model_architectures_training}

Task-specific models were used for field reconstruction, property prediction
and cross-modal retrieval. ED and ESP reconstruction employed separate
instances of a probe-based E(3)-equivariant network, trained using an \(L_1\) loss between
predicted and reference field values.

For property prediction, CIF structures were processed by
CGCNN~\cite{xie2018crystal}, whereas ED and ESP point clouds
used separate instances of the same point-cloud backbone.
Regression heads mapped the resulting material representations
to scalar properties. Multimodal prediction was evaluated using
three fusion strategies: residual-gated fusion, symmetric gated
fusion and attention-gated fusion. Each strategy combined
128-dimensional modality features and was evaluated for CIF+ED,
CIF+ESP and CIF+ED+ESP across all 13 property targets.
All multimodal models were trained end to end using a normalized
mean squared error loss.

For cross-modal retrieval, modality-specific encoders and projection heads
produced 256-dimensional, \(L_2\)-normalized embeddings. CIF--ED, CIF--ESP
and ED--ESP models were trained independently using a symmetric
contrastive loss with in-batch negatives and a temperature of
\(0.07\)~\cite{radford2021learning}. Candidates were ranked by cosine
similarity.

Reconstruction, property-prediction and cross-modal retrieval models were optimized using
AdamW or Adam. The reconstruction, property-prediction and retrieval models were
trained for 15, 50 and 200 epochs, respectively. Model training was performed on a GPU cluster equipped with NVIDIA
GeForce RTX 4090, RTX A6000 and L40S GPUs. Paired modalities were
assigned to the same material-level data partitions. Checkpoints were
selected using validation performance, and final evaluation used the
held-out test sets. Detailed encoder configurations and task-specific
training settings are provided in the Supplementary Information.

\section*{Data availability}

The data supporting this study will be publicly released as
MOFFieldBench-Data upon publication. The release will include aligned
crystallographic structures, electron-density and electrostatic-potential
fields, task-specific property labels, and material-level training,
validation and code. Dataset download links and documentation
will be provided through the project repository at
\url{https://github.com/hzlhzlhzlhzl/MOFFieldBench-Data}.

\section*{Competing Interests}
The authors declare no competing interests.

\bibliography{reference}

\newpage
\begin{appendix}

\captionsetup[figure]{labelformat=empty}
\captionsetup[table]{labelformat=empty}
\renewcommand{\thefigure}{Supplementary Figure \arabic{figure}}
\renewcommand{\thetable}{Supplementary Table \arabic{table}}
\setcounter{figure}{0}
\setcounter{table}{0}

\section*{Supplementary Information}


\section{Dataset construction and electronic-structure calculations}
\label{sec:supp_dataset_dft}

\subsection{Dataset sources and benchmark construction}

MOFFieldBench contains 45,909 metal--organic frameworks with aligned crystallographic structures, electron-density (ED) fields and electrostatic-potential (ESP) fields. The collection comprises 24,908 structures from ARC-MOF~\cite{burner2023arc}, 15,435 from QMOF~\cite{rosen2021machine}, 688 from CoRE MOF~\cite{chung2019advances}, 1,955 from the Cambridge Structural Database (CSD)~\cite{groom2016cambridge}, 1,736 from the database of approximately one million hypothetical
MOFs constructed using PORMAKE~\cite{lee2021computational,kang2023multi}, and 1,187 from the Ultrastable MOF database~\cite{nandy2023database}. The source composition is summarized in Supplementary Table~\ref{tab:dataset_composition}. For each retained framework, the CIF specifies the atomic species, coordinates and lattice geometry, and the corresponding ED and ESP fields describe the same periodic unit cell.

Candidate structures were checked for valid lattice geometry, identifiable atomic species and complete atomic coordinates before conversion to CP2K inputs. A structure was retained only after successful SCF convergence and generation of both electronic fields. Field files were checked for complete volumetric grids, and their cell geometry was compared with that of the corresponding CIF. Records with failed calculations, missing or malformed field outputs, or inconsistent unit-cell information were excluded. Each retained CIF--ED--ESP triplet was linked by a common material identifier.

The unimodal and multimodal property-prediction experiments used the
ARC-MOF and QMOF collection, with property labels inherited directly from
the original source records. Each target was evaluated on the subset
with the corresponding labels available. Retaining these labels maintains
traceable target provenance and consistency with the original property
definitions and calculation conditions, while providing the same reference
values for all unimodal and multimodal comparisons within each task.
Corresponding labels for the benchmark targets were unavailable for the
additional 5,566 frameworks, which were therefore excluded from property
prediction. Cross-modal retrieval also used the 40,343-framework
ARC-MOF/QMOF collection to provide a common material basis for the
downstream benchmarks.


\begin{table}[!htbp]
\centering
\caption{Composition of MOFFieldBench by structural source.
Counts refer to retained frameworks with aligned CIF, ED and ESP representations.
All 45,909 frameworks were included in dataset characterization.
The ARC-MOF and QMOF collection was used for the unimodal and multimodal property-prediction and cross-modal retrieval experiments, with target-specific labelled subsets used for property prediction.}
\label{tab:dataset_composition}
\begin{tabular}{lr}
\toprule
Structural source & Retained frameworks \\
\midrule
ARC-MOF & 24,908 \\
QMOF & 15,435 \\
CoRE MOF & 688 \\
CSD & 1,955 \\
PORMAKE & 1,736 \\
Ultrastable MOF & 1,187 \\
\midrule
Total & 45,909 \\
\bottomrule
\end{tabular}
\end{table}

\subsection{Physical definitions and calculation principles}

ED and ESP describe the electron distribution and the electrostatic environment of a framework, respectively. Both fields were obtained from periodic Kohn--Sham density-functional theory (DFT)~\cite{KohnSham1965}. In the pseudopotential representation used here, the Kohn--Sham equations can be written schematically in atomic units as
\begin{equation}
\left[
-\frac{1}{2}\nabla^2
+\hat v_{\mathrm{ion}}
+v_{\mathrm H}^{\mathrm{el}}[\rho](\mathbf r)
+v_{\mathrm{xc}}^{\sigma}
[\rho_{\uparrow},\rho_{\downarrow}](\mathbf r)
\right]\psi_{n\sigma}(\mathbf r)
=
\varepsilon_{n\sigma}\psi_{n\sigma}(\mathbf r),
\label{eq:supp_ks_equation}
\end{equation}
where $n$ and $\sigma$ index the orbital and spin channel, respectively. The operator $\hat v_{\mathrm{ion}}$ contains the local and non-local ionic pseudopotential contributions, $v_{\mathrm H}^{\mathrm{el}}$ is the electron--electron Hartree potential, and $v_{\mathrm{xc}}^{\sigma}$ is the exchange-correlation potential. The electron--electron Hartree term in this equation is distinct from the CP2K potential output used to construct ESP, which also includes ionic charges.

At the $\Gamma$ point, the spin-summed ED is
\begin{equation}
\rho(\mathbf r)
=
\rho_{\uparrow}(\mathbf r)+\rho_{\downarrow}(\mathbf r)
=
\sum_{\sigma}\sum_n
f_{n\sigma}
\left|\psi_{n\sigma}(\mathbf r)\right|^2,
\label{eq:supp_ed_definition}
\end{equation}
where $f_{n\sigma}$ is the occupation of a spin orbital. Here, $\rho$ is an electron-number density. Its integral over the periodic cell $\Omega$ satisfies
$\int_{\Omega}\rho(\mathbf r)\,\mathrm d^3\mathbf r
=N_{\mathrm e}^{\mathrm{val}}$. The quantity $N_{\mathrm e}^{\mathrm{val}}$ counts the electrons explicitly treated by the pseudopotentials. ED therefore corresponds to a valence-electron pseudodensity, including any semicore electrons in the chosen valence space; frozen core electrons are not reconstructed.

The Gaussian and plane-wave (GPW) method implemented in CP2K combines atom-centred Gaussian orbital basis functions with auxiliary grids for the density and electrostatic calculations~\cite{VandeVondele2005,Kuhne2020}. The orbitals are expanded as
$\psi_{n\sigma}(\mathbf r)
=\sum_{\mu}C_{\mu n\sigma}\chi_{\mu}(\mathbf r)$,
where $\chi_{\mu}$ denotes a Gaussian basis function. The resulting density is represented on regular real-space grids, and Fourier transforms connect the real-space and reciprocal-space representations used by the periodic Coulomb solver. The Gaussian basis and auxiliary plane-wave cutoffs thus control different parts of the numerical representation.

ESP is defined using the physical charge-sign convention. For the total effective charge density, the periodic Poisson equation is
\begin{equation}
\begin{aligned}
\varrho_{\mathrm{tot}}(\mathbf r)
&=
\varrho_{\mathrm{ion}}(\mathbf r)-\rho(\mathbf r),\\
\nabla^2 V(\mathbf r)
&=
-4\pi\varrho_{\mathrm{tot}}(\mathbf r),
\end{aligned}
\label{eq:supp_esp_poisson}
\end{equation}
where $\varrho_{\mathrm{ion}}$ is the effective positive ionic charge density used in the GPW electrostatic treatment. It represents the nuclei and frozen core electrons through smooth effective ionic charges. For the neutral calculations used here,
$\int_{\Omega}\varrho_{\mathrm{tot}}(\mathbf r)\,
\mathrm d^3\mathbf r=0$.
For each non-zero reciprocal-lattice vector $\mathbf G$, the Fourier coefficients satisfy
\begin{equation}
V(\mathbf G)
=
\frac{4\pi}{|\mathbf G|^2}
\varrho_{\mathrm{tot}}(\mathbf G),
\qquad \mathbf G\ne\mathbf 0.
\label{eq:supp_esp_reciprocal}
\end{equation}
The zero-frequency component specifies an arbitrary additive constant and requires a reference convention, defined below. Because ESP depends on the charge distribution throughout the periodic system, it can vary substantially within pore regions where ED is small. The ESP stored in MOFFieldBench contains the electrostatic contributions from the effective electronic and ionic charges. It excludes the exchange-correlation potential and non-local pseudopotential operator and therefore does not represent the full Kohn--Sham effective potential.

\subsection{Electronic-structure calculations and electronic-field generation}

All electronic fields were generated using periodic single-point DFT calculations in CP2K v2025.1. Calculations used the Quickstep module with the GPW method and \texttt{RUN\_TYPE ENERGY}, with atomic species, coordinates and lattice geometry supplied by the source CIFs. Exchange and correlation were described by the Perdew--Burke--Ernzerhof (PBE) generalized-gradient approximation~\cite{perdew1996generalized}. D3 and D4 dispersion corrections were not included. Element-specific \texttt{DZVP-MOLOPT-SR-GTH-q*} Gaussian basis sets from the \texttt{BASIS\_MOLOPT} library~\cite{vandevondele2007} were paired with GTH-PBE pseudopotentials~\cite{goedecker1996gth,krack2005pseudopotentials}.

The auxiliary plane-wave cutoff was $350\,\mathrm{Ry}$, and the relative cutoff was $50\,\mathrm{Ry}$. The global numerical threshold was set to \texttt{EPS\_DEFAULT} $=10^{-11}$. Three-dimensional periodic boundary conditions were applied to both the electronic-structure calculation and the Poisson solver using \texttt{PERIODIC XYZ}. Calculations used $\Gamma$-point sampling and an overall neutral charge. Spin multiplicities of one or two were specified during input preparation, with unrestricted Kohn--Sham calculations enabled when required. The calculated fields describe self-consistent electronic states at the supplied geometries and imposed spin multiplicities.

SCF calculations used diagonalization with Broyden mixing, a mixing parameter of \texttt{ALPHA} $=0.4$ and a history length of \texttt{NBROYDEN} $=8$. The maximum number of iterations was \texttt{MAX\_SCF} $=128$, and the convergence threshold was \texttt{EPS\_SCF} $=5\times10^{-6}$. Orbital transformation and electronic smearing were not enabled. Fresh calculations used an atomic-density initial guess. When an existing wavefunction restart file was available for a continuation calculation, it was read using \texttt{WFN\_RESTART\_FILE\_NAME} and \texttt{SCF\_GUESS RESTART}, with the same final electronic-structure settings.

After SCF convergence, the spin-summed ED was exported using \texttt{E\_DENSITY\_CUBE} with \texttt{STRIDE 1}, retaining the complete output grid without subsampling during CUBE generation. ED values were stored in $\mathrm{e\,Bohr^{-3}}$. The electrostatic potential was exported using \texttt{V\_HARTREE\_CUBE}, also with \texttt{STRIDE 1}, and stored in Hartree after the sign and reference transformations described below.

The raw CP2K potential has the opposite sign to the physical ESP convention adopted here. Its sign was therefore reversed, and its mean over the complete unit cell was subtracted to establish a common reference. For raw values $V_{\mathrm H}(\mathbf r_i)$ on a regular grid of $N_{\mathrm{grid}}$ points, the stored ESP was defined by
\begin{equation}
\begin{aligned}
\overline{V}_{\mathrm H}
&=
\frac{1}{N_{\mathrm{grid}}}
\sum_{i=1}^{N_{\mathrm{grid}}}
V_{\mathrm H}(\mathbf r_i),\\
V(\mathbf r_i)
&=
-\left[
V_{\mathrm H}(\mathbf r_i)-\overline{V}_{\mathrm H}
\right].
\end{aligned}
\label{eq:supp_esp_reference}
\end{equation}
This transformation sets the cell-averaged ESP to zero while preserving its spatial variation under the adopted sign convention. The reference was calculated from the complete native grid before compression or point-cloud sampling.

ED and ESP were stored as regular three-dimensional grids spanning the periodic unit cell. Their dimensions $(N_x,N_y,N_z)$ varied with the lattice geometry and the CP2K grid construction. Grid dimensions, origins and step vectors were parsed directly from the CUBE headers. For zero-based indices, the Cartesian coordinate of a grid point is
\begin{equation}
\mathbf r_{ijk}
=
\mathbf r_0
+i\mathbf a_g
+j\mathbf b_g
+k\mathbf c_g,
\label{eq:supp_cube_coordinates}
\end{equation}
where $\mathbf r_0$ is the grid origin and $\mathbf a_g$, $\mathbf b_g$ and $\mathbf c_g$ are the three grid-step vectors, expressed in a common length unit. The indices satisfy $0\leq i<N_x$, $0\leq j<N_y$ and $0\leq k<N_z$. Using the native grid geometry maintains the spatial registration of each field with its corresponding crystallographic unit cell.

\section{Electronic-field statistics and compression}
\label{sec:supp_field_compression}

\subsection{Additional electronic-field statistics}

Additional statistics were calculated for a 2,000-MOF mixed analysis subset to characterize the native volumetric grids and field-value distributions. Each quantity was first calculated for an individual structure and then summarized across the subset (Supplementary Table~\ref{tab:field_statistics}). The number of grid points per field ranges from approximately $2.88\times10^{5}$ to $5.90\times10^{7}$, with a median of $3.50\times10^{6}$ and an interquartile range of $1.94\times10^{6}$--$4.50\times10^{6}$. The structures contain 17--712 atoms per unit cell, with a median of 87 atoms. These summaries characterize the selected analysis subset.

To compare volumetric resolution across structures, an effective linear grid spacing was defined as
\begin{equation}
d_{\mathrm{eff}}
=
\left(
\frac{V_{\mathrm{cell}}}{N_{\mathrm{grid}}}
\right)^{1/3},
\label{eq:supp_effective_grid_spacing}
\end{equation}
where $V_{\mathrm{cell}}$ is the unit-cell volume and $N_{\mathrm{grid}}$ is the number of native grid points. This quantity expresses the cube root of the volume per point and summarizes the effective volumetric resolution. The individual grid-step lengths and angles depend on the cell geometry. The median $d_{\mathrm{eff}}$ is $0.0845\,\text{\AA}$, with an interquartile range of $0.0831$--$0.0859\,\text{\AA}$ and an overall range of $0.0713$--$0.0896\,\text{\AA}$, indicating similar effective resolution within this subset.

ED exhibits a broad and asymmetric spatial distribution. The median of the structure-wise mean densities is $0.168\,\mathrm{e}\,\text{\AA}^{-3}$, whereas the median of the structure-wise median densities is $3.84\times10^{-3}\,\mathrm{e}\,\text{\AA}^{-3}$. The medians of the structure-wise standard deviations and maxima are $0.601$ and $23.50\,\mathrm{e}\,\text{\AA}^{-3}$, respectively. The separation between the mean and median densities is consistent with localized high-density regions occupying a relatively small fraction of the cell volume.

ESP statistics were calculated using the sign and cell-average reference convention defined in Supplementary Section~\ref{sec:supp_dataset_dft}. The median of the structure-wise median potentials is $0.107\,\mathrm{Hartree}$, and the median structure-wise standard deviation is $0.453\,\mathrm{Hartree}$. The medians of the structure-wise minima and maxima are $-12.83$ and $0.187\,\mathrm{Hartree}$, respectively. The cell-averaged potentials remain close to zero under the adopted reference convention. Positive spatial medians can coexist with a near-zero mean because the potential distributions are asymmetric.

\begin{figure*}[!htbp]
\centering
\includegraphics[width=0.98\textwidth]{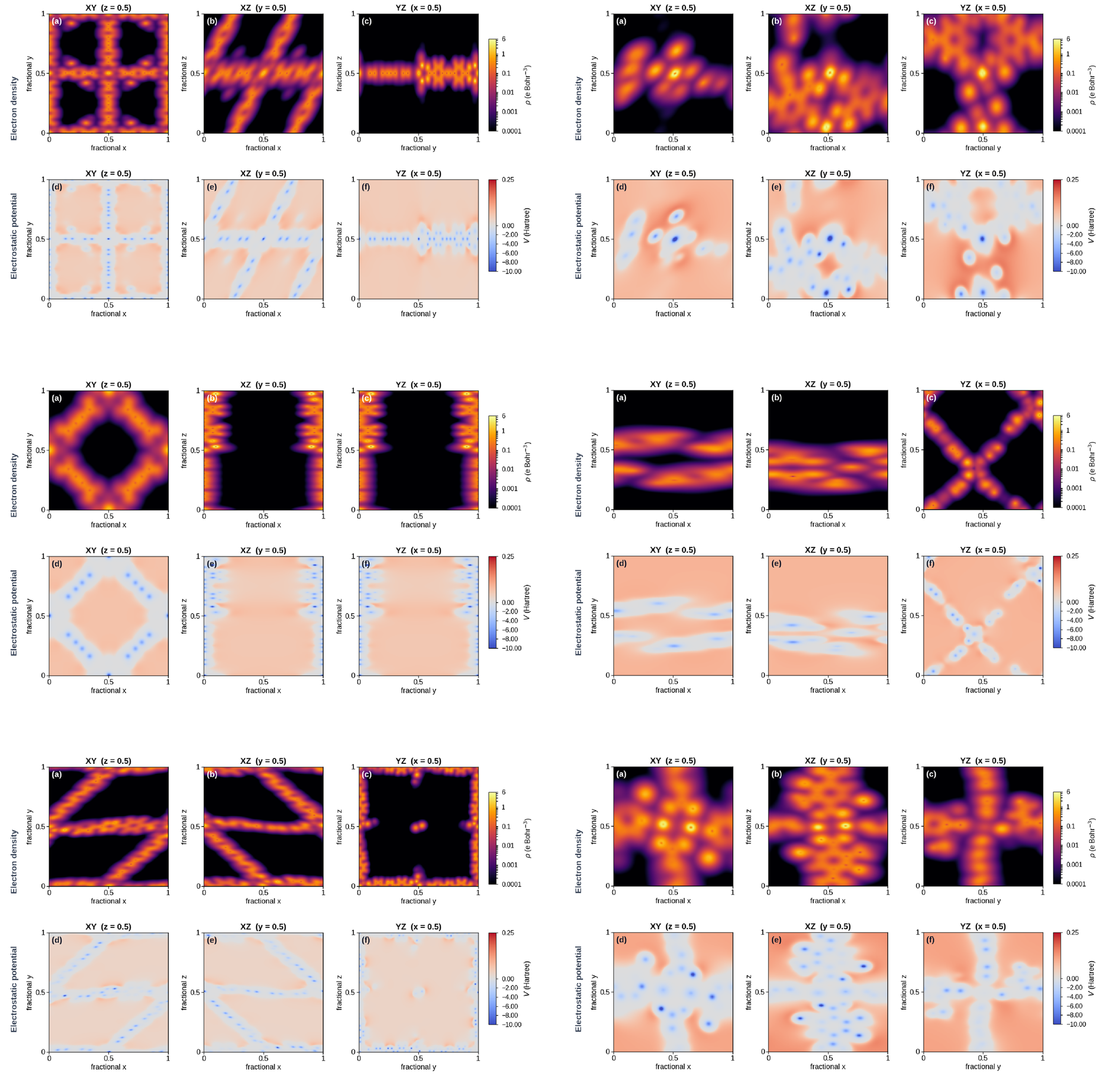}
\caption{
\textbf{Electronic-field sections for six MOF examples.}
ED and ESP are shown on the $xy$, $xz$ and $yz$ planes at fractional coordinates $z=0.5$, $y=0.5$ and $x=0.5$, respectively.
ED is expressed in $\mathrm{e\,Bohr^{-3}}$ and ESP in Hartree.
The sections illustrate the distribution of electron density and electrostatic potential across the framework and pore regions.
}
\label{fig:field_slices}
\end{figure*}

\begin{figure*}[!htbp]
\centering
\includegraphics[width=0.98\textwidth]{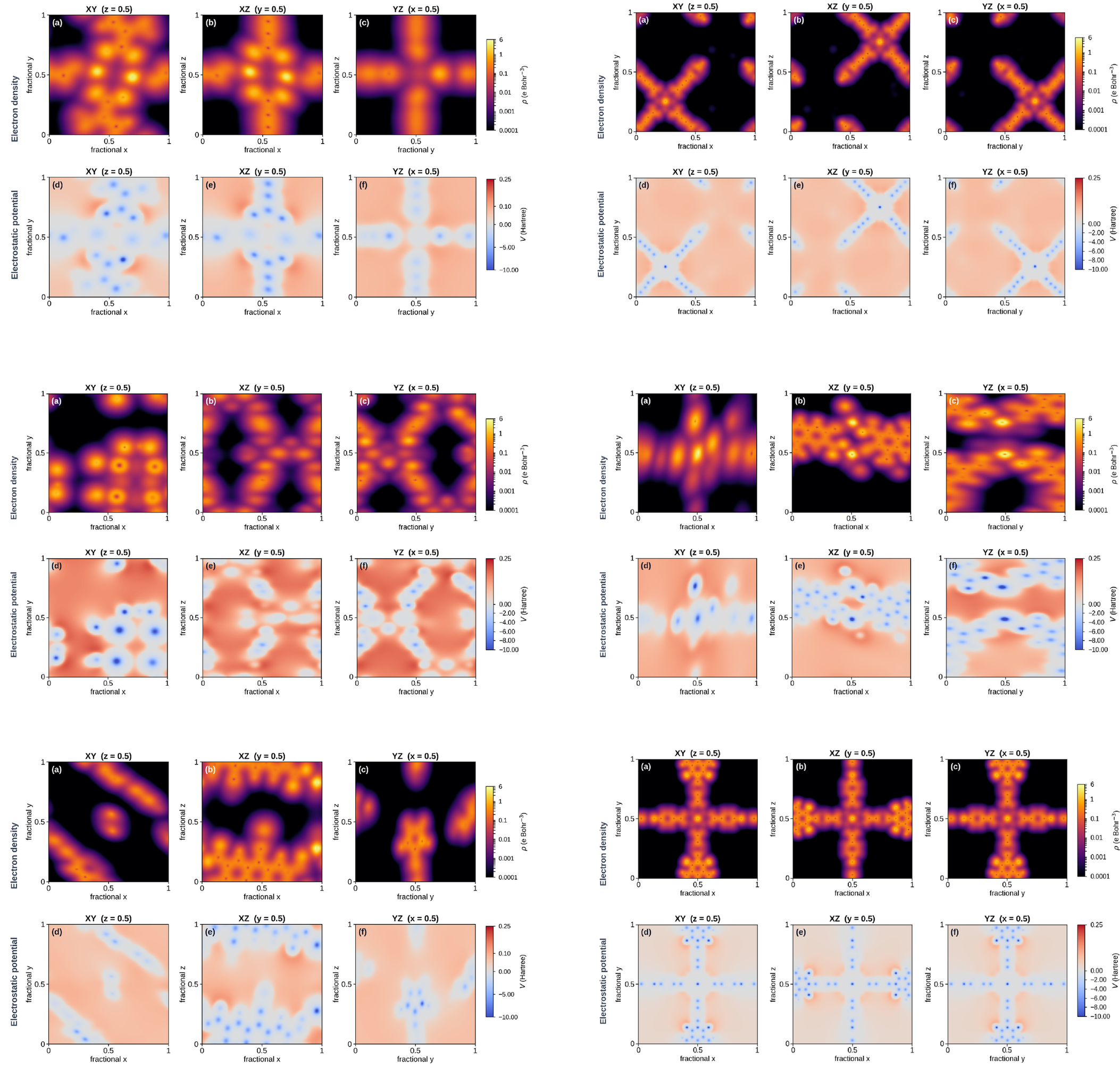}
\caption{
\textbf{Electronic-field sections for six additional MOF examples.}
ED and ESP are shown on the $xy$, $xz$ and $yz$ planes at fractional coordinates $z=0.5$, $y=0.5$ and $x=0.5$, respectively.
ED is expressed in $\mathrm{e\,Bohr^{-3}}$ and ESP in Hartree.
These examples complement Supplementary Fig.~\ref{fig:field_slices} by showing field distributions in additional framework geometries.
}
\label{fig:field_slices_additional}
\end{figure*}

Supplementary Figs.~\ref{fig:field_slices} and \ref{fig:field_slices_additional} show ED and ESP sections for twelve MOF examples. The $xy$, $xz$ and $yz$ sections are taken at fractional coordinates $z=0.5$, $y=0.5$ and $x=0.5$, respectively. ED highlights the spatial arrangement of the framework and low-density voids, whereas ESP shows electrostatic variation over the same planes. These examples illustrate field heterogeneity across different structures.

ED statistics in Supplementary Table~\ref{tab:field_statistics} are reported in $\mathrm{e}\,\text{\AA}^{-3}$, whereas the field-section figures and compression errors use $\mathrm{e\,Bohr^{-3}}$. The conversion between these units is
$1\,\mathrm{e\,Bohr^{-3}}
\simeq6.7483\,\mathrm{e}\,\text{\AA}^{-3}$.
ESP is reported in Hartree throughout.

\begin{table}[!htbp]
\centering
\caption{Additional volumetric-grid and electronic-field statistics for the 2,000 MOF analysis subset. Each row summarizes a structure-level quantity across the subset. The median and interquartile range are calculated across structures; the interquartile range denotes the 25th--75th percentiles, and the range denotes the minimum and maximum.}
\label{tab:field_statistics}
\small
\setlength{\tabcolsep}{5pt}
\begin{tabular}{lccc}
\toprule
Statistic & Median & Interquartile range & Range \\
\midrule
Grid points ($10^6$) & 3.499 & 1.944--4.500 & 0.288--58.982 \\
Atoms per unit cell & 87 & 73--108 & 17--712 \\
Effective grid spacing (\AA) & 0.0845 & 0.0831--0.0859 & 0.0713--0.0896 \\
Mean ED ($\mathrm{e}\,\text{\AA}^{-3}$) & 0.168 & 0.138--0.284 & 0.019--0.577 \\
Median ED ($\mathrm{e}\,\text{\AA}^{-3}$) & 0.00384 & 0.00096--0.0513 & $1.47\times10^{-10}$--0.156 \\
ED standard deviation ($\mathrm{e}\,\text{\AA}^{-3}$) & 0.601 & 0.542--0.688 & 0.200--1.484 \\
Maximum ED ($\mathrm{e}\,\text{\AA}^{-3}$) & 23.50 & 18.68--31.51 & 2.58--46.64 \\
Mean ESP (Hartree) & $-1.6\times10^{-8}$
& $[-4.6\times10^{-7},\,4.4\times10^{-7}]$
& $[-3.7\times10^{-6},\,4.6\times10^{-6}]$ \\
Median ESP (Hartree) & 0.107 & 0.088--0.168 & 0.013--0.233 \\
ESP standard deviation (Hartree) & 0.453 & 0.409--0.557 & 0.140--0.798 \\
Minimum ESP (Hartree) & $-12.83$ & $[-16.46,\,-12.60]$ & $[-26.35,\,-6.79]$ \\
Maximum ESP (Hartree) & 0.187 & 0.163--0.256 & 0.035--0.358 \\
\bottomrule
\end{tabular}
\end{table}

\subsection{CP3D compression performance and numerical fidelity}

CP3D compresses volumetric electronic fields using the predictive quantization procedure described in the main Methods. Quantized prediction residuals are stored primarily as signed 16-bit integers. Residuals outside the representable range are handled by storing explicit anchor values and their grid positions. Signed residual codes are transformed using ZigZag encoding, and the codes, anchor information, grid metadata and reconstruction parameters are packed in a lossless NPZ/ZIP container. Quantization introduces the controlled numerical approximation, while container packing preserves the quantized representation. Compression is applied to the native volumetric fields before Fourier downsampling and point-cloud construction.

Storage efficiency and numerical fidelity were evaluated on a sampled set of MOF electronic fields. For each evaluated field, the original CUBE file was compared with its CP3D representation, and the decompressed values were compared with the original grid values. Storage ratios therefore refer to the on-disk CP3D and original CUBE files in this sampled evaluation set.

For a field $j$, let $S_j^{\mathrm{CUBE}}$ and $S_j^{\mathrm{CP3D}}$ denote the original and compressed file sizes. The per-field compression ratio and aggregate compression ratio are
\begin{equation}
C_j
=
\frac{S_j^{\mathrm{CUBE}}}{S_j^{\mathrm{CP3D}}},
\qquad
C_{\mathrm{agg}}
=
\frac{\sum_j S_j^{\mathrm{CUBE}}}
{\sum_j S_j^{\mathrm{CP3D}}}.
\label{eq:supp_compression_ratios}
\end{equation}
The mean per-field ratio is the arithmetic mean of $C_j$, while the aggregate ratio describes the reduction in total storage. The retained storage fraction is $100/C_{\mathrm{agg}}$, expressed as a percentage.

For the sampled ED fields, total storage decreases from $58.191$ to $2.519\,\mathrm{GiB}$, corresponding to an aggregate compression ratio of $23.10\times$. For ESP, storage decreases from $57.769$ to $1.859\,\mathrm{GiB}$, giving an aggregate ratio of $31.08\times$. Across the sampled ED and ESP files, $115.961\,\mathrm{GiB}$ is reduced to $4.378\,\mathrm{GiB}$, corresponding to an aggregate compression ratio of $26.49\times$ and a storage reduction of $96.225\%$ (Supplementary Table~\ref{tab:compression_performance}).

The mean original and compressed file sizes are $59.59$ and $2.58\,\mathrm{MiB}$ for ED, and $59.21$ and $1.91\,\mathrm{MiB}$ for ESP. The mean per-field compression ratios are $22.43\times$ and $30.40\times$, respectively. Both the mean per-field and aggregate ratios indicate greater storage reduction for ESP in this evaluation set. Compression performance can vary between individual fields, so the two summaries describe complementary aspects of storage efficiency.

\begin{table}[!htbp]
\centering
\caption{Storage performance of CP3D on the sampled electronic fields.
Mean file sizes are calculated across individual fields.
The aggregate compression ratio is the total original CUBE size divided by the total CP3D size.
The retained fraction is its inverse expressed as a percentage, and storage reduction is 100\% minus the retained fraction.
MiB and GiB denote $2^{20}$ and $2^{30}$ bytes, respectively.
Totals and aggregate ratios are calculated from unrounded byte counts.}
\label{tab:compression_performance}
\begin{tabular}{lrrr}
\toprule
Quantity & ED & ESP & Overall \\
\midrule
Mean CUBE size per field (MiB) & 59.59 & 59.21 & 59.40 \\
Mean CP3D size per field (MiB) & 2.58 & 1.91 & 2.24 \\
Retained fraction (\%) & 4.329 & 3.217 & 3.775 \\
Storage reduction (\%) & 95.671 & 96.783 & 96.225 \\
Aggregate compression ratio & $23.10\times$ & $31.08\times$ & $26.49\times$ \\
\bottomrule
\end{tabular}
\end{table}

Numerical fidelity was evaluated on the same sampled fields using nominal absolute tolerances of $10^{-6}\,\mathrm{e\,Bohr^{-3}}$ for ED and $10^{-4}\,\mathrm{Hartree}$ for ESP. For a field with $N$ grid points, the reconstruction errors are
\begin{equation}
\begin{aligned}
\mathrm{MAE}
&=
\frac{1}{N}\sum_{i=1}^{N}
|\hat f_i-f_i|,\\
\mathrm{RMSE}
&=
\left[
\frac{1}{N}\sum_{i=1}^{N}
(\hat f_i-f_i)^2
\right]^{1/2},\\
E_{\max}
&=
\max_{1\leq i\leq N}|\hat f_i-f_i|,
\end{aligned}
\label{eq:supp_compression_errors}
\end{equation}
where $f_i$ and $\hat f_i$ denote the original and decompressed values, respectively. Errors were calculated separately for each field and then summarized using arithmetic means and medians across fields.

For ED, the mean per-field MAE and RMSE are $4.775\times10^{-7}$ and $5.610\times10^{-7}\,\mathrm{e\,Bohr^{-3}}$, respectively. For ESP, the corresponding values are $5.000\times10^{-5}$ and $5.779\times10^{-5}\,\mathrm{Hartree}$ (Supplementary Table~\ref{tab:compression_error}). The mean MAE and RMSE correspond to $47.75\%$ and $56.10\%$ of the ED tolerance, and $50.00\%$ and $57.79\%$ of the ESP tolerance, respectively.

The largest recorded pointwise errors are $1.013279\times10^{-6}\,\mathrm{e\,Bohr^{-3}}$ for ED and $1.001358\times10^{-4}\,\mathrm{Hartree}$ for ESP. These values exceed the nominal tolerances by approximately $1.33\%$ and $0.14\%$, respectively. The measured maximum errors are therefore reported alongside the nominal quantization tolerances to characterize the implemented compression procedure. The evaluation quantifies numerical reconstruction of the native fields; approximation introduced by subsequent downsampling or point-cloud sampling is outside this compression-error assessment.

\begin{table}[!htbp]
\centering
\caption{Numerical fidelity of CP3D reconstruction on the sampled fields used for storage evaluation.
Means and medians summarize per-field error metrics.
ED errors are expressed in $\mathrm{e\,Bohr^{-3}}$ and ESP errors in Hartree.
Maximum-error statistics are taken from the recorded per-field verification results.
Nominal tolerances and measured reconstruction errors are reported separately.}
\label{tab:compression_error}
\begin{tabular}{lcc}
\toprule
Metric & ED & ESP \\
\midrule
Nominal absolute tolerance & $1.0\times10^{-6}$ & $1.0\times10^{-4}$ \\
Mean MAE & $4.775\times10^{-7}$ & $5.000\times10^{-5}$ \\
Median MAE & $4.784\times10^{-7}$ & $5.000\times10^{-5}$ \\
Mean RMSE & $5.610\times10^{-7}$ & $5.779\times10^{-5}$ \\
Median RMSE & $5.606\times10^{-7}$ & $5.778\times10^{-5}$ \\
Median maximum absolute error & $1.013279\times10^{-6}$ & $1.001358\times10^{-4}$ \\
Largest maximum absolute error & $1.013279\times10^{-6}$ & $1.001358\times10^{-4}$ \\
\bottomrule
\end{tabular}
\end{table}

\section{Electronic-field reconstruction}
\label{sec:supp_reconstruction}

\subsection{Implementation and evaluation protocol}

ED and ESP were reconstructed using the probe-based E(3)-equivariant architecture described in the main Methods~\cite{koker2024charge3net}. Separate models were trained for the two fields using the same architecture: three atomic interaction layers, three probe-message layers, a representation multiplicity of 500 and a maximum angular degree of $l_{\max}=4$. Radial features were represented by 20 Gaussian basis functions and processed by a radial network with dimensions $20\rightarrow100$. A common radial cutoff was used for atom--atom and atom--probe interactions. Architecture and training settings are summarized in Supplementary Table~\ref{tab:reconstruction_hyperparameters}.

The reconstruction subset comprised 40,343 structures, with 36,310 assigned to training, 2,016 to validation and 2,017 to testing, giving an approximately 90\%/5\%/5\% split. During training, 500 native grid positions were sampled uniformly with replacement each time a structure was accessed. Validation and test evaluation used 1,000 uniformly sampled positions per structure, also sampled with replacement and held fixed.

Training minimized the element-wise L1 loss over the sampled probes. The objective contained only the field-value regression loss. Adam optimization used the learning-rate schedule
\begin{equation}
\eta(s)
=
5\times10^{-3}\times0.96^{s/3000},
\label{eq:supp_reconstruction_lr}
\end{equation}
where $s$ is the global optimization step. Distributed data-parallel training used one MOF per GPU and a maximum budget of $10^6$ optimization steps. Weight decay, gradient clipping and automatic mixed precision were not enabled. The checkpoint with the lowest validation normalized MAE was selected for evaluation.

\begin{table}[!htbp]
\centering
\caption{Architecture, optimization and probe-sampling settings for electronic-field reconstruction. ED and ESP models were trained separately using the same architecture. The maximum number of optimization steps specifies the training budget; checkpoint selection was based on validation normalized MAE.}
\label{tab:reconstruction_hyperparameters}
\small
\begin{tabular}{ll}
\toprule
Setting & Value \\
\midrule
Architecture & Probe-based E(3)-equivariant network \\
Atomic interaction layers & 3 \\
Probe-message layers & 3 \\
Representation multiplicity & 500 \\
Maximum angular degree & $l_{\max}=4$ \\
Radial basis & 20 Gaussian basis functions \\
Radial network & $20\rightarrow100$ \\
Training loss & Mean absolute error (L1) \\
Optimizer & Adam \\
Initial learning rate & $5\times10^{-3}$ \\
Learning-rate schedule & Eq.~\ref{eq:supp_reconstruction_lr} \\
\bottomrule
\end{tabular}
\end{table}

 For each MOF $m$ and evaluation regime $\mathcal R$, let
  $\mathcal P_{m,\mathcal R}$ denote the set of evaluated probe points.
  The per-MOF mean absolute error and normalized mean absolute error
  were defined as
  \begin{equation}
  \begin{aligned}
  \mathrm{MAE}_{m,\mathcal R}
  &=
  \frac{1}{|\mathcal P_{m,\mathcal R}|}
  \sum_{i\in\mathcal P_{m,\mathcal R}}
  |\hat y_{m,i}-y_{m,i}|,\\
  \mathrm{nMAE}_{m,\mathcal R}(\%)
  &=
  100\,
  \frac{
  \sum_{i\in\mathcal P_{m,\mathcal R}}
  |\hat y_{m,i}-y_{m,i}|
  }{
  \sum_{i\in\mathcal P_{m,\mathcal R}}
  |y_{m,i}|
  },\\
  \overline{\mathrm{nMAE}}_{\mathcal R}(\%)
  &=
  \frac{1}{|\mathcal M_{\mathcal R}|}
  \sum_{m\in\mathcal M_{\mathcal R}}
  \mathrm{nMAE}_{m,\mathcal R}(\%),
  \end{aligned}
  \label{eq:supp_reconstruction_metrics}
  \end{equation}
  where $y_{m,i}$ and $\hat y_{m,i}$ are the reference and predicted
  field values, respectively, and $\mathcal M_{\mathcal R}$ denotes
  the set of MOFs with valid evaluation points and a nonzero
  normalization denominator in regime $\mathcal R$.
  The reported nMAE is the arithmetic mean of the per-MOF nMAEs,
  with each eligible structure contributing equally.
  For whole-cell test evaluation, 2,000 probe points were sampled
  uniformly from the native volumetric grid of each MOF across the
  entire periodic unit cell. This sampling covers the whole-cell
  domain without exhaustively evaluating every grid point.
  For regional evaluation, normalization was performed separately
  within each MOF using the points evaluated in the corresponding
  reference-field regime.
  The denominator uses absolute reference values to accommodate
  signed ESP. Consequently, nMAE depends on both the absolute
  prediction error and the reference-field magnitude.
  ED values and density thresholds are expressed in
  $e\,\mathrm{Bohr}^{-3}$.
  Pearson's correlation coefficient $r$ and Spearman's rank
  correlation coefficient $r_{\mathrm S}$ were also calculated
  over the reference--prediction pairs to quantify linear
  association and rank agreement.

  For full-grid inference, the periodic atomic graph and atomic
  representations were computed once per structure and reused for
  successive batches of probes. Query coordinates were constructed
  from the grid dimensions, origin and step vectors recorded in
  the field metadata. Predictions were evaluated in chunks and
  assembled into the original three-dimensional grid.
  Full-grid inference was distinct from the sampled test-set
  evaluation described above.
  Model inputs comprised the atomic structure and query positions;
  reference scalar field values were used for supervision and
  evaluation.

\subsection{Spatial reconstruction and regional error analysis}

\begin{figure}[p]
\centering
\includegraphics[width=\linewidth]{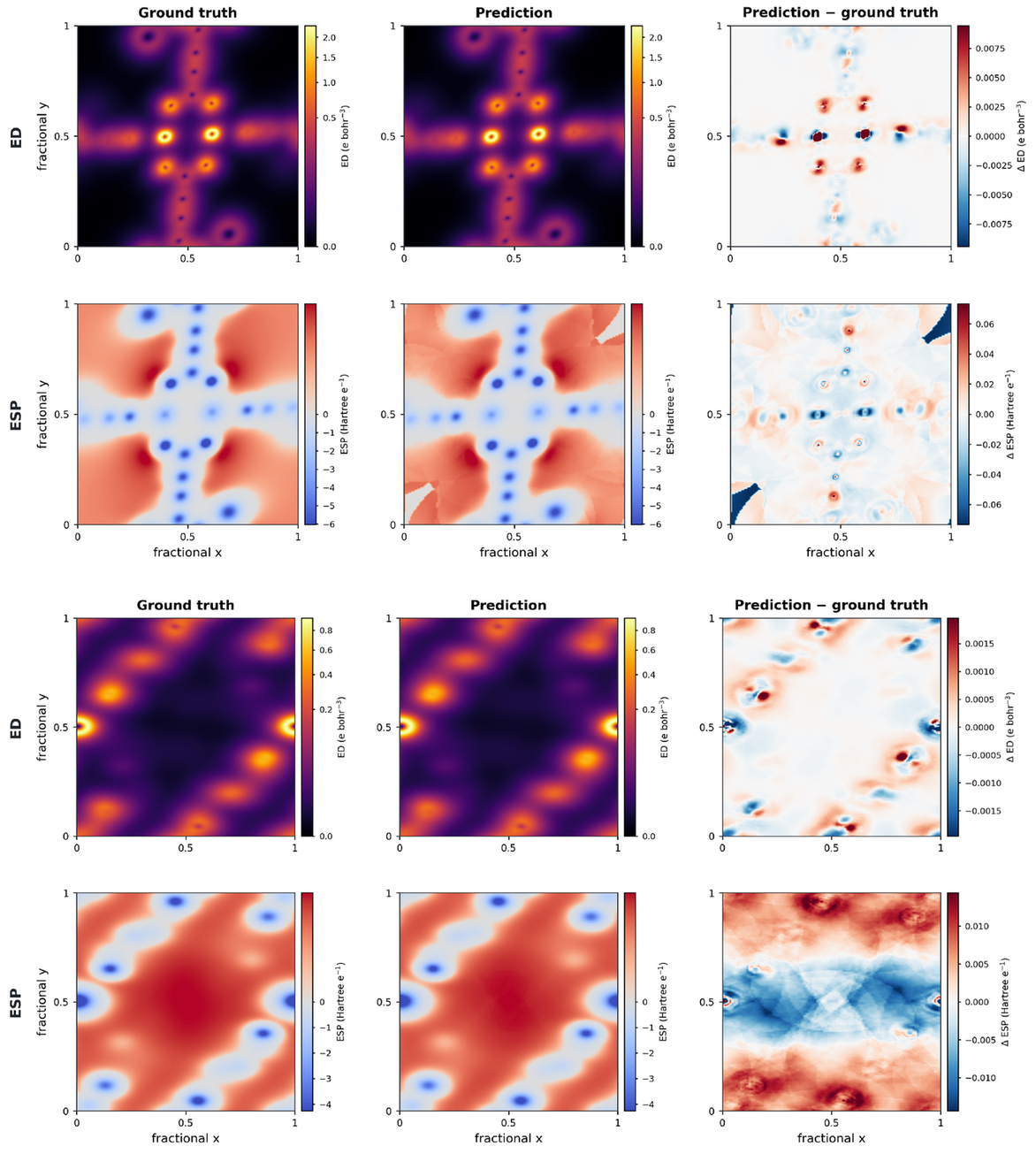}
\caption{
\textbf{Fixed-plane examples of ED and ESP reconstruction.}
\textbf{Top,} Validation structure \texttt{DB0-m2\_o21\_o25\_f0\_pcu.sym.29}.
\textbf{Bottom,} Validation structure \texttt{qmof-1a19916}.
Each example shows a plane at fractional $z=0.5$, with ED in the upper row and ESP in the lower row.
The two in-plane axes are the first and second fractional coordinates.
Columns show the reference field, prediction and prediction-minus-reference residual.
Reference and prediction share a colour scale within each row; residual scales are centred on zero.
ED is expressed in $\mathrm{e\,Bohr^{-3}}$ and ESP in atomic units of potential, $\mathrm{Hartree}\,\mathrm{e}^{-1}$.
Lighter colours indicate higher ED; blue and red indicate lower and higher ESP, respectively.
In residual maps, blue and red indicate negative and positive differences.
The reported nMAEs are calculated over the displayed planes.
}
\label{fig:reconstruction_fixed_z_examples}
\end{figure}

Supplementary Fig.~\ref{fig:reconstruction_fixed_z_examples} compares reference fields, predictions and signed residuals on fractional-coordinate planes at $z=0.5$ for two validation structures, \texttt{DB0-m2\_o21\_o25\_f0\_pcu.sym.29} and \texttt{qmof-1a19916}. All three maps were evaluated at the same spatial positions. Reference and prediction share a colour scale within each field, while residual maps use a separate scale centred on zero. Residuals are defined as prediction minus reference.

For these examples, ED predictions reproduce the principal density features and intervening low-density regions, with the most visible residuals near sharp density features. ESP predictions capture the main potential patterns, while residuals include localized differences and broader spatial variations. The plane-specific nMAEs are 0.56\% for ED and 1.55\% for ESP in \texttt{DB0-m2\_o21\_o25\_f0\_pcu.sym.29}, and 0.37\% for ED and 1.75\% for ESP in \texttt{qmof-1a19916}. These metrics apply to the displayed planes.

\begin{figure}[p]
\centering
\includegraphics[width=\linewidth]{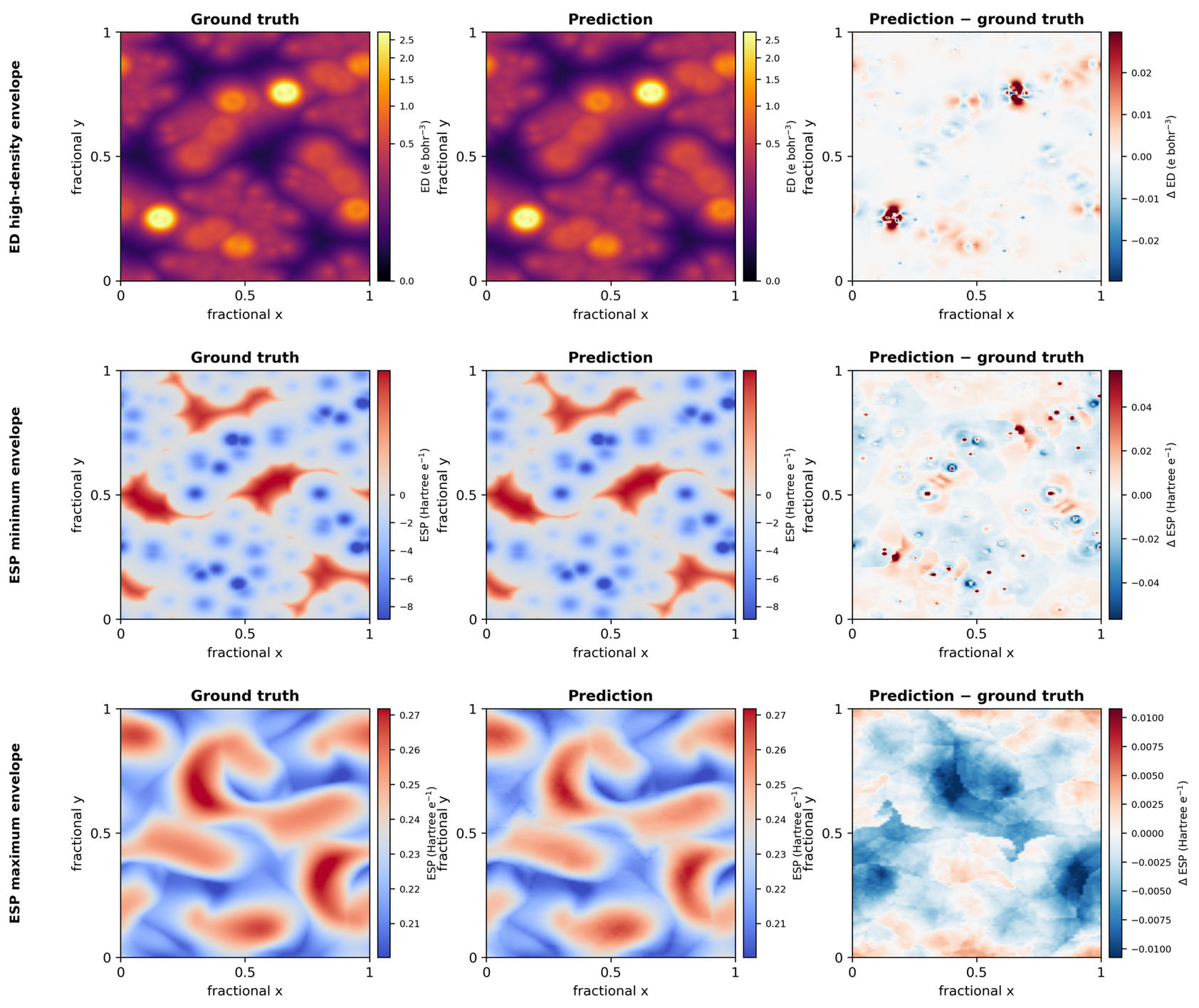}
\caption{
\textbf{Extremum-envelope comparison for full-grid reconstruction of \texttt{qmof-46fadab}.}
Rows show the ED maximum envelope, ESP minimum envelope and ESP maximum envelope along the third grid direction.
For each pair of in-plane indices, the reference field determines the extremum location, and the prediction and residual are evaluated at the same three-dimensional position.
Columns show the CP2K reference, reconstruction and prediction-minus-reference residual.
Reference and prediction share colour scales within each row; residuals use separate scales centred on zero.
ED is expressed in $\mathrm{e\,Bohr^{-3}}$ and ESP in $\mathrm{Hartree}\,\mathrm{e}^{-1}$.
The reported nMAEs are calculated over the complete three-dimensional grid.
}
\label{fig:reconstruction_full_grid_example}
\end{figure}

Supplementary Fig.~\ref{fig:reconstruction_full_grid_example} visualizes full-grid reconstruction of \texttt{qmof-46fadab} using extremum envelopes along the third grid direction. At each pair of in-plane indices $(i,j)$, the reference ED selects the location of its maximum, and the reference ESP selects the locations of its minimum and maximum. Predictions and residuals are evaluated at these same reference-selected positions. Each envelope therefore combines positions across the third grid direction. The ED maximum envelope and ESP minimum and maximum envelopes capture the principal field features, with localized ED residuals and more extended ESP deviations. Evaluation over the complete three-dimensional grid gives nMAEs of 0.55\% for ED and 1.18\% for ESP for this structure.

\begin{figure}[p]
\centering
\includegraphics[width=\linewidth]{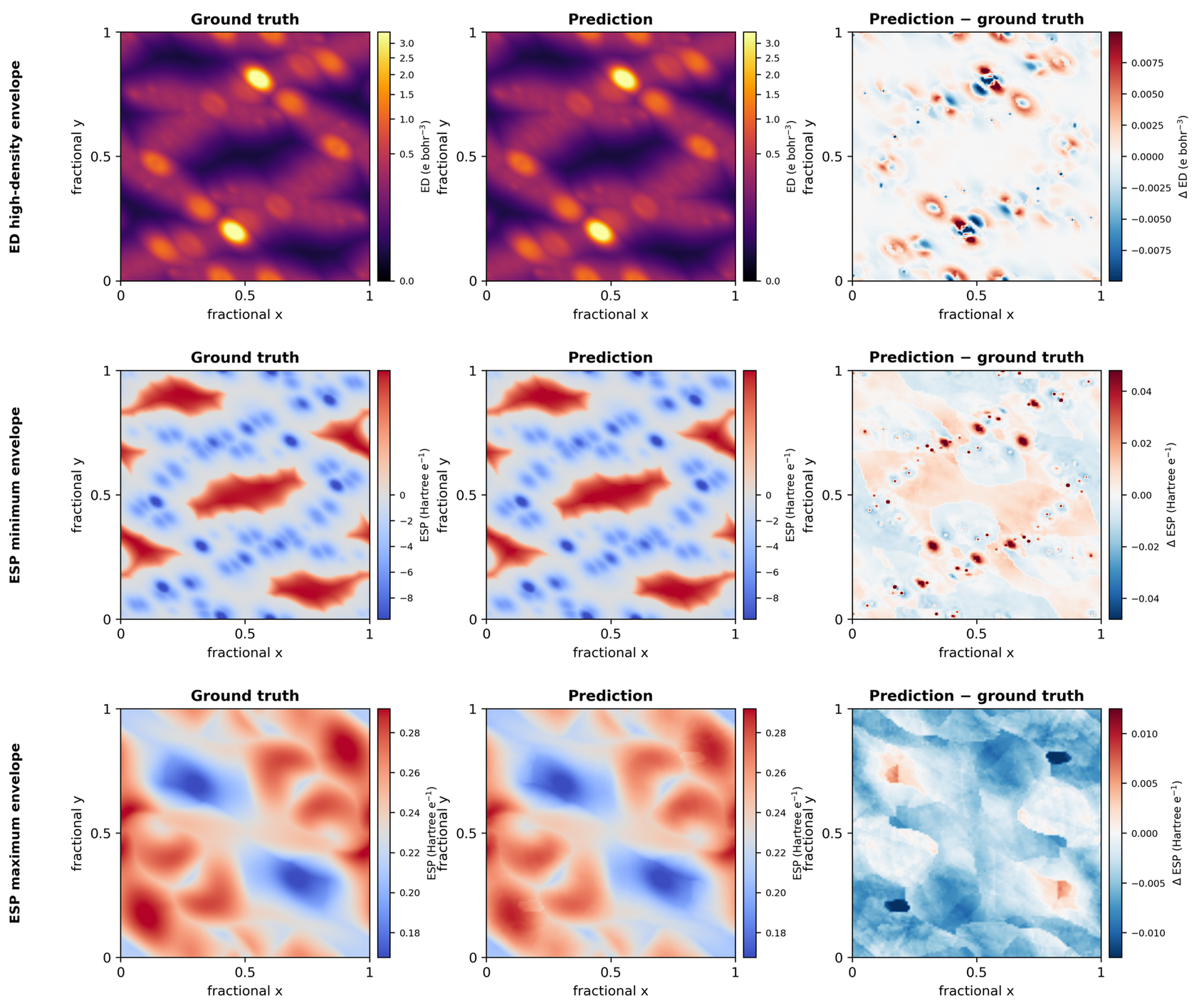}
\caption{
\textbf{Additional comparison of reference and reconstructed electronic fields.}
Reference fields, model predictions and prediction-minus-reference residuals provide complementary views of ED and ESP reconstruction.
Reference and prediction use matching colour scales within each field, while residual scales are centred on zero.
}
\label{fig:reconstruction_additional_example}
\end{figure}

Quantitative test-set evaluation used uniform probes,
reference-field-defined regions and nearest-atom distance intervals
(Supplementary Table~\ref{tab:reconstruction_field_regions}).
Uniform evaluation used 2,000 fixed probes from each of the 2,017
test structures, yielding 4,034,000 evaluated points in total.
nMAE was calculated separately for each MOF and then averaged
across structures, with equal weight assigned to each MOF.
The resulting mean nMAEs were 0.508\% for ED and 8.414\% for ESP.
Pearson's $r$ values were 0.99997 and 0.99791, and Spearman's
$r_{\mathrm S}$ values were 0.98578 and 0.86044, respectively.
These metrics capture complementary aspects of reconstruction
accuracy: strong linear association can coexist with field-value
discrepancies and weaker rank agreement.



Nearest-atom distance intervals further resolve the spatial dependence of reconstruction error. ED nMAEs in the 0--1, 1--2, 2--3 and $3$--$4\,\text{\AA}$ intervals were 0.422\%, 0.980\%, 5.809\% and 42.421\%, respectively. Absolute ED MAEs decrease across these intervals, while normalized errors increase as reference densities become smaller. ESP nMAEs were 1.254\%, 6.601\%, 6.018\% and 9.890\%, showing a different spatial dependence. The first two intervals each contained 2,018 eligible structures, compared with 2,016 and 1,538 in the latter two intervals. Comparisons across intervals therefore also reflect differences in the eligible structure populations. Both fields show reduced linear and rank agreement in more distant regions; in the $3$--$4\,\text{\AA}$ interval, Pearson's $r$ and Spearman's $r_{\mathrm S}$ were 0.93707 and 0.70857 for ED, and 0.62586 and 0.60455 for ESP.

\begin{table}[p]
\centering
\small
\setlength{\tabcolsep}{5pt}
\caption{Reconstruction errors and correlations across evaluation populations.
Distance denotes the distance to the nearest atom.
Each eligible structure contributes 2,000 probes to the corresponding evaluation population.
nMAE is pooled over probes according to Eq.~\ref{eq:supp_reconstruction_metrics}.
Pearson's $r$ measures linear association, and Spearman's $r_{\mathrm S}$ measures rank agreement.
Eligible structure counts for the distance intervals are reported in the text.}
\label{tab:reconstruction_field_regions}
\begin{tabular}{llrrr}
\toprule
Field & Evaluation population & nMAE (\%) & Pearson $r$ & Spearman $r_{\mathrm S}$ \\
\midrule
ED & Uniform & 0.508 & 0.99997 & 0.98578 \\
ED & Distance: 0--1~\AA & 0.422 & 0.99995 & 0.99995 \\
ED & Distance: 1--2~\AA & 0.980 & 0.99985 & 0.99960 \\
ED & Distance: 2--3~\AA & 5.809 & 0.99457 & 0.99200 \\
ED & Distance: 3--4~\AA & 42.421 & 0.93707 & 0.70857 \\
\midrule
ESP & Uniform & 8.414 & 0.99791 & 0.86044 \\
ESP & Distance: 0--1~\AA & 1.254 & 0.99996 & 0.99995 \\
ESP & Distance: 1--2~\AA & 6.601 & 0.99349 & 0.98061 \\
ESP & Distance: 2--3~\AA & 6.018 & 0.89226 & 0.87035 \\
ESP & Distance: 3--4~\AA & 9.890 & 0.62586 & 0.60455 \\
\bottomrule
\end{tabular}
\end{table}

\subsection{Predicted-ED initialization for SCF calculations}

Predicted ED was evaluated as an initial guess for periodic SCF calculations. For each MOF in the SCF evaluation subset, the reconstructed density replaced the standard atomic-density initial guess. The density was subsequently updated through SCF iteration. Paired calculations used the same crystallographic geometry, electronic-structure settings and convergence criterion, allowing the effects of initialization on iteration count and calculation time to be compared.

Let $N_i^{\mathrm{std}}$ and $N_i^{\mathrm{pred}}$ denote the SCF iteration counts for standard and predicted-ED initialization, and let $T_i^{\mathrm{std}}$ and $T_i^{\mathrm{pred}}$ denote the corresponding wall-clock times. Per-MOF percentage reductions were defined as
\begin{equation}
\begin{aligned}
\Delta N_i(\%)
&=
100\left(
1-\frac{N_i^{\mathrm{pred}}}{N_i^{\mathrm{std}}}
\right),\\
\Delta T_i(\%)
&=
100\left(
1-\frac{T_i^{\mathrm{pred}}}{T_i^{\mathrm{std}}}
\right).
\end{aligned}
\label{eq:supp_scf_reductions}
\end{equation}
Positive values indicate a reduction relative to standard initialization; negative values indicate an increase. Reported mean reductions are arithmetic means of the per-MOF percentages.

\begin{figure*}[!htbp]
\centering
\includegraphics[width=0.98\textwidth]{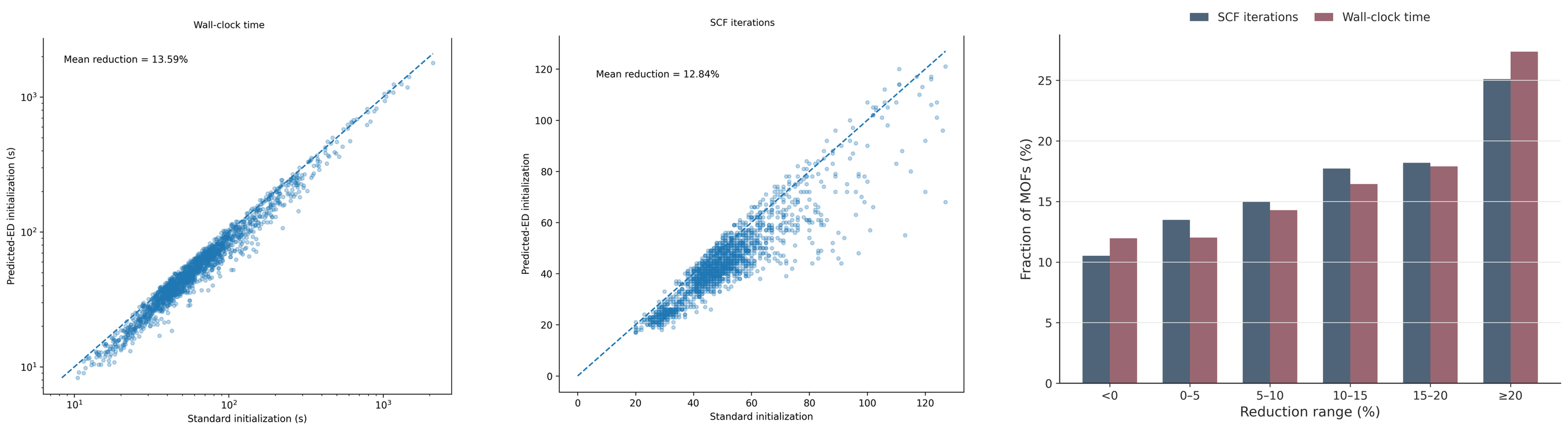}
\caption{
\textbf{Paired evaluation of predicted-ED initialization for SCF calculations.}
Left, wall-clock times with standard initialization and predicted-ED initialization, shown on logarithmic axes.
Middle, the corresponding SCF iteration counts.
Each point represents paired calculations for the same MOF, and dashed lines denote the equality relation $y=x$.
Points below the dashed lines indicate shorter calculation times or fewer iterations with predicted-ED initialization.
The annotated mean reductions are 13.59\% for wall-clock time and 12.84\% for SCF iterations.
Right, fractions of evaluated MOFs in the indicated reduction intervals, shown separately for iteration count and wall-clock time.
Reductions are calculated relative to standard initialization according to Eq.~\ref{eq:supp_scf_reductions}; negative values indicate an increase.
}
\label{fig:supp_scf_initialization}
\end{figure*}

Supplementary Fig.~\ref{fig:supp_scf_initialization} compares the paired calculation times and iteration counts and summarizes their reduction distributions. Predicted-ED initialization gives mean reductions of 12.84\% in SCF iterations and 13.59\% in wall-clock time for the evaluated cohort. Most paired points lie below the equality lines, but the reduction distributions also contain negative values. The results therefore demonstrate average gains with material-dependent variation, including cases in which predicted-ED initialization increases the iteration count or calculation time.

\section{Electronic-field point-cloud construction}
\label{sec:supp_pointcloud}

\subsection{Implementation settings}

This section specifies the preprocessing settings and provides additional visualizations of the point-cloud construction procedure described in the main Methods.

The adaptive Fourier-downsampling settings are listed in Supplementary Table~\ref{tab:fourier_downsampling}. For each native grid dimension $N_{\alpha}$, the target dimension was calculated as
\begin{equation}
N'_{\alpha}
=
\operatorname{round}
\left(
\frac{N_{\alpha}}{s}
\right),
\qquad \alpha\in\{x,y,z\},
\label{eq:pointcloud_target_grid}
\end{equation}
where $s$ is the per-axis factor selected according to the native grid-point count. Rounding was performed independently along each axis. The resized grid retained the original unit cell and grid origin, with updated grid-step vectors. The retained fractions in the table refer to grid points after resizing and before candidate extraction or FPS.

\begin{table}[!htbp]
\centering
\caption{Adaptive Fourier-downsampling settings.
$N_{\mathrm{grid}}$ denotes the native grid-point count.
The factor $s$ is applied along each of the three grid axes.
Retained fractions are approximate because target dimensions are rounded independently.
Grids containing at most $13\times10^6$ points are used at their native resolution.}
\label{tab:fourier_downsampling}
\begin{tabular}{lcc}
\toprule
Native grid points & Per-axis factor $s$ & Retained grid fraction \\
\midrule
$N_{\mathrm{grid}}\leq13\times10^6$ & 1.00 & 100.0\% \\
$13\times10^6<N_{\mathrm{grid}}\leq25\times10^6$ & 1.25 & 51.2\% \\
$25\times10^6<N_{\mathrm{grid}}\leq50\times10^6$ & 1.50 & 29.6\% \\
$N_{\mathrm{grid}}>50\times10^6$ & 2.00 & 12.5\% \\
\bottomrule
\end{tabular}
\end{table}

The candidate-extraction settings and point budgets considered are summarized in Supplementary Table~\ref{tab:pointcloud_settings}. ED thresholds were applied in $\mathrm{e\,Bohr^{-3}}$ to the working grid. ESP fractions were specified separately for the lower and upper tails of each structure's working-grid potential distribution. A fraction of 3\%, 5\% or 7\% per tail therefore gives a combined candidate pool of approximately 6\%, 10\% or 14\% of the working grid, respectively.

FPS used Cartesian Euclidean distances within the primary unit cell. Candidate distances to the nearest selected point were cached and updated incrementally after each selection. The final budget $K$ specifies the total number of points per electronic field. For ESP, FPS was applied to the combined candidate pool, without imposing an equal allocation between the two tails. All candidate pools in the examples below contained at least $K$ positions.

\begin{table}[!htbp]
\centering
\caption{Candidate-extraction and sampling settings considered for point-cloud construction.
ED thresholds and ESP tail fractions are applied before FPS.
ESP fractions are specified per tail, while $K$ denotes the total number of points in the final cloud for one field.
The listed values specify the configurations considered rather than a single default configuration.}
\label{tab:pointcloud_settings}
\begin{tabular}{lp{0.57\linewidth}}
\toprule
Setting & Value \\
\midrule
ED candidate thresholds &
$0.10$, $0.08$, $0.05$, $0.03$ and $0.02\,\mathrm{e\,Bohr^{-3}}$ \\
ESP candidate fractions &
Lowest and highest 3\%, 5\% or 7\% each \\
Combined ESP candidate fraction &
Approximately 6\%, 10\% or 14\% before FPS \\
FPS distance &
Cartesian Euclidean distance within the primary cell, without minimum-image wrapping \\
Point budget $K$ &
4,096; 6,144; 8,192; or 10,240 per field \\
ESP allocation after FPS &
Determined by sampling from the combined candidate pool \\
\bottomrule
\end{tabular}
\end{table}

\subsection{Point-cloud examples}

\begin{figure}[p]
\centering
\includegraphics[width=\linewidth,height=0.78\textheight,keepaspectratio]{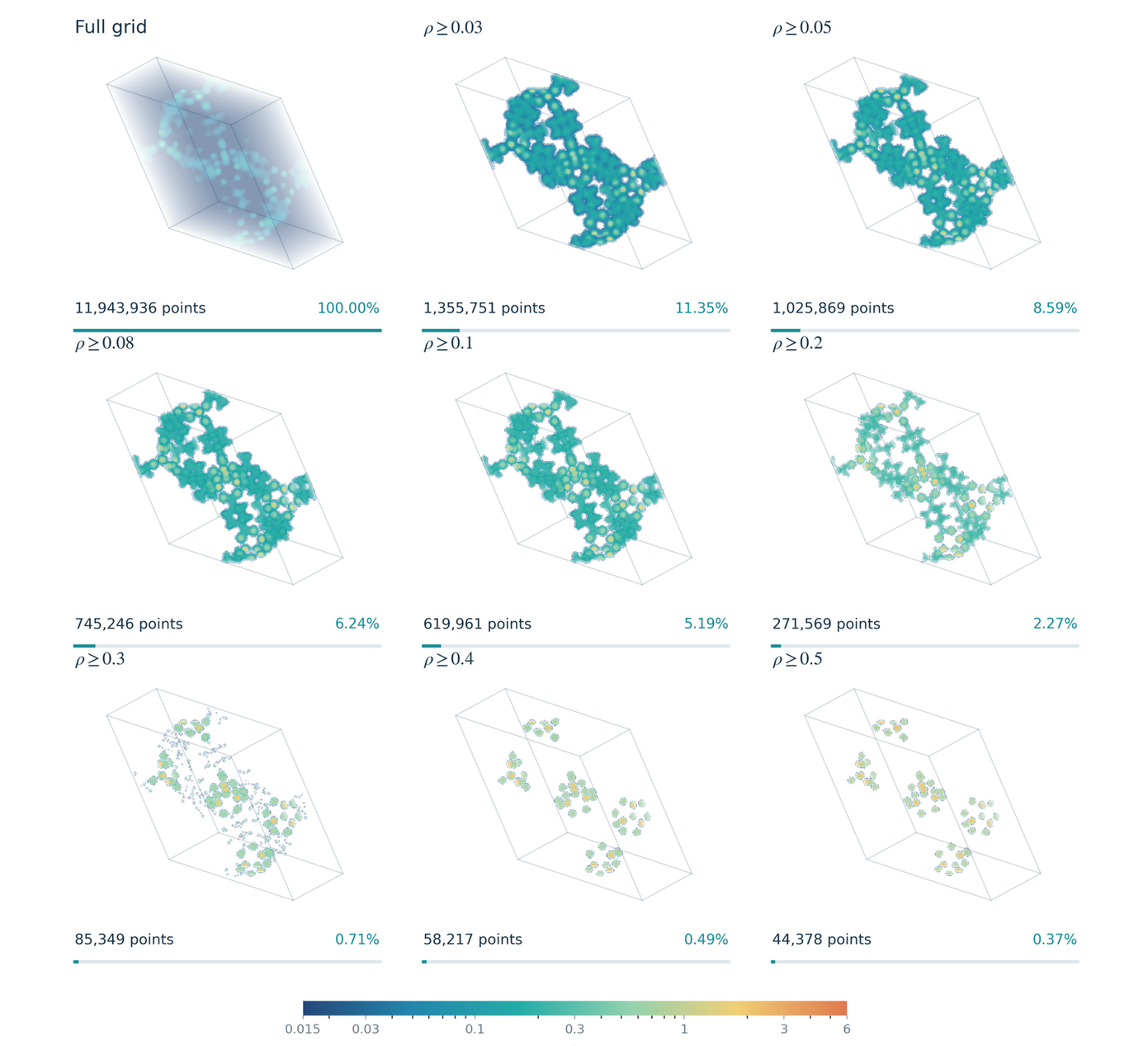}
\caption{
\textbf{ED candidate regions at different density thresholds.}
Panels show the native grid and candidate regions satisfying the indicated thresholds, from $\rho\geq0.03$ to $\rho\geq0.50\,\mathrm{e\,Bohr^{-3}}$.
Point counts and percentages refer to candidate positions relative to the native grid of 11,943,936 points.
Panels use the same cell geometry, viewing direction and logarithmic ED colour scale.
Thresholds above $0.10\,\mathrm{e\,Bohr^{-3}}$ extend the illustration beyond the settings in Supplementary Table~\ref{tab:pointcloud_settings}.
Counts describe the candidate pools before FPS.
}
\label{fig:pointcloud_ed_candidates}
\end{figure}

Supplementary Fig.~\ref{fig:pointcloud_ed_candidates} illustrates ED candidate extraction for a field containing 11,943,936 native grid points, below the Fourier-downsampling trigger. Increasing the density threshold from $0.03$ to $0.10\,\mathrm{e\,Bohr^{-3}}$ reduces the candidate pool from 1,355,751 to 619,961 points, corresponding to 11.35\% and 5.19\% of the native grid. The higher thresholds shown in the figure extend the visualization beyond the preprocessing settings listed in Supplementary Table~\ref{tab:pointcloud_settings}. At $0.50\,\mathrm{e\,Bohr^{-3}}$, the illustrative candidate pool contains 44,378 points.

\begin{figure}[p]
\centering
\includegraphics[width=\linewidth,height=0.78\textheight,keepaspectratio]{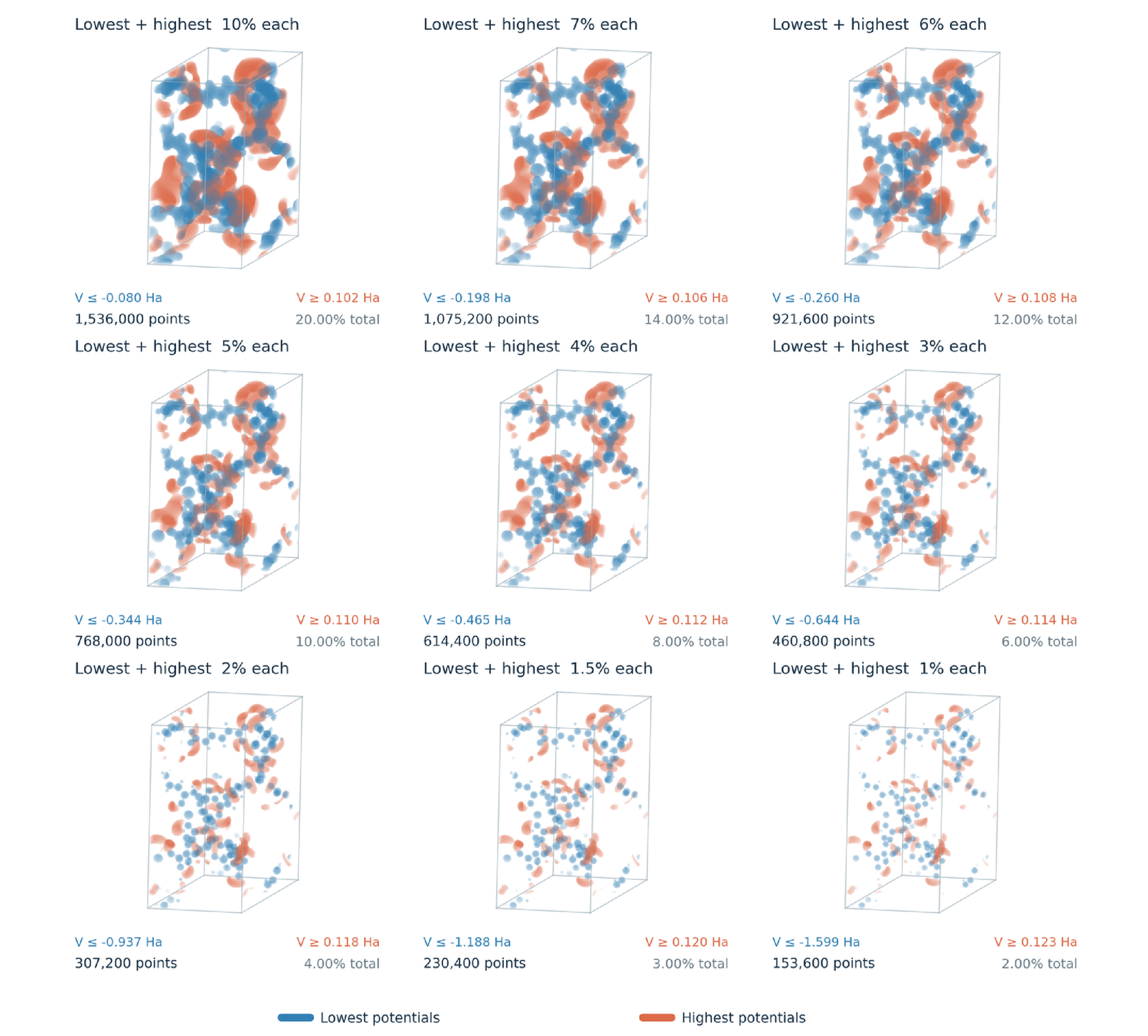}
\caption{
\textbf{ESP candidate regions at different tail fractions.}
Each panel retains the lowest and highest fractions indicated in its title.
Blue and orange identify the lower and upper tails, respectively; their colours can blend where projected points overlap.
Potential cutoffs are structure-specific and expressed in Hartree.
Counts refer to the combined candidate pool before FPS, whose fraction is twice the stated per-tail fraction in these examples.
Panels use the same cell geometry and viewing direction.
Fractions beyond 3\%, 5\% and 7\% extend the visualization beyond the listed preprocessing settings.
}
\label{fig:pointcloud_esp_candidates}
\end{figure}

Supplementary Fig.~\ref{fig:pointcloud_esp_candidates} shows the corresponding effect of varying the ESP tail fraction. Selecting 3\%, 5\% or 7\% from each tail gives combined pools of 460,800, 768,000 and 1,075,200 points, respectively. For the 3\% setting, the structure-specific cutoffs are $V\leq-0.644\,\mathrm{Hartree}$ and $V\geq0.114\,\mathrm{Hartree}$. The cutoffs for the other fractions are shown in the figure. Their unequal magnitudes reflect the asymmetric reference-potential distribution of this example.

\begin{figure}[p]
\centering
\includegraphics[width=\linewidth]{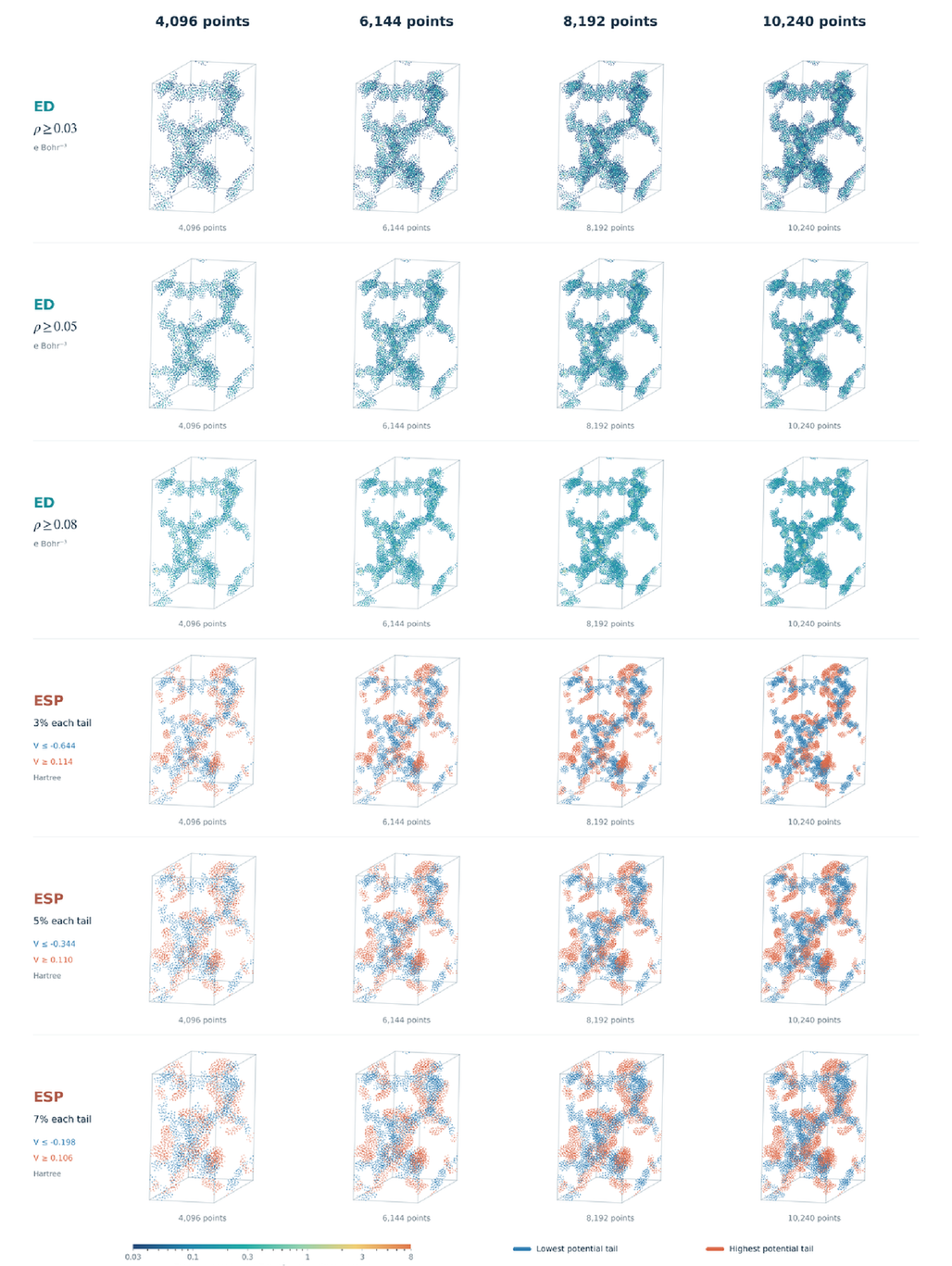}
\caption{
\textbf{ED and ESP point clouds at different sampling budgets: example 1.}
A representative MOF illustrates the effects of filtering and point count.
Rows 1--3 show ED clouds with thresholds of
$\rho\geq0.03$, $0.05$ and $0.08\,\mathrm{e\,Bohr^{-3}}$,
respectively.
Rows 4--6 show ESP clouds retaining candidates from the lowest and
highest 3\%, 5\% and 7\% of potentials, respectively.
Columns correspond to $N=4096$, $6144$, $8192$ and $10240$ sampled
points per field, selected by Cartesian Euclidean FPS.
ED colours indicate density on a common logarithmic scale;
blue and orange identify the lower and upper ESP tails.
ESP cutoffs are expressed in Hartree, and $N$ includes both tails.
All sampled points are displayed, with cell geometry and viewing
direction fixed across panels.
}
\label{fig:pointcloud_fps_ed}
\end{figure}

\begin{figure}[p]
\centering
\includegraphics[width=\linewidth]{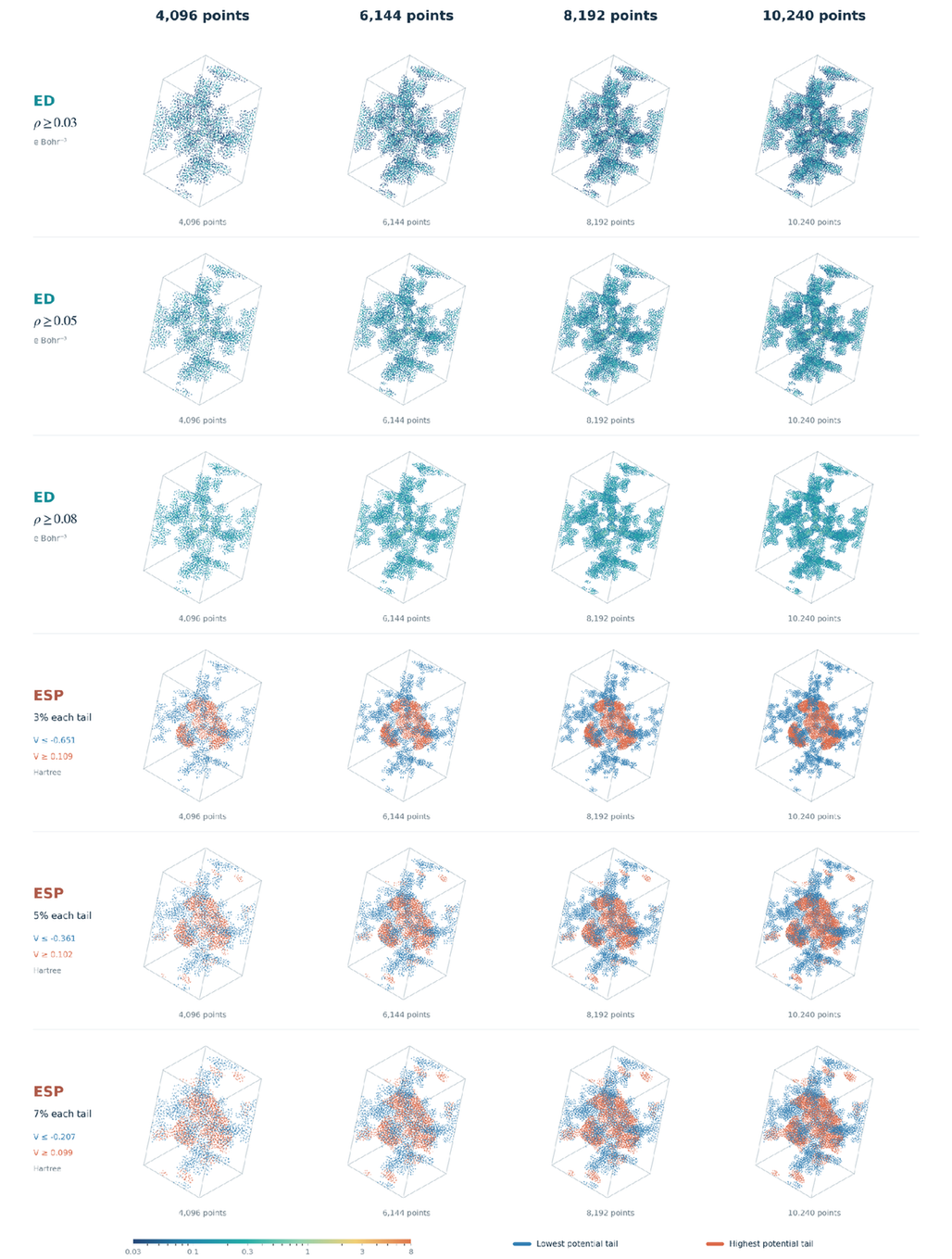}
\caption{
\textbf{ED and ESP point clouds at different sampling budgets: example 2.}
A second representative MOF illustrates the same sampling settings
for a different framework.
Rows 1--3 show ED clouds with thresholds of
$\rho\geq0.03$, $0.05$ and $0.08\,\mathrm{e\,Bohr^{-3}}$,
respectively.
Rows 4--6 show ESP clouds retaining candidates from the lowest and
highest 3\%, 5\% and 7\% of potentials, respectively.
Columns correspond to $N=4096$, $6144$, $8192$ and $10240$ sampled
points per field, selected by Cartesian Euclidean FPS.
ED colours indicate density on a common logarithmic scale;
blue and orange identify the lower and upper ESP tails.
ESP cutoffs are expressed in Hartree, and $N$ includes both tails.
All sampled points are displayed, with cell geometry and viewing
direction fixed across panels.
}
\label{fig:pointcloud_fps_esp}
\end{figure}

Supplementary Figs.~\ref{fig:pointcloud_fps_ed} and
\ref{fig:pointcloud_fps_esp} show ED and ESP point-cloud construction
for two different representative MOFs, illustrating the effects of
candidate filtering and final point count. ED rows use thresholds of $0.03$, $0.05$ and $0.08\,\mathrm{e\,Bohr^{-3}}$, while ESP rows use 3\%, 5\% and 7\% per tail. Within each row, increasing $K$ increases the spatial sampling density under a fixed candidate criterion. Across rows, the extraction criterion changes the eligible spatial regions. Candidate counts in the extraction figures refer to the pools before FPS; the point counts in the FPS figures refer to the final learning representations.

\section{MOF property prediction}
\label{sec:supp_property_prediction}

\subsection{Implementation and selected configurations}

Property-prediction experiments used target-specific labelled subsets of the 40,359 ARC-MOF and QMOF structures. The 13 prediction tasks are summarized in Supplementary Table~\ref{tab:property_tasks}, and the corresponding training, validation and test sample counts are reported in Supplementary Table~\ref{tab:property_splits}. For each target, CIF, ED and ESP models used identical material-level partitions. Reference target distributions are shown in Supplementary Fig.~\ref{fig:property_target_distributions}.

\begin{table*}[htbp]
\centering
\small
\setlength{\tabcolsep}{4pt}
\caption{
Property-prediction tasks included in MOFFieldBench. Selectivity targets are
dimensionless; uptake and working-capacity targets follow the units of the
corresponding source datasets.
}
\label{tab:property_tasks}
\begin{tabular}{
>{\raggedright\arraybackslash}p{0.25\textwidth}
>{\raggedright\arraybackslash}p{0.12\textwidth}
>{\raggedright\arraybackslash}p{0.16\textwidth}
>{\raggedright\arraybackslash}p{0.37\textwidth}}
\toprule
Task & Category & Target type & Physical interpretation \\
\midrule
Pore-limiting diameter (PLD/Df)
& Geometry
& Pore size
& Limiting diameter accessible through the pore network \\

Largest cavity diameter (LCD/Di)
& Geometry
& Pore size
& Diameter of the largest accessible pore cavity \\

Density
& Geometry
& Framework property
& Crystal/framework density \\

Band gap
& Quantum
& Electronic property
& Electronic band gap \\

Post-combustion selectivity
& Adsorption
& Selectivity
& Gas-mixture separation under post-combustion conditions \\

Natural-gas selectivity
& Adsorption
& Selectivity
& Gas-mixture separation under natural-gas purification conditions \\

Landfill-gas selectivity
& Adsorption
& Selectivity
& Gas-mixture separation under landfill-gas conditions \\

Pre-combustion selectivity
& Adsorption
& Selectivity
& Gas-mixture separation under pre-combustion conditions \\

Pre-combustion working capacity
& Adsorption
& Working capacity
& Deliverable capacity under pre-combustion conditions \\

Methane-storage PSA uptake
& Adsorption
& Uptake
& Methane uptake under storage conditions \\

Methane-storage PSA working capacity
& Adsorption
& Working capacity
& Deliverable methane-storage capacity \\

Post-combustion VSA working capacity
& Adsorption
& Working capacity
& Deliverable capacity under post-combustion conditions \\

Natural-gas purification uptake
& Adsorption
& Uptake
& Gas uptake under natural-gas purification conditions \\
\bottomrule
\end{tabular}
\end{table*}

\begin{table*}[htbp]
\centering
\small
\setlength{\tabcolsep}{6pt}
\caption{
Number of structures in the training, validation and test partitions for each
property-prediction task. Partitions are shared across CIF, ED and ESP for a given target.
}
\label{tab:property_splits}
\begin{tabular}{>{\raggedright\arraybackslash}p{0.54\textwidth}rrr}
\toprule
Task & Training & Validation & Test \\
\midrule
PLD / Df & 32,274 & 4,034 & 4,035 \\
LCD / Di & 32,274 & 4,034 & 4,035 \\
Density & 32,273 & 4,034 & 4,035 \\
Band gap & 12,347 & 1,543 & 1,545 \\
Post-combustion VSA selectivity & 19,877 & 2,484 & 2,486 \\
Natural-gas purification selectivity & 19,777 & 2,470 & 2,477 \\
Landfill-gas VPSA selectivity & 19,777 & 2,470 & 2,472 \\
Pre-combustion selectivity & 19,877 & 2,484 & 2,486 \\
Methane-storage PSA uptake & 19,877 & 2,484 & 2,486 \\
Methane-storage PSA working capacity & 19,877 & 2,484 & 2,486 \\
Post-combustion VSA working capacity & 19,877 & 2,484 & 2,486 \\
Natural-gas purification uptake & 19,877 & 2,484 & 2,486 \\
Pre-combustion working capacity & 19,877 & 2,484 & 2,486 \\
\bottomrule
\end{tabular}
\end{table*}

 ED and ESP models used an X-3D-based point-cloud encoder~\cite{sun2024x}. Both field modalities used the same encoder family within each task and were optimized with AdamW. Their point-cloud configurations, learning rates, and weight decay are reported together in Supplementary Table~\ref{tab:property_pointcloud_configurations}.

All regression models were trained for 50 epochs using MSE loss. Checkpoints were selected by the lowest validation MAE for CIF and the lowest validation RMSE for ED and ESP. Representation configurations were selected using validation performance before evaluation on the held-out test sets. The regression objective and evaluation metrics are defined in the main Methods.

\begin{table*}[htbp]
\centering
\small
\setlength{\tabcolsep}{2pt}
\renewcommand{\arraystretch}{1.1}
\caption{
Selected point-cloud and optimization settings for ED- and ESP-based
property prediction. ED thresholds $\tau_{\rho}$ are in
$\mathrm{e\,Bohr^{-3}}$, and ESP fractions $q$ apply to each tail before
FPS. $K$ denotes the point count, $k$ the Fourier-encoding order, LR
the learning rate and WD the weight decay. Dashes indicate unreported
batch sizes.
}
\label{tab:property_pointcloud_configurations}
\begin{tabular*}{\textwidth}{@{\extracolsep{\fill}}>{\raggedright\arraybackslash}p{0.25\textwidth}*{12}{r}@{}}
\toprule
& \multicolumn{6}{c}{ED} & \multicolumn{6}{c}{ESP} \\
\cmidrule(lr){2-7}\cmidrule(lr){8-13}
Task & $\tau_{\rho}$ & $K$ & $k$ & LR & WD & Batch
     & $q$ & $K$ & $k$ & LR & WD & Batch \\
\midrule
PLD / Df
& 0.02 & 4,096 & 1 & $3\times10^{-4}$ & $10^{-4}$ & 16
& 3\% & 4,096 & 1 & $3\times10^{-4}$ & $10^{-4}$ & 16 \\
LCD / Di
& 0.02 & 4,096 & 3 & $3\times10^{-4}$ & $10^{-4}$ & 16
& 3\% & 4,096 & 1 & $3\times10^{-4}$ & $10^{-4}$ & 16 \\
Density
& 0.03 & 8,192 & 2 & $2\times10^{-4}$ & $10^{-4}$ & 8
& 3\% & 8,192 & 2 & $2\times10^{-4}$ & $10^{-4}$ & 8 \\
Band gap
& 0.03 & 10,240 & 2 & $3\times10^{-4}$ & $10^{-4}$ & --
& 3\% & 8,192 & 2 & $1.52\times10^{-3}$ & $2.19\times10^{-4}$ & -- \\
Post-combustion VSA selectivity
& 0.02 & 4,096 & 2 & $3\times10^{-4}$ & $10^{-4}$ & 16
& 3\% & 8,192 & 2 & $3\times10^{-4}$ & $10^{-4}$ & 8 \\
Natural-gas purification selectivity
& 0.02 & 4,096 & 2 & $2\times10^{-4}$ & $10^{-4}$ & 12
& 5\% & 8,192 & 2 & $2\times10^{-4}$ & $10^{-4}$ & 8 \\
Landfill-gas VPSA selectivity
& 0.02 & 8,192 & 2 & $3\times10^{-4}$ & $10^{-4}$ & 8
& 5\% & 8,192 & 2 & $3\times10^{-4}$ & $10^{-4}$ & 8 \\
Pre-combustion selectivity
& 0.08 & 4,096 & 2 & $3\times10^{-4}$ & $10^{-4}$ & 16
& 7\% & 8,192 & 2 & $3\times10^{-4}$ & $10^{-4}$ & 8 \\
Methane-storage PSA uptake
& 0.02 & 8,192 & 1 & $3\times10^{-4}$ & $10^{-4}$ & 8
& 3\% & 8,192 & 1 & $3\times10^{-4}$ & $10^{-4}$ & 8 \\
Methane-storage PSA working capacity
& 0.02 & 8,192 & 3 & $3\times10^{-4}$ & $10^{-4}$ & 8
& 5\% & 10,240 & 1 & $3\times10^{-4}$ & $10^{-4}$ & 6 \\
Post-combustion VSA working capacity
& 0.05 & 4,096 & 1 & $3\times10^{-4}$ & $10^{-4}$ & 16
& 5\% & 8,192 & 1 & $3\times10^{-4}$ & $10^{-4}$ & 8 \\
Natural-gas purification uptake
& 0.02 & 4,096 & 2 & $3\times10^{-4}$ & $10^{-4}$ & 16
& 3\% & 8,192 & 2 & $3\times10^{-4}$ & $10^{-4}$ & 8 \\
Pre-combustion working capacity
& 0.02 & 4,096 & 2 & $3\times10^{-4}$ & $10^{-4}$ & 16
& 5\% & 10,240 & 2 & $3\times10^{-4}$ & $10^{-4}$ & 6 \\
\bottomrule
\end{tabular*}
\end{table*}

\begin{figure}[p]
\centering
\includegraphics[width=\linewidth,height=0.72\textheight,keepaspectratio]{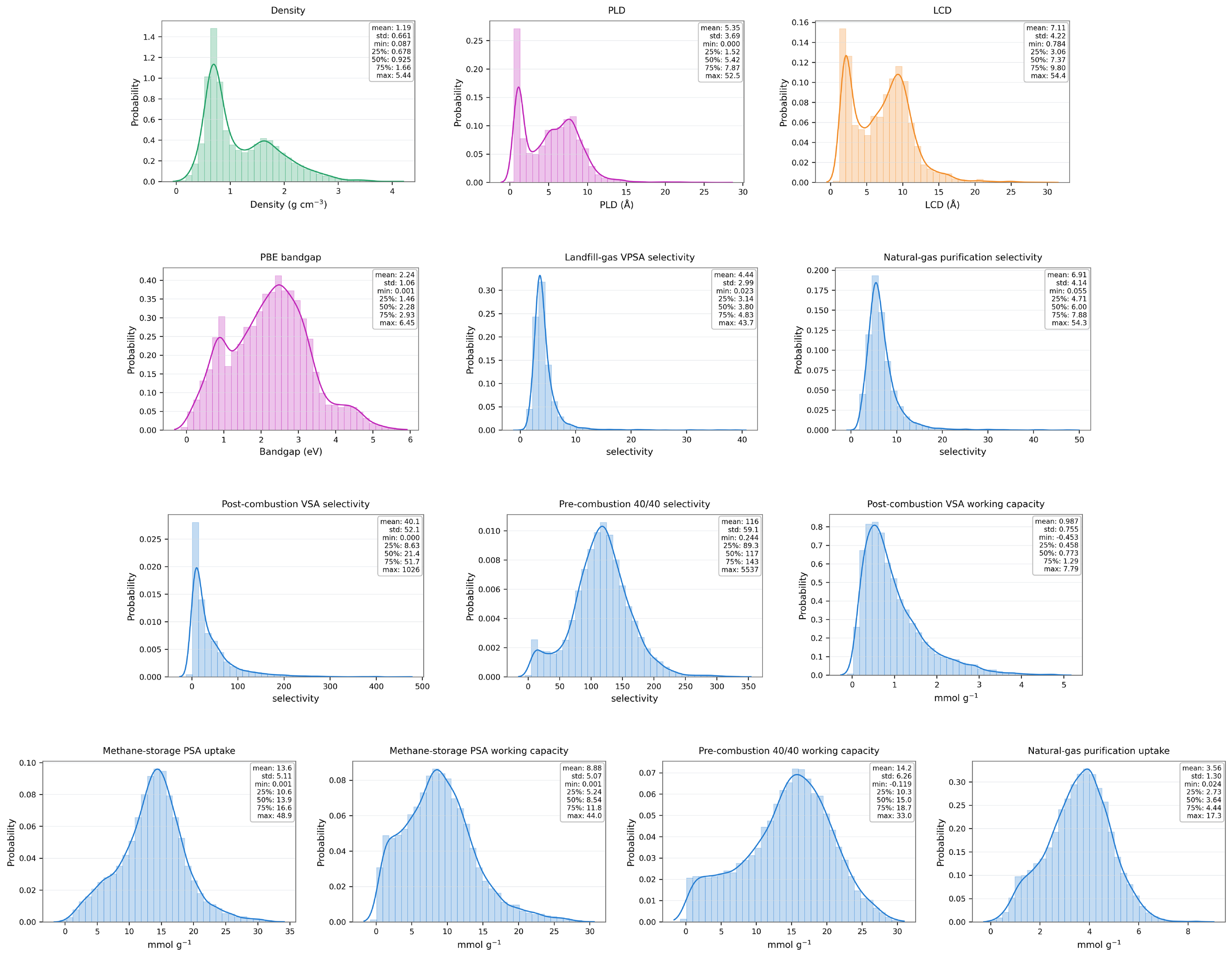}
\caption{
\textbf{Reference target distributions.}
Distributions of reference property values in the labelled subsets used for the corresponding prediction tasks. Targets are expressed in their respective original units.
}
\label{fig:property_target_distributions}
\end{figure}

\subsection{Representation sensitivity}

The point budgets and candidate-extraction ranges follow Supplementary Table~\ref{tab:pointcloud_settings}. Periodic encoding orders $k=0$, 1, 2 and 3 were also examined; $k=0$ omits the explicit Fourier positional features while retaining the Cartesian geometry input.

Point-budget analyses are shown in Supplementary Figs.~\ref{fig:property_point_count} and \ref{fig:property_point_count_additional}, and encoding-order analyses in Supplementary Figs.~\ref{fig:property_pe_frequency} and \ref{fig:property_pe_frequency_additional}. Within the examined configurations, increasing the point count or encoding order does not produce a consistent improvement across targets. Preferred settings differ between tasks and field modalities.

Sensitivity to candidate extraction is shown for ED in Supplementary Fig.~\ref{fig:property_extraction} and for ESP in Supplementary Figs.~\ref{fig:property_extraction_esp} and \ref{fig:property_extraction_esp_mae}. These analyses complement the selected configurations by showing how predictive performance varies over the examined thresholds and tail fractions. They support selecting representation settings for each target and modality using validation performance.

\begin{figure}[p]
\centering
\includegraphics[width=\linewidth,height=0.72\textheight,keepaspectratio]{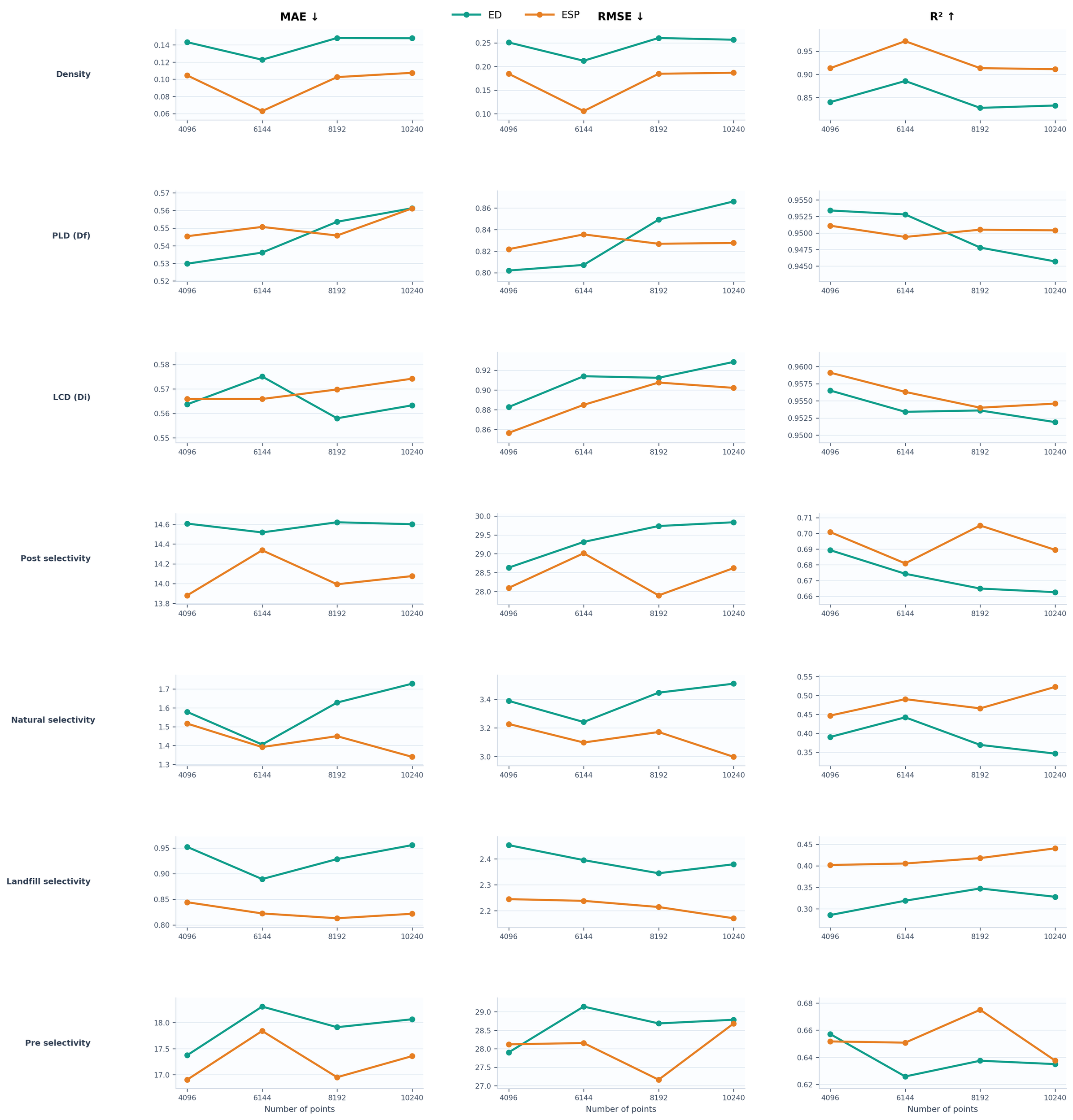}
\caption{
\textbf{Sensitivity to point-cloud size: tasks 1--7.}
ED and ESP prediction performance at point budgets of 4,096, 6,144, 8,192 and 10,240 for the first seven benchmark tasks, as identified in the panels. Panels report the indicated prediction metrics; error metrics retain the original units of each target.
}
\label{fig:property_point_count}
\end{figure}
\begin{figure}[p]
\centering
\includegraphics[width=\linewidth,height=0.72\textheight,keepaspectratio]{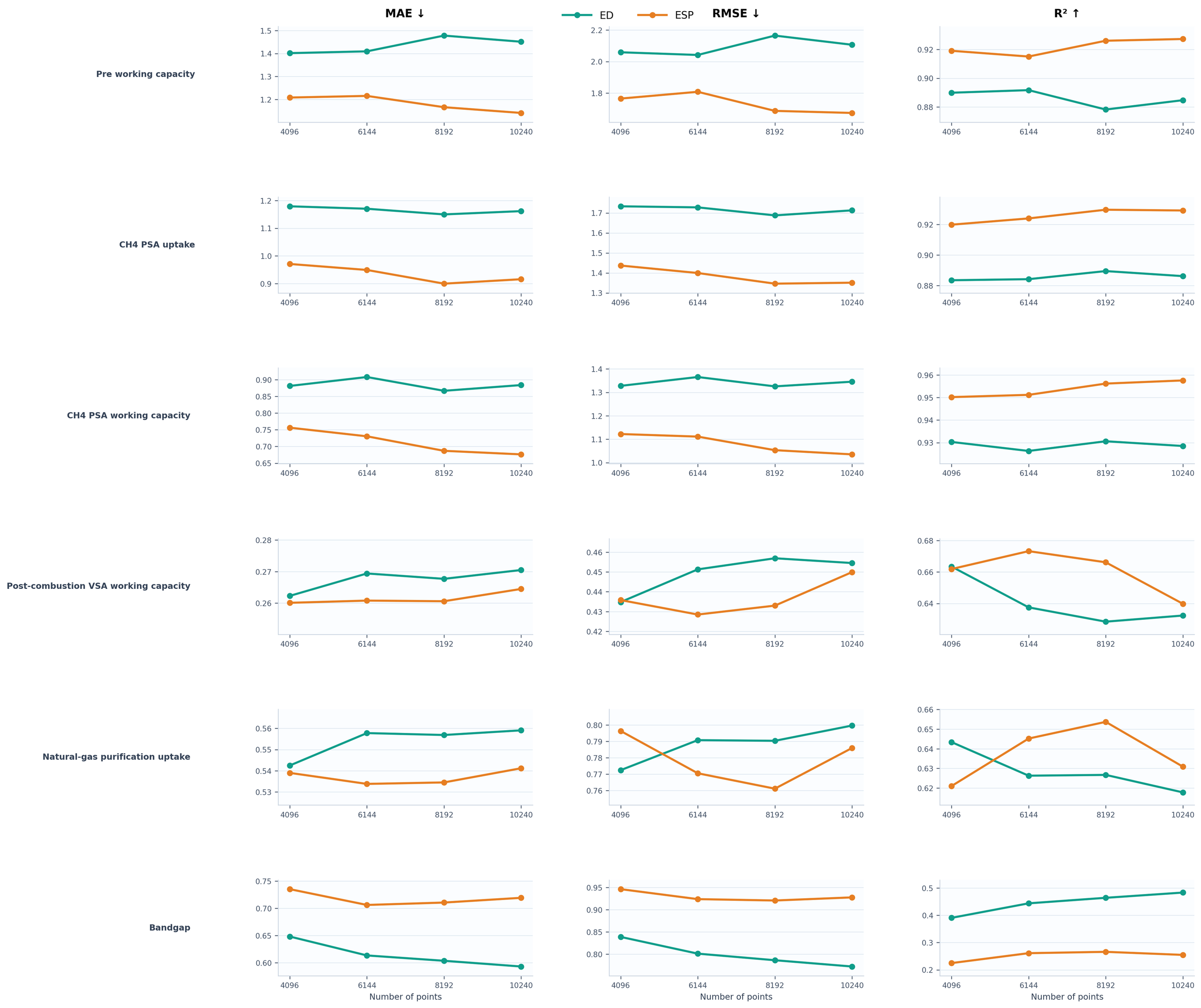}
\caption{
\textbf{Sensitivity to point-cloud size: tasks 8--13.}
ED and ESP prediction performance at point budgets of 4,096, 6,144, 8,192 and 10,240 for the remaining six benchmark tasks. Panel labels identify the targets and metrics. These results complement Supplementary Fig.~\ref{fig:property_point_count}.
}
\label{fig:property_point_count_additional}
\end{figure}
\begin{figure}[p]
\centering
\includegraphics[width=\linewidth,height=0.72\textheight,keepaspectratio]{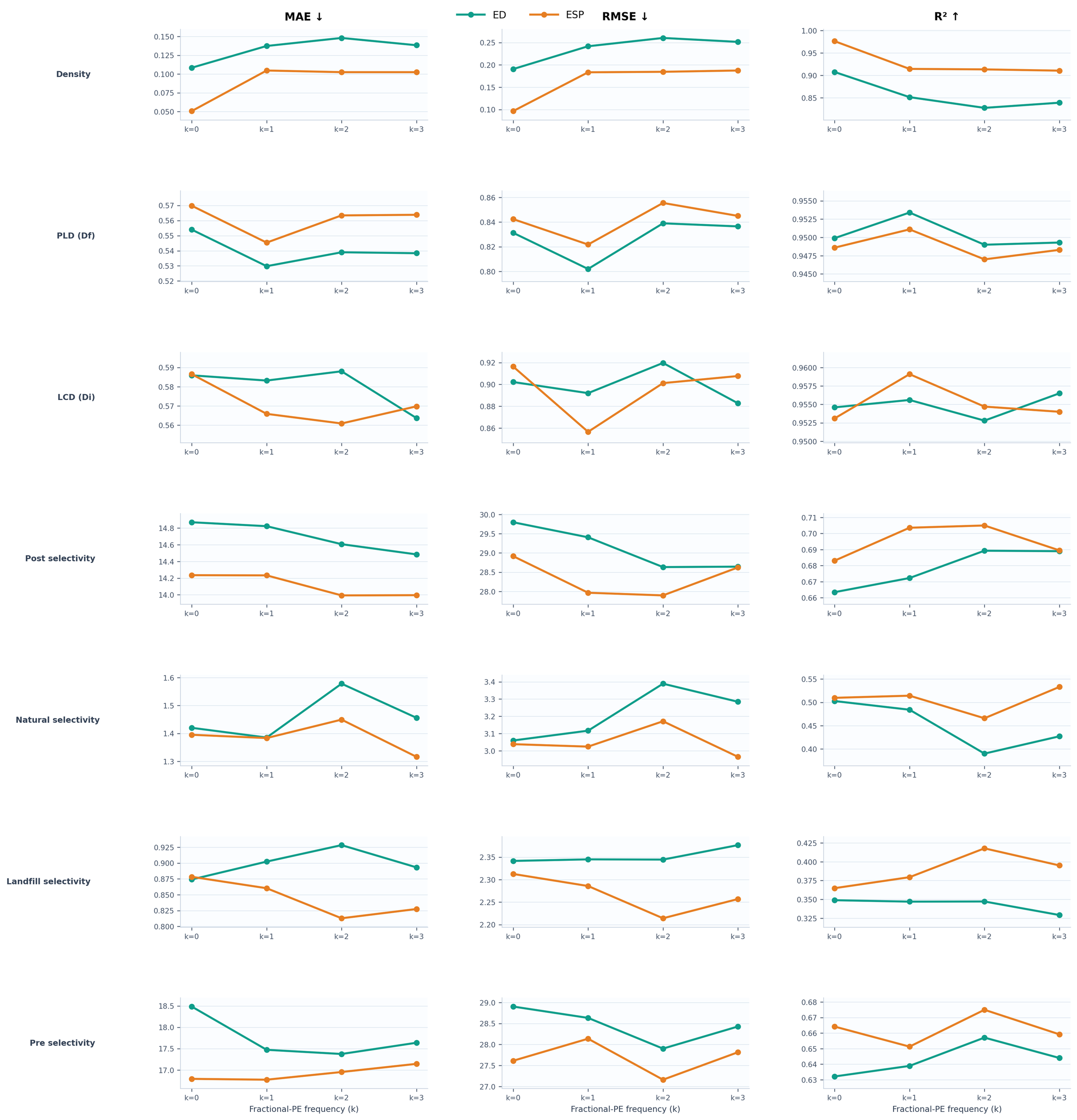}
\caption{
\textbf{Sensitivity to periodic encoding order: tasks 1--7.}
ED and ESP prediction performance for $k=0$, 1, 2 and 3, where $k$ is the number of Fourier frequency components per fractional-coordinate axis. The $k=0$ setting omits explicit Fourier positional features. Panels identify the properties and prediction metrics.
}
\label{fig:property_pe_frequency}
\end{figure}
\begin{figure}[p]
\centering
\includegraphics[width=\linewidth,height=0.72\textheight,keepaspectratio]{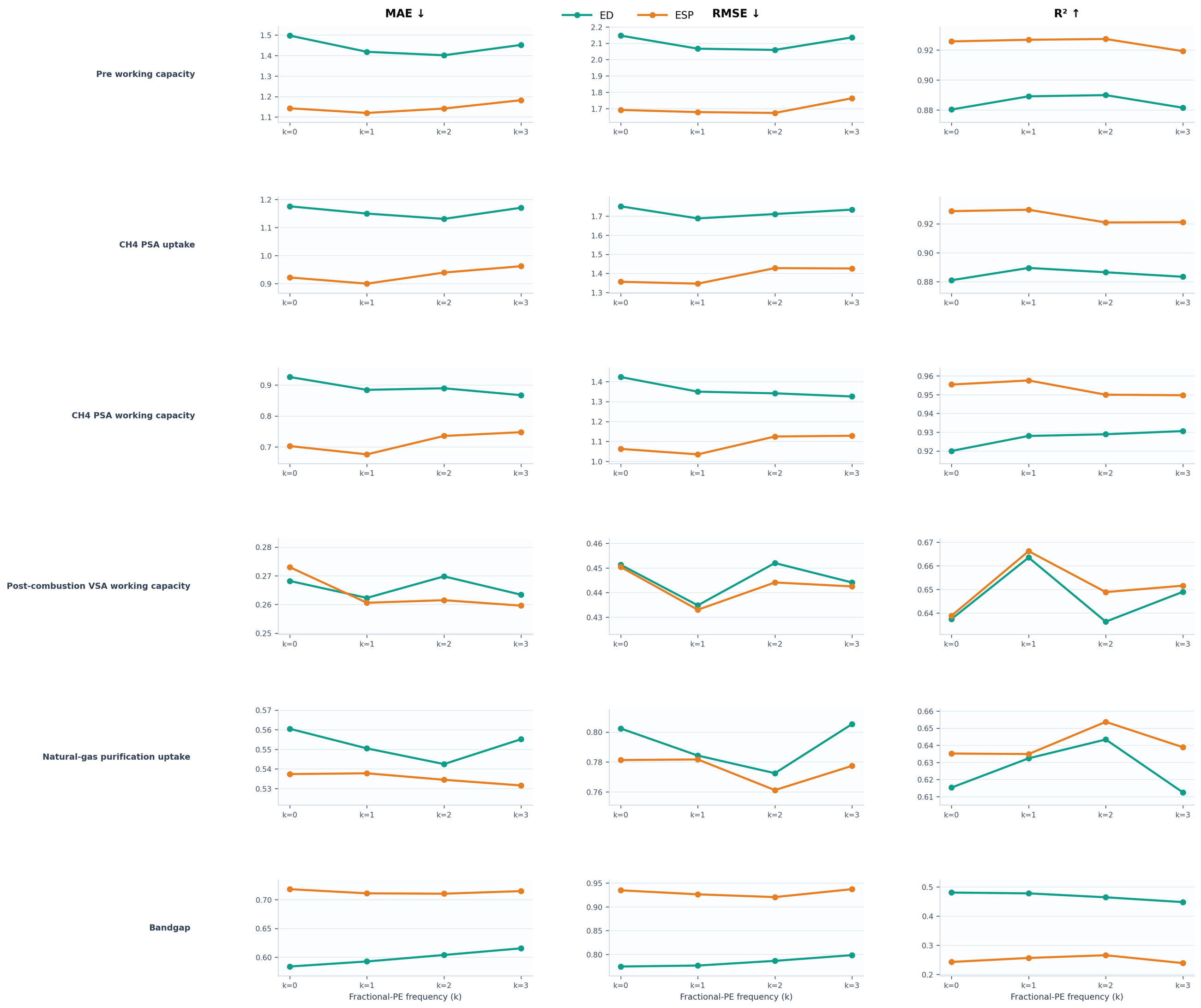}
\caption{
\textbf{Sensitivity to periodic encoding order: tasks 8--13.}
ED and ESP prediction performance for the remaining six tasks at encoding orders $k=0$, 1, 2 and 3. Panel labels identify the targets and metrics. These results complement Supplementary Fig.~\ref{fig:property_pe_frequency}.
}
\label{fig:property_pe_frequency_additional}
\end{figure}
\begin{figure}[p]
\centering
\includegraphics[width=\linewidth,height=0.72\textheight,keepaspectratio]{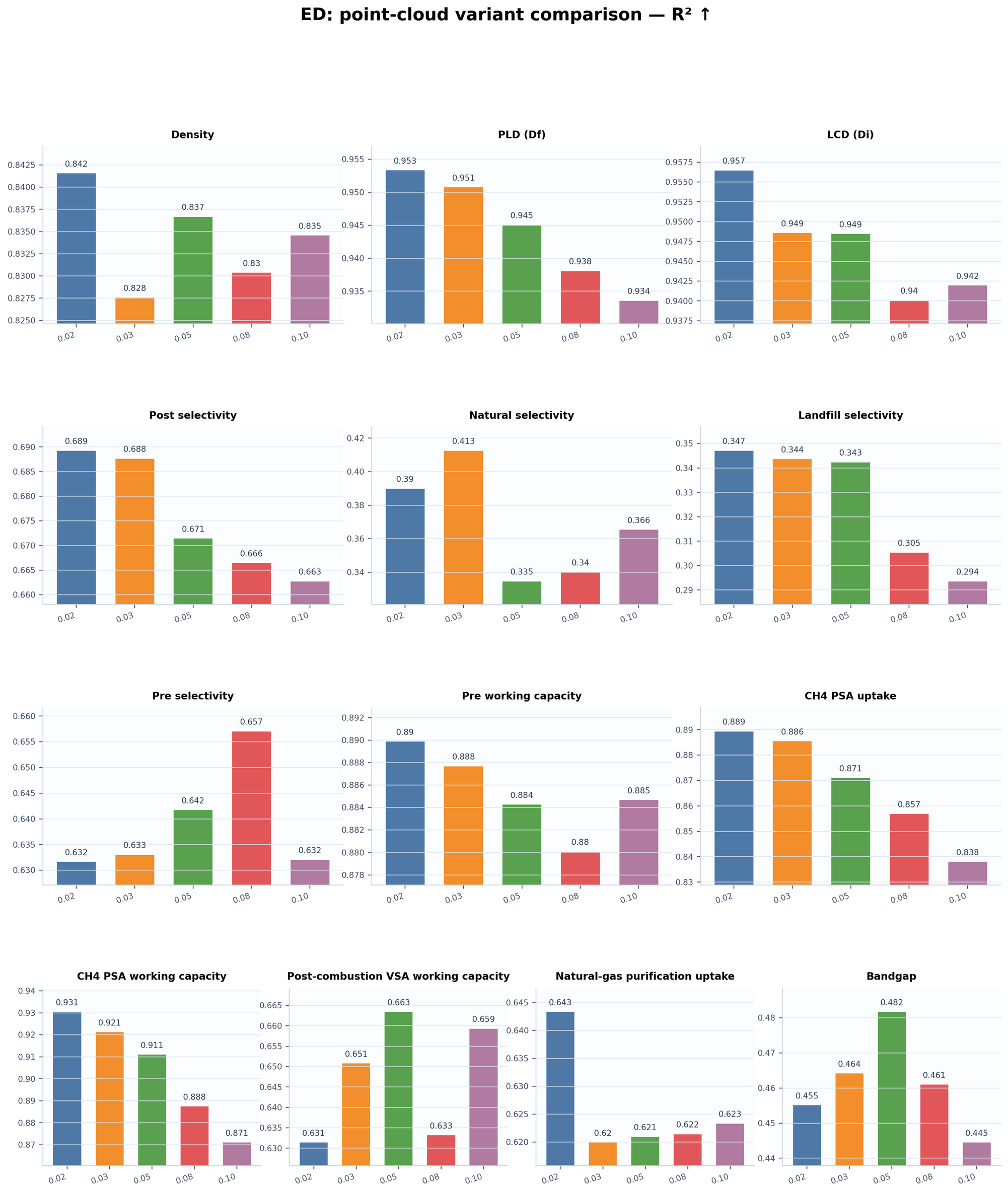}
\caption{
\textbf{Sensitivity of ED prediction to the candidate threshold.}
Bars show $R^2$ across the 13 prediction tasks for ED thresholds of $0.02$, $0.03$, $0.05$, $0.08$ and $0.10\,\mathrm{e\,Bohr^{-3}}$, applied before FPS. Higher $R^2$ indicates better predictive performance.
}
\label{fig:property_extraction}
\end{figure}
\begin{figure}[p]
\centering
\includegraphics[width=\linewidth,height=0.72\textheight,keepaspectratio]{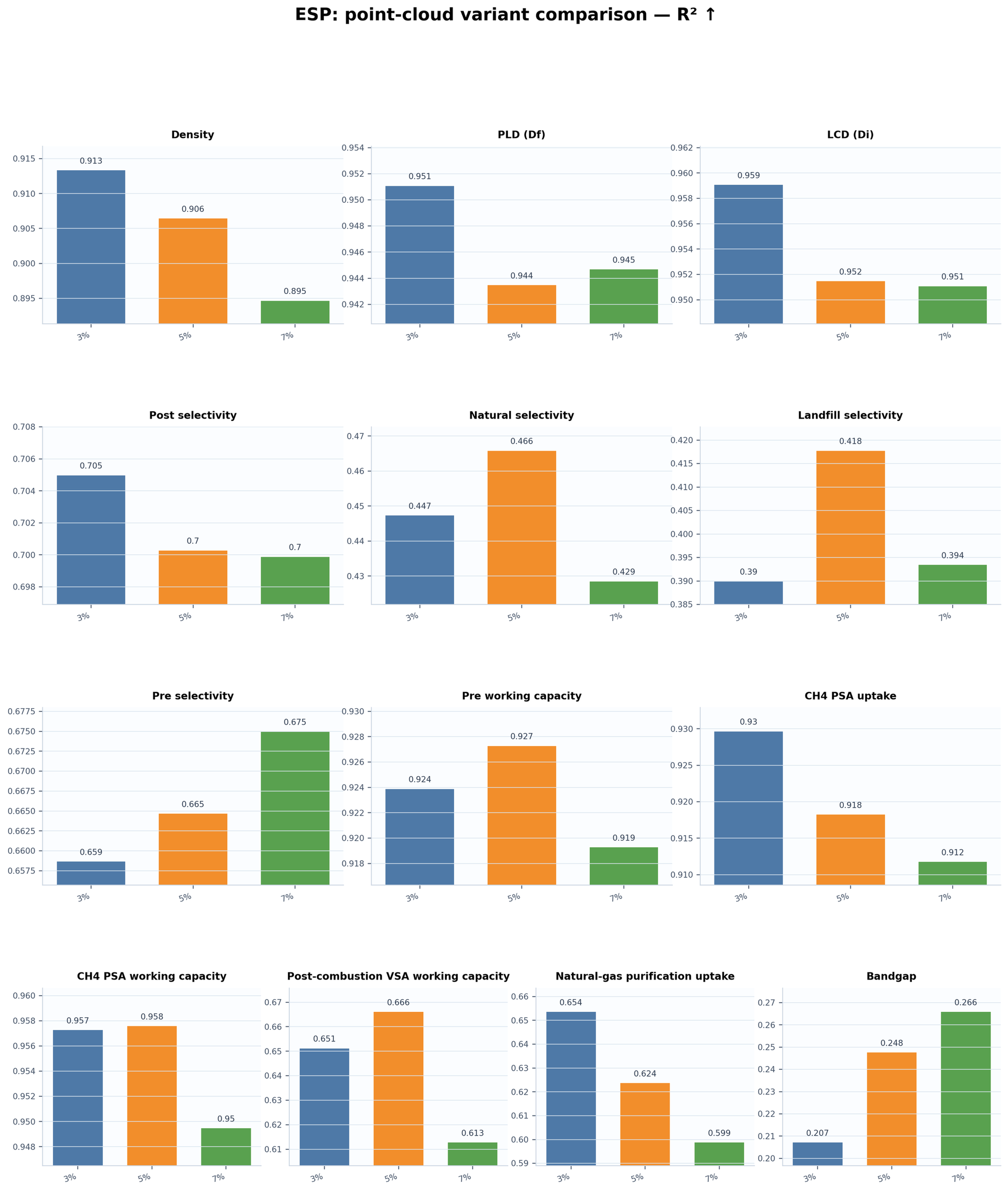}
\caption{
\textbf{Sensitivity of ESP prediction to the candidate tail fraction.}
Bars show $R^2$ across the 13 prediction tasks for candidate pools containing the lowest and highest 3\%, 5\% or 7\% of potentials each. Fractions are specified per tail before FPS. Higher $R^2$ indicates better predictive performance.
}
\label{fig:property_extraction_esp}
\end{figure}
\begin{figure}[p]
\centering
\includegraphics[width=\linewidth,height=0.72\textheight,keepaspectratio]{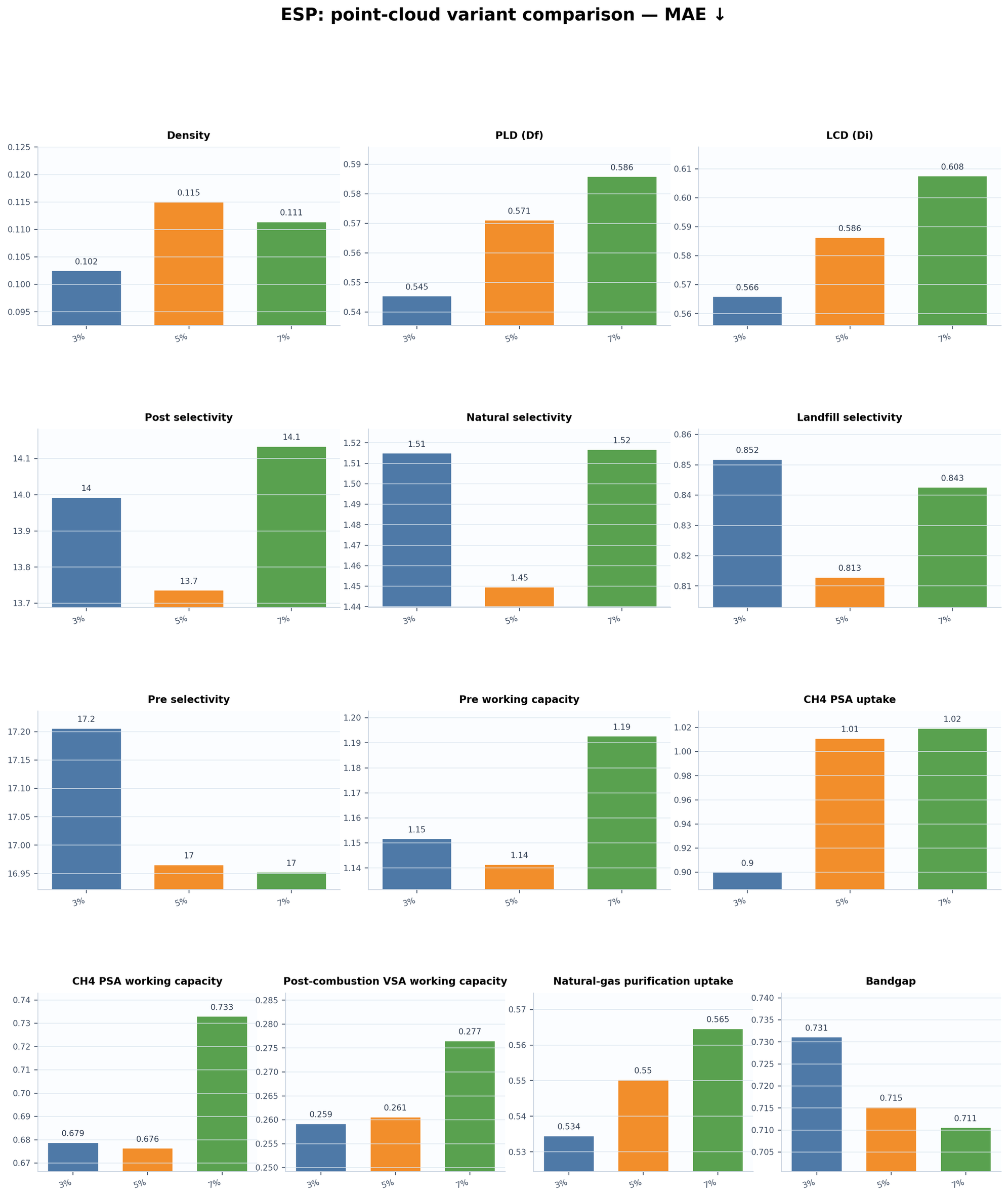}
\caption{
\textbf{ESP prediction errors at different candidate tail fractions.}
Bars show MAE across the 13 prediction tasks for candidate pools containing the lowest and highest 3\%, 5\% or 7\% of potentials each. MAE is reported in the original units of each target, with lower values indicating better accuracy.
}
\label{fig:property_extraction_esp_mae}
\end{figure}

\subsection{Complete results and embedding visualizations}

Complete test-set metrics for the selected unimodal models are reported in Supplementary Table~\ref{tab:unimodal_property_results}. MAE, RMSE and $R^2$ describe the same selected model for each target and modality. Bold MAE entries identify the unimodal reference used for the corresponding task; they do not designate the best value of every metric.

\begin{table*}[!htbp]
\centering
\small
\caption{Complete unimodal property-prediction results.
Entries report MAE / RMSE / $R^2$ for the selected models on the held-out test set.
Bold MAE values indicate the lowest MAE among CIF, ED and ESP for each target.
Pore diameters are expressed in \AA, density in $\mathrm{g\,cm^{-3}}$, band gap in eV, and uptake and working capacity in $\mathrm{mmol\,g^{-1}}$.
Selectivity is dimensionless.
MAE and RMSE use the corresponding target units.}
 \label{tab:unimodal_property_results}
  \resizebox{\textwidth}{!}{
  \begin{tabular}{lccc}
  \toprule
  Property & CIF & ED & ESP \\
  \midrule
  PLD / Df & 0.9396 / 1.3533 / 0.8673 & \textbf{0.5298} / 0.8020 / 0.9534 & 0.5454 / 0.8218 / 0.9511 \\
  LCD / Di & 0.9864 / 1.3949 / 0.8914 & \textbf{0.5637} / 0.8827 / 0.9565 & 0.5659 / 0.8565 / 0.9591 \\
  Density & 0.0713 / 0.0882 / 0.9802 & 0.1032 / 0.1872 / 0.9109 & \textbf{0.0508} / 0.0916 / 0.9787 \\
  Band gap & \textbf{0.2974} / 0.4056 / 0.8575 & 0.5930 / 0.7720 / 0.4830 & 0.7110 / 0.9200 / 0.2660 \\
  Post-combustion VSA selectivity & 15.124 / 31.507 / 0.6238 & 14.571 / 28.382 / 0.6947 & \textbf{13.992} / 27.898 / 0.7050 \\
  Natural-gas purification selectivity & 1.4362 / 3.0797 / 0.4964 & 1.4050 / 3.1875 / 0.4605 & \textbf{1.3510} / 2.9496 / 0.5380 \\
  Landfill-gas VPSA selectivity & 0.8932 / 2.2699 / 0.3881 & 0.9283 / 2.3445 / 0.3472 & \textbf{0.8129} / 2.2142 / 0.4178 \\
  Pre-combustion selectivity & 19.414 / 31.438 / 0.5647 & 17.374 / 27.902 / 0.6571 & \textbf{16.952} / 27.163 / 0.6750 \\
  Methane-storage PSA uptake & 1.2642 / 1.5762 / 0.9036 & 1.1498 / 1.6878 / 0.8895 & \textbf{0.9000} / 1.3463 / 0.9297 \\
  Methane-storage PSA working capacity & 1.4276 / 1.7150 / 0.8838 & 0.8669 / 1.3256 / 0.9306 & \textbf{0.6764} / 1.0354 / 0.9576 \\
  Post-combustion VSA working capacity & 0.2682 / 0.5279 / 0.5039 & 0.2623 / 0.4348 / 0.6635 & \textbf{0.2606} / 0.4330 / 0.6662 \\
  Natural-gas purification uptake & 0.5690 / 0.7952 / 0.6221 & 0.5425 / 0.7725 / 0.6434 & \textbf{0.5344} / 0.7612 / 0.6537 \\
  Pre-combustion working capacity & 2.5415 / 2.9406 / 0.7754 & 1.4016 / 2.0591 / 0.8899 & \textbf{1.1414} / 1.6735 / 0.9273 \\
  \bottomrule
  \end{tabular}
  }
  \end{table*}

\begin{figure}[p]
\centering
\includegraphics[width=\linewidth,height=0.72\textheight,keepaspectratio]{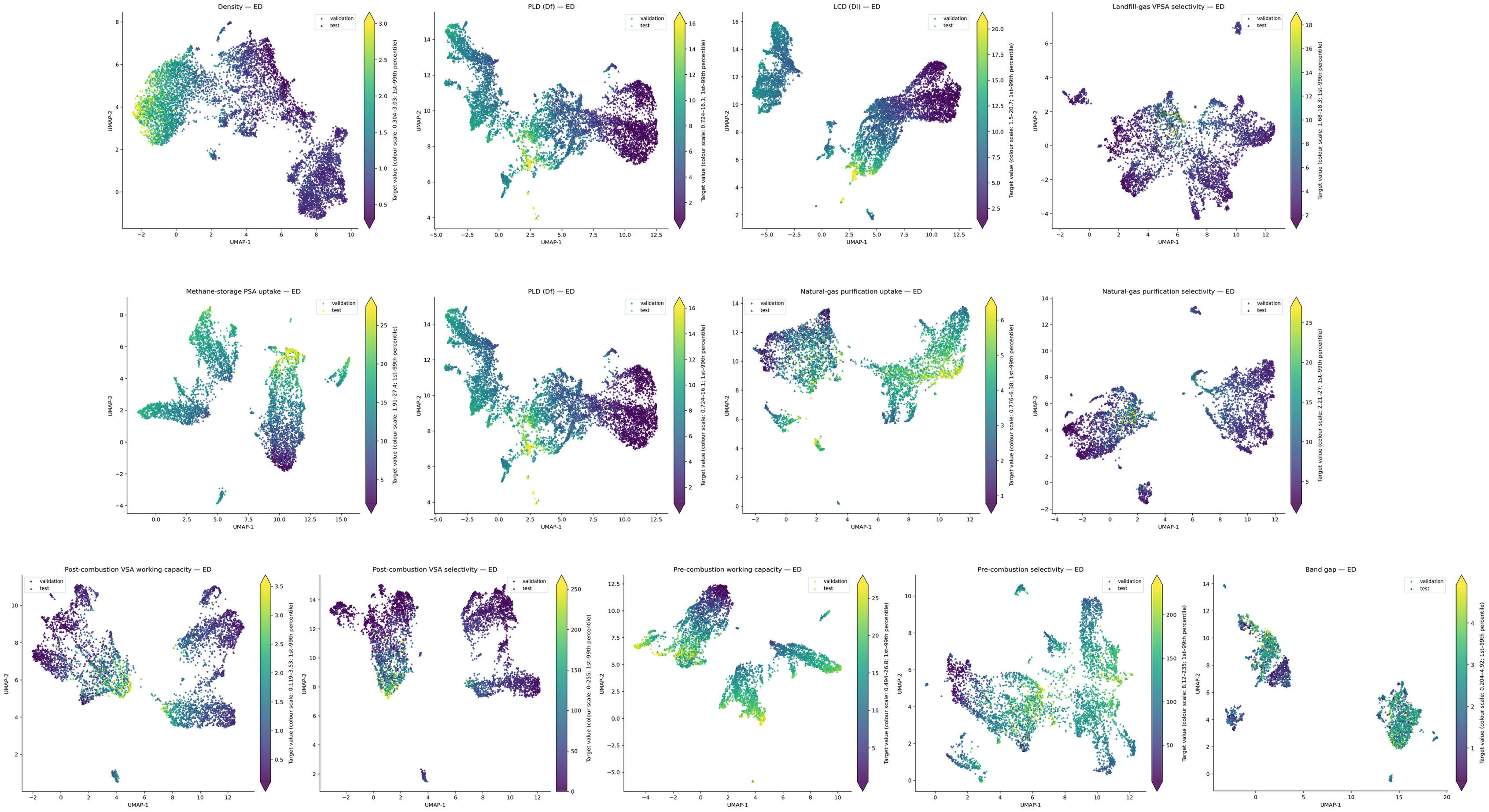}
\caption{
\textbf{UMAP views of learned ED  property representations.}
For each property and modality identified in the panels, material-level embeddings from the validation and test sets are projected jointly using UMAP. Marker symbols distinguish validation and test samples, and colour indicates target-value percentile within the property. Each model has an independently fitted projection. These maps visualize the learned embeddings; predictive metrics are calculated from the regression outputs.
}
\label{fig:property_umap}
\end{figure}

\begin{figure}[p]
\centering
\includegraphics[width=\linewidth,height=0.72\textheight,keepaspectratio]{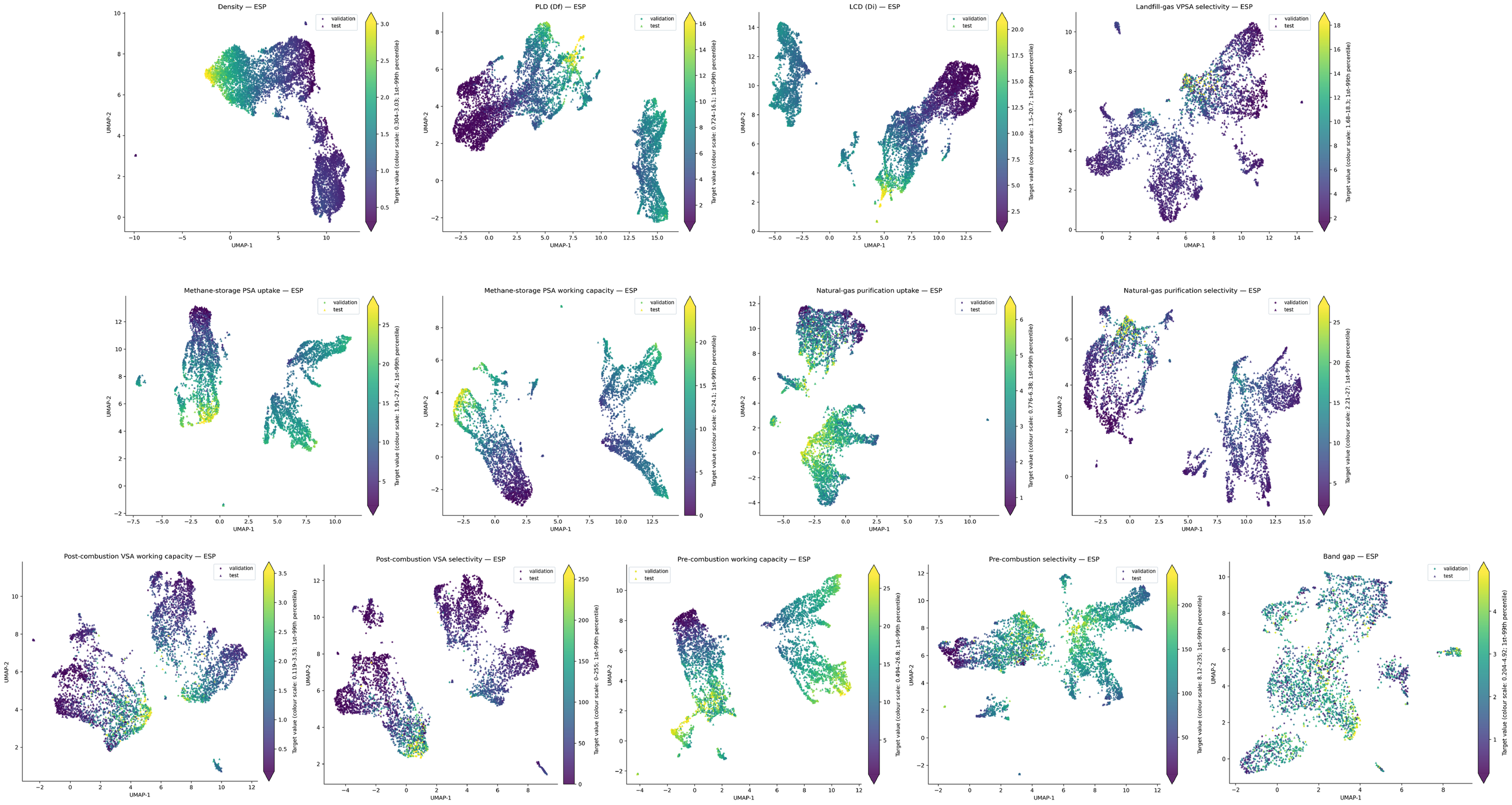}
\caption{
\textbf{UMAP views of learned  ESP property representations.}
 Validation and test embeddings are projected jointly within each task and modality. Marker symbols identify the split, and colour denotes target-value percentile.
}
\label{fig:property_umap_additional}
\end{figure}

Material-level embeddings were additionally visualized using UMAP~\cite{mcinnes2018umap} (Supplementary Figs.~\ref{fig:property_umap} and \ref{fig:property_umap_additional}). For each task and modality, validation and test embeddings from the trained property model were combined in a two-dimensional projection. Marker symbols distinguish the splits, and colour denotes the percentile of the corresponding target value. The projections provide a qualitative view of sample overlap and property-related organization within each learned representation. Because projections are fitted independently for different models, their absolute positions and orientations are not directly comparable across tasks or modalities. Predictive performance is assessed using the test-set metrics above.


\section{Multimodal learning}
\label{sec:supp_multimodal}

Multimodal property prediction and cross-modal retrieval used the
40,359-framework ARC-MOF and QMOF collection. Property prediction used
the target-specific labelled subsets and material-level partitions
reported in Supplementary Table~\ref{tab:property_splits}. Representations
of the same MOF were assigned to the same partition for both prediction
and retrieval.

\subsection{Multimodal property prediction}

Three fusion strategies, residual-gated fusion, symmetric gated fusion
and attention-gated fusion, were evaluated for CIF+ED, CIF+ESP and
CIF+ED+ESP across all 13 targets, giving 117 configurations.
The modality-specific encoders followed Supplementary
Section~\ref{sec:supp_property_prediction}. Encoders supplied 128-dimensional
features to the fusion modules. All models were trained end to end
using the normalized MSE regression loss and AdamW for 50 epochs,
with checkpoints selected using validation performance.

Let \(\mathbf h_m\in\mathbb R^{128}\) denote the embedding of
modality \(m\), and let \(M=2\) or \(3\) denote the number of
input modalities.

Residual-gated fusion used learnable gates to combine modality
features and predict an additive residual
correction:
\begin{equation}
\hat y=\hat y_0+\Delta\hat y_{\mathrm{gated}},
\end{equation}
where \(\hat y_0\) is the prediction-branch output and
\(\Delta\hat y_{\mathrm{gated}}\) is the correction obtained
from gated multimodal features and cross-modal interactions.

Symmetric gated fusion combined modality embeddings through
a learned weighted sum:
\begin{equation}
\begin{aligned}
\boldsymbol\alpha
&=\operatorname{softmax}(a_1,\ldots,a_M),\\
\mathbf h_{\mathrm{fuse}}
&=\sum_{m=1}^{M}\alpha_m\mathbf h_m,\\
\hat y&=H(\mathbf h_{\mathrm{fuse}}).
\end{aligned}
\end{equation}
Each score \(a_m\) was obtained by applying the same scalar
scoring function to the layer-normalized modality embedding.
The weights were normalized across modalities and varied
between samples.

Attention-gated fusion used a learned query to aggregate modality
embeddings through multi-head attention.
The trainable query supplied \(\mathbf Q\), while the modality
embeddings provided the key and value inputs, \(\mathbf K\)
and \(\mathbf V\):
\begin{equation}
\begin{aligned}
\mathbf h_{\mathrm{fuse}}
&=\operatorname{LN}\!\left(
\operatorname{MHA}(\mathbf Q,\mathbf K,\mathbf V)
\right),\\
\hat y&=H(\mathbf h_{\mathrm{fuse}}).
\end{aligned}
\end{equation}
Here, \(\operatorname{LN}\) denotes layer normalization
and \(H\) is the regression head.


For each target, the unimodal model with the lowest MAE among
CIF, ED and ESP provided the reference MAE and \(R^2\).
Relative multimodal MAE reductions and \(\Delta R^2\) were
calculated against this reference, as defined in the main Methods.

Test-set MAE and \(R^2\) for all 117 configurations are reported
in Supplementary Tables~\ref{tab:multimodal_results} and
\ref{tab:multimodal_fusion_r2}, with graphical comparisons in
Supplementary Figs.~\ref{fig:supp_multimodal_fusion_mae} and
\ref{fig:supp_multimodal_fusion_r2}.
For each target and modality combination, the main-text comparison
used the fusion model with the lowest MAE among the three
evaluated strategies, together with its corresponding \(R^2\).
Selection was based on unrounded MAE values.

Fusion-method selection was separate from selection of the unimodal
reference. For each target, the unimodal reference was the
lowest-MAE configuration among CIF, ED and ESP, as defined in the
main Methods. Both the relative multimodal MAE reduction and
$\Delta R^2$ were calculated against that same unimodal configuration.

The preferred fusion method depended on the target and modality
combination. Across the 39 target--modality combinations,
residual-gated, symmetric-gated and attention-gated fusion achieved
the lowest MAE in 23, 9 and 7 cases, respectively. After this
MAE-based selection, CIF+ED+ESP gave the lowest MAE among the three
multimodal input combinations for 10 of the 13 targets. CIF+ED
performed best for density and band gap, whereas CIF+ESP performed
best for landfill-gas VPSA selectivity.



\begin{table*}[htbp]
\centering
\footnotesize
\setlength{\tabcolsep}{2pt}
\renewcommand{\arraystretch}{1.15}
\caption{Test-set MAE for all 13 targets, three modality combinations and three fusion methods.
RG, SG and AG denote residual-gated, symmetric-gated and attention-gated fusion,
respectively. Bold identifies the lowest MAE within each target and modality
combination using unrounded values. MAE is reported in \AA{} for
PLD and LCD, $\mathrm{g\,cm^{-3}}$ for density, eV for band gap and
$\mathrm{mmol\,g^{-1}}$ for uptake and working capacity; selectivity is
dimensionless. Values are rounded to five decimal places.}
\label{tab:multimodal_results}
\begin{tabular*}{\textwidth}{@{\extracolsep{\fill}}>{\raggedright\arraybackslash}p{0.28\textwidth}*{9}{r}@{}}
\toprule
& \multicolumn{3}{c}{CIF+ED}
& \multicolumn{3}{c}{CIF+ESP}
& \multicolumn{3}{c}{CIF+ED+ESP} \\
\cmidrule(lr){2-4}\cmidrule(lr){5-7}\cmidrule(lr){8-10}
Property & RG & SG & AG & RG & SG & AG & RG & SG & AG \\
\midrule
PLD / Df
& 0.51735 & \textbf{0.51691} & 0.52174 & 0.54919 & \textbf{0.54905} & 0.56229 & \textbf{0.49219} & 0.50087 & 0.50329 \\
LCD / Di
& \textbf{0.53504} & 0.53717 & 0.54323 & \textbf{0.55336} & 0.56059 & 0.55494 & 0.49524 & \textbf{0.49518} & 0.50714 \\
Density
& 0.05227 & 0.04877 & \textbf{0.04743} & 0.05461 & 0.05495 & \textbf{0.05382} & 0.05256 & 0.05272 & \textbf{0.05091} \\
Band gap
& 0.28102 & 0.28534 & \textbf{0.27616} & 0.29846 & \textbf{0.28764} & 0.29088 & 0.29047 & \textbf{0.28090} & 0.28585 \\
Post-combustion VSA selectivity
& \textbf{12.36059} & 12.54644 & 13.36152 & \textbf{12.61161} & 12.62780 & 13.06572 & 12.31888 & 12.48667 & \textbf{12.06999} \\
Natural-gas purification selectivity
& \textbf{1.30344} & 1.33402 & 1.32574 & 1.34655 & 1.34767 & \textbf{1.29696} & 1.29640 & \textbf{1.28511} & 1.32439 \\
Landfill-gas VPSA selectivity
& \textbf{0.75831} & 0.77531 & 0.77520 & \textbf{0.72736} & 0.74771 & 0.76099 & \textbf{0.75410} & 0.76035 & 0.77311 \\
Pre-combustion selectivity
& \textbf{15.59178} & 16.46841 & 16.25034 & \textbf{15.45089} & 15.56311 & 15.48638 & \textbf{15.09362} & 15.57537 & 15.48957 \\
Methane-storage PSA uptake
& \textbf{0.61624} & 0.61857 & 0.62956 & 0.57738 & \textbf{0.57589} & 0.61286 & 0.56796 & 0.56325 & \textbf{0.55619} \\
Methane-storage PSA working capacity
& \textbf{0.58634} & 0.58912 & 0.61314 & 0.50492 & \textbf{0.50295} & 0.54945 & \textbf{0.49636} & 0.50807 & 0.50883 \\
Post-combustion VSA working capacity
& \textbf{0.21996} & 0.23030 & 0.23365 & \textbf{0.22797} & 0.22992 & 0.22931 & \textbf{0.21169} & 0.21779 & 0.23006 \\
Natural-gas purification uptake
& 0.43109 & \textbf{0.42746} & 0.43253 & \textbf{0.42298} & 0.42786 & 0.42954 & \textbf{0.40673} & 0.41890 & 0.42081 \\
Pre-combustion working capacity
& \textbf{0.94948} & 0.95904 & 1.00044 & \textbf{0.82659} & 0.84322 & 0.89658 & \textbf{0.81453} & 0.82837 & 0.87075 \\
\bottomrule
\end{tabular*}
\end{table*}

\begin{table*}[htbp]
\centering
\footnotesize
\setlength{\tabcolsep}{2pt}
\renewcommand{\arraystretch}{1.15}
\caption{Test-set $R^2$ for the same configurations as Supplementary
Table~\ref{tab:multimodal_results}. RG, SG and AG denote residual-gated,
symmetric-gated and attention-gated fusion. Bold identifies the $R^2$ of
the MAE-selected fusion method within each target and modality combination;
it does not necessarily indicate the highest $R^2$. Values are rounded to
four decimal places.}
\label{tab:multimodal_fusion_r2}
\begin{tabular*}{\textwidth}{@{\extracolsep{\fill}}>{\raggedright\arraybackslash}p{0.28\textwidth}*{9}{r}@{}}
\toprule
& \multicolumn{3}{c}{CIF+ED}
& \multicolumn{3}{c}{CIF+ESP}
& \multicolumn{3}{c}{CIF+ED+ESP} \\
\cmidrule(lr){2-4}\cmidrule(lr){5-7}\cmidrule(lr){8-10}
Property & RG & SG & AG & RG & SG & AG & RG & SG & AG \\
\midrule
PLD / Df
& 0.9530 & \textbf{0.9535} & 0.9534 & 0.9504 & \textbf{0.9489} & 0.9467 & \textbf{0.9581} & 0.9564 & 0.9545 \\
LCD / Di
& \textbf{0.9594} & 0.9587 & 0.9569 & \textbf{0.9584} & 0.9548 & 0.9556 & 0.9658 & \textbf{0.9652} & 0.9639 \\
Density
& 0.9839 & 0.9857 & \textbf{0.9866} & 0.9827 & 0.9827 & \textbf{0.9837} & 0.9848 & 0.9845 & \textbf{0.9856} \\
Band gap
& 0.8662 & 0.8663 & \textbf{0.8744} & 0.8612 & \textbf{0.8697} & 0.8626 & 0.8612 & \textbf{0.8688} & 0.8661 \\
Post-combustion VSA selectivity
& \textbf{0.7431} & 0.7487 & 0.7159 & \textbf{0.7458} & 0.7416 & 0.7191 & 0.7407 & 0.7494 & \textbf{0.7597} \\
Natural-gas purification selectivity
& \textbf{0.4852} & 0.4896 & 0.4925 & 0.4849 & 0.5029 & \textbf{0.5171} & 0.5100 & \textbf{0.5067} & 0.4883 \\
Landfill-gas VPSA selectivity
& \textbf{0.4503} & 0.4189 & 0.4653 & \textbf{0.5095} & 0.4991 & 0.4990 & \textbf{0.4547} & 0.4494 & 0.4387 \\
Pre-combustion selectivity
& \textbf{0.7037} & 0.6539 & 0.6731 & \textbf{0.6905} & 0.6891 & 0.6944 & \textbf{0.7091} & 0.7055 & 0.7112 \\
Methane-storage PSA uptake
& \textbf{0.9712} & 0.9708 & 0.9703 & 0.9750 & \textbf{0.9749} & 0.9722 & 0.9759 & 0.9765 & \textbf{0.9762} \\
Methane-storage PSA working capacity
& \textbf{0.9723} & 0.9721 & 0.9700 & 0.9803 & \textbf{0.9809} & 0.9774 & \textbf{0.9816} & 0.9806 & 0.9805 \\
Post-combustion VSA working capacity
& \textbf{0.7394} & 0.7279 & 0.7197 & \textbf{0.7297} & 0.7193 & 0.7295 & \textbf{0.7763} & 0.7515 & 0.7184 \\
Natural-gas purification uptake
& 0.7683 & \textbf{0.7760} & 0.7748 & \textbf{0.7820} & 0.7820 & 0.7669 & \textbf{0.8016} & 0.7760 & 0.7809 \\
Pre-combustion working capacity
& \textbf{0.9521} & 0.9522 & 0.9480 & \textbf{0.9652} & 0.9631 & 0.9577 & \textbf{0.9665} & 0.9648 & 0.9604 \\
\bottomrule
\end{tabular*}
\end{table*}

\begin{figure*}[p]
\centering
\includegraphics[width=0.98\textwidth]{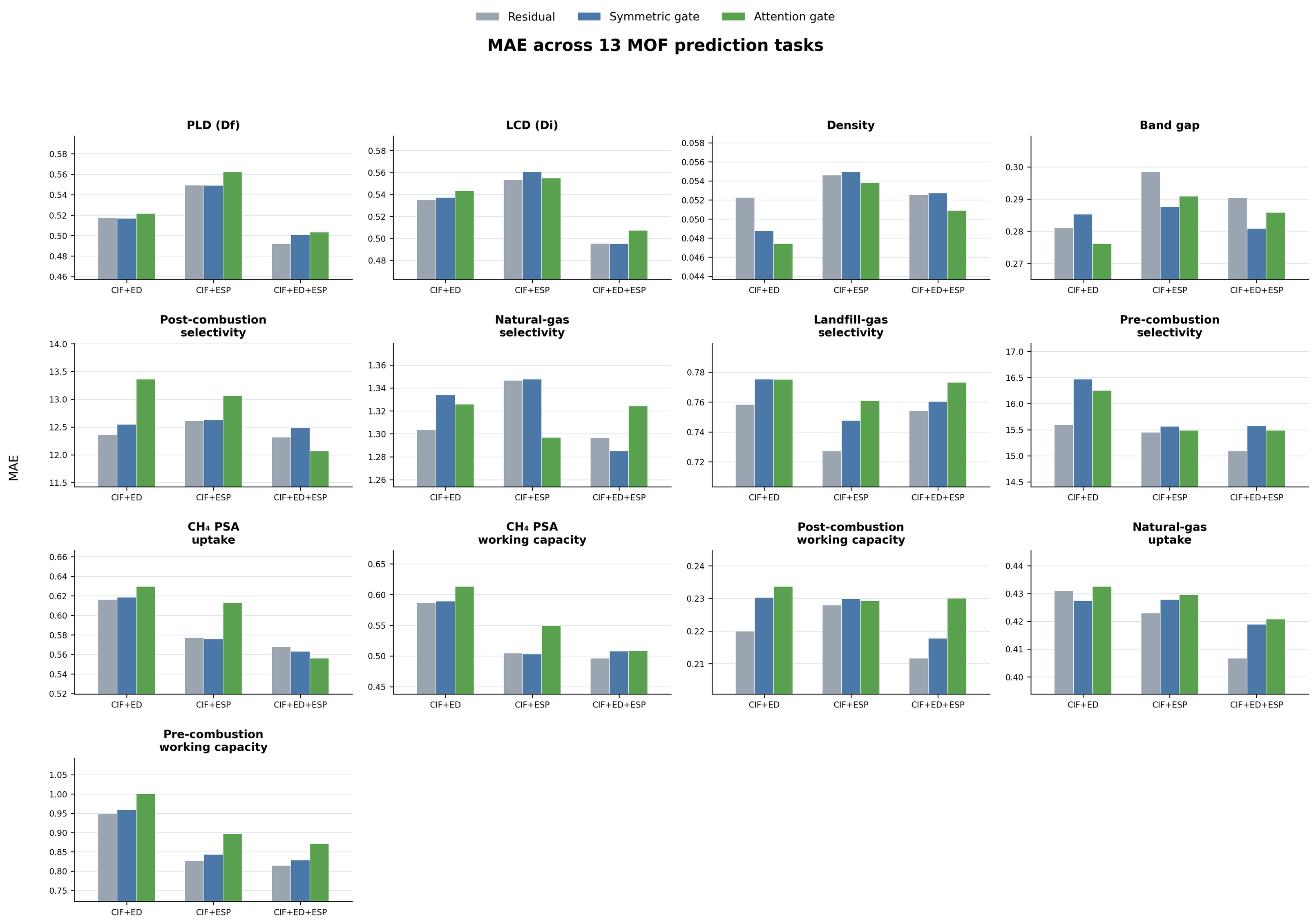}
\caption{
\textbf{Comparison of fusion methods across all 13 targets using MAE.}
Each panel shows one property. Bar groups represent CIF+ED, CIF+ESP and
CIF+ED+ESP; grey, blue and green bars indicate residual-gated,
symmetric-gated and attention-gated fusion, respectively.
Lower values indicate smaller errors. Units are specified on the
vertical axes, and vertical limits vary between panels.
}
\label{fig:supp_multimodal_fusion_mae}
\end{figure*}

\begin{figure*}[p]
\centering
\includegraphics[width=0.98\textwidth]{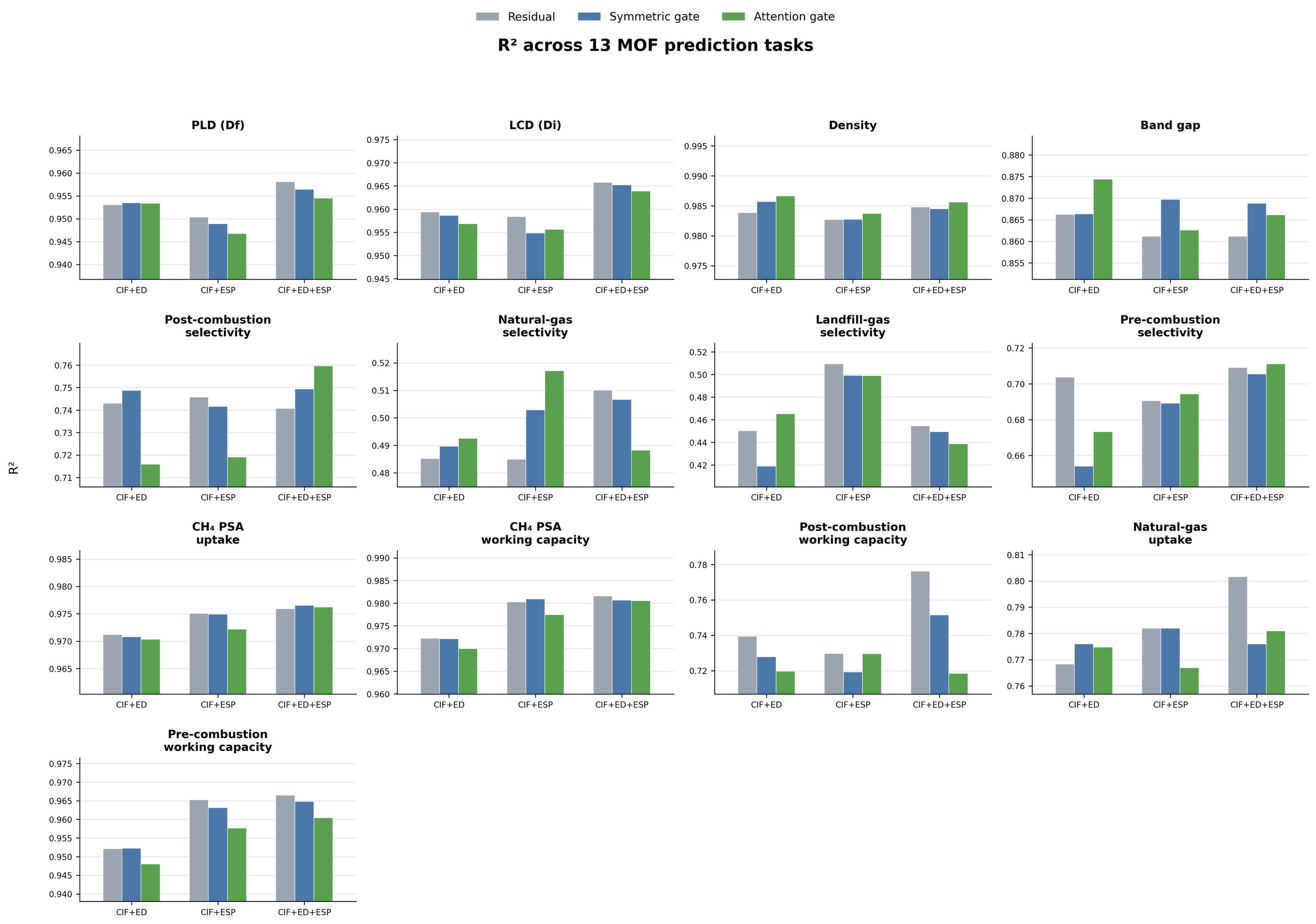}
\caption{
\textbf{Comparison of fusion methods across all 13 targets using $R^2$.}
Panels and bar groups correspond to Supplementary
Fig.~\ref{fig:supp_multimodal_fusion_mae}, with the same colour scheme.
Higher values indicate better agreement with the reference properties.
Vertical limits vary between panels.
}
\label{fig:supp_multimodal_fusion_r2}
\end{figure*}

\subsection{Cross-modal retrieval}
\label{sec:supp_crossmodal_retrieval}

Cross-modal retrieval evaluated whether a MOF could be identified from
its representation in another modality. CIF--ED, CIF--ESP and ED--ESP
models were trained independently using the symmetric contrastive
objective described in the main Methods. Modality-specific encoders
and projection heads produced 256-dimensional, $L_2$-normalized
embeddings, with negative pairs drawn from the same minibatch.
The selected retrieval configurations were trained for 200 epochs
using AdamW and a contrastive temperature of $0.07$.
Implementation settings for property prediction and retrieval are
summarized in Supplementary
Table~\ref{tab:multimodal_implementation}.

Each electronic-field point supplied its Cartesian coordinates and
scalar field value, supplemented by Fourier features of its fractional
coordinates. The selected retrieval configurations used two Fourier
frequencies per fractional-coordinate axis ($k=2$), with pair-specific
point counts and filtering settings
(Supplementary Table~\ref{tab:retrieval_pointcloud_configurations}).
For ED--ESP retrieval, the prescribed number of points was sampled
separately for each field. Here, $N$ denotes the number of sampled
points per field, whereas $K$ in Recall@$K$ denotes retrieval depth.

\begin{table*}[htbp]
\centering
\small
\setlength{\tabcolsep}{5pt}
\caption{
Implementation settings for multimodal property prediction and the
selected retrieval configurations. Feature dimensions refer to fusion
features for property prediction and projected embeddings for retrieval.
Learning rates are initial values.
}
\label{tab:multimodal_implementation}
\begin{tabular}{
>{\raggedright\arraybackslash}p{0.21\textwidth}
>{\raggedright\arraybackslash}p{0.34\textwidth}
>{\raggedright\arraybackslash}p{0.34\textwidth}}
\toprule
Setting & Property prediction & Cross-modal retrieval \\
\midrule
Training objective
& Normalized MSE regression loss
& Symmetric contrastive loss \\
Feature dimension
& 128
& 256 \\
Feature normalization
& Layer normalization
& $L_2$ normalization \\
Optimizer
& AdamW
& AdamW \\
Initial learning rate
& $3\times10^{-4}$
& $3\times10^{-4}$ \\
Weight decay
& $1\times10^{-4}$
& $1\times10^{-4}$ \\
Batch size
& Task-specific: 8 or 16
& 48 for CIF--ED; 64 for CIF--ESP and ED--ESP \\
Training duration
& 50 epochs
& 200 epochs \\
Contrastive temperature
& --
& 0.07 \\
\bottomrule
\end{tabular}
\end{table*}

\begin{table*}[htbp]
\centering
\small
\setlength{\tabcolsep}{6pt}
\caption{
Selected electronic-field inputs for cross-modal retrieval.
$\tau_{\rho}$ is the ED threshold in $\mathrm{e\,Bohr^{-3}}$,
$q$ is the fraction retained from each ESP tail before FPS, $N$ is
the point count per field and $k$ is the Fourier-encoding order.
For ED--ESP, $N$ applies separately to each field.
Dashes indicate settings that do not apply.
}
\label{tab:retrieval_pointcloud_configurations}
\begin{tabular}{lrrrr}
\toprule
Modality pair & $\tau_{\rho}$ & $q$ & $N$ & $k$ \\
\midrule
CIF--ED  & 0.02 & --  & 6,144 & 2 \\
CIF--ESP & --   & 3\% & 4,096 & 2 \\
ED--ESP  & 0.03 & 3\% & 4,096 & 2 \\
\bottomrule
\end{tabular}
\end{table*}

At evaluation, each query was ranked by cosine similarity against the
complete candidate set from the paired target modality. Directional
metrics and their arithmetic means are reported in Supplementary
Table~\ref{tab:crossmodal_retrieval}. ED--ESP achieved bidirectionally
averaged Recall@1 of $0.914$ and MRR of $0.952$, compared with $0.552$
and $0.688$ for CIF--ED and $0.520$ and $0.653$ for CIF--ESP.
Its Recall@5 and Recall@10 reached $0.998$ and $0.999$, respectively.
Recall across retrieval depths is shown in Supplementary
Fig.~\ref{fig:retrieval_recall_curve}.

\begin{table*}[htbp]
\centering
\small
\setlength{\tabcolsep}{8pt}
\caption{
Test-set cross-modal retrieval performance for the selected configurations.
Queries are ranked against the complete candidate set in the target modality.
The mean row reports the arithmetic average of the two retrieval directions.
}
\label{tab:crossmodal_retrieval}
\begin{tabular}{llrrrr}
\toprule
Modality pair & Retrieval direction & Recall@1 & Recall@5 & Recall@10 & MRR \\
\midrule
\multirow{3}{*}{CIF $\leftrightarrow$ ED}
& CIF $\rightarrow$ ED & 0.547 & 0.867 & 0.925 & 0.686 \\
& ED $\rightarrow$ CIF & 0.558 & 0.856 & 0.919 & 0.689 \\
& Mean                 & 0.552 & 0.862 & 0.922 & 0.688 \\
\midrule
\multirow{3}{*}{CIF $\leftrightarrow$ ESP}
& CIF $\rightarrow$ ESP & 0.523 & 0.823 & 0.890 & 0.656 \\
& ESP $\rightarrow$ CIF & 0.517 & 0.811 & 0.885 & 0.649 \\
& Mean                  & 0.520 & 0.817 & 0.887 & 0.653 \\
\midrule
\multirow{3}{*}{ED $\leftrightarrow$ ESP}
& ED $\rightarrow$ ESP & 0.913 & 0.998 & 0.999 & 0.952 \\
& ESP $\rightarrow$ ED & 0.914 & 0.999 & 0.999 & 0.953 \\
& Mean                 & 0.914 & 0.998 & 0.999 & 0.952 \\
\bottomrule
\end{tabular}
\end{table*}

\begin{figure*}[htbp]
\centering
\includegraphics[width=0.86\textwidth]
{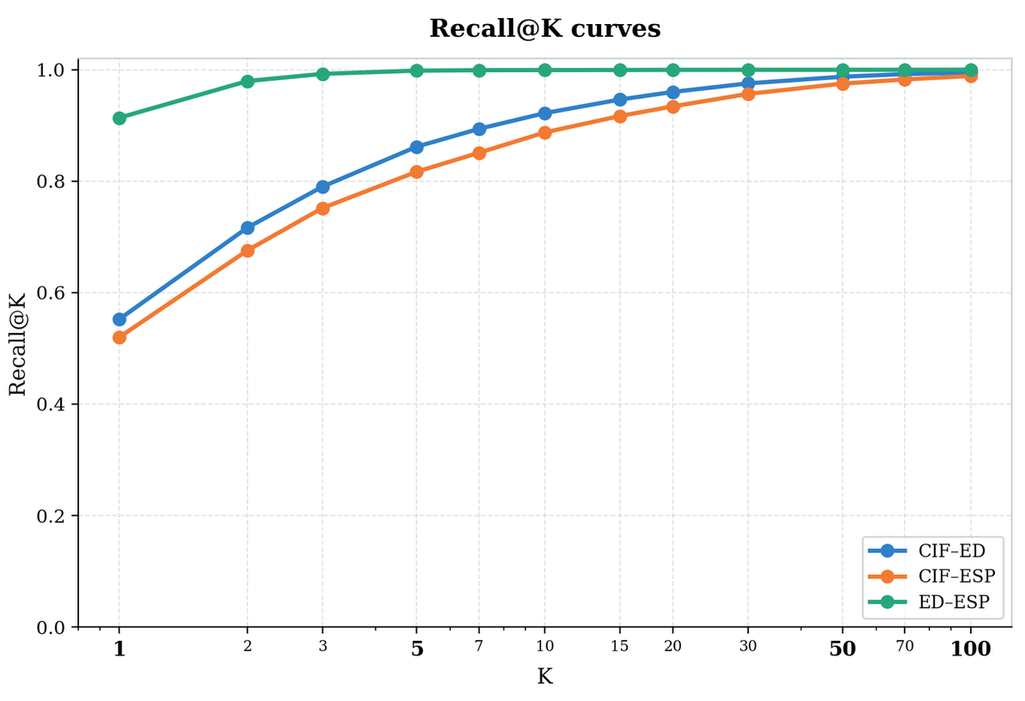}
\caption{
\textbf{Recall@$K$ curves for cross-modal retrieval.}
Bidirectionally averaged Recall@$K$ for CIF--ED, CIF--ESP and ED--ESP
as the number of top-ranked candidates $K$ increases.
}
\label{fig:retrieval_recall_curve}
\end{figure*}

Retrieval sensitivity was examined by varying the point count over
$N\in\{4096,6144,8192,10240\}$ and the periodic encoding order over
$k\in\{0,1,2,3\}$. Each modality pair was trained independently for
200 epochs, with the remaining training settings held fixed within
each analysis. All reported metrics were averaged over the two
retrieval directions.

\begin{figure*}[htbp]
\centering
\includegraphics[width=\textwidth]
{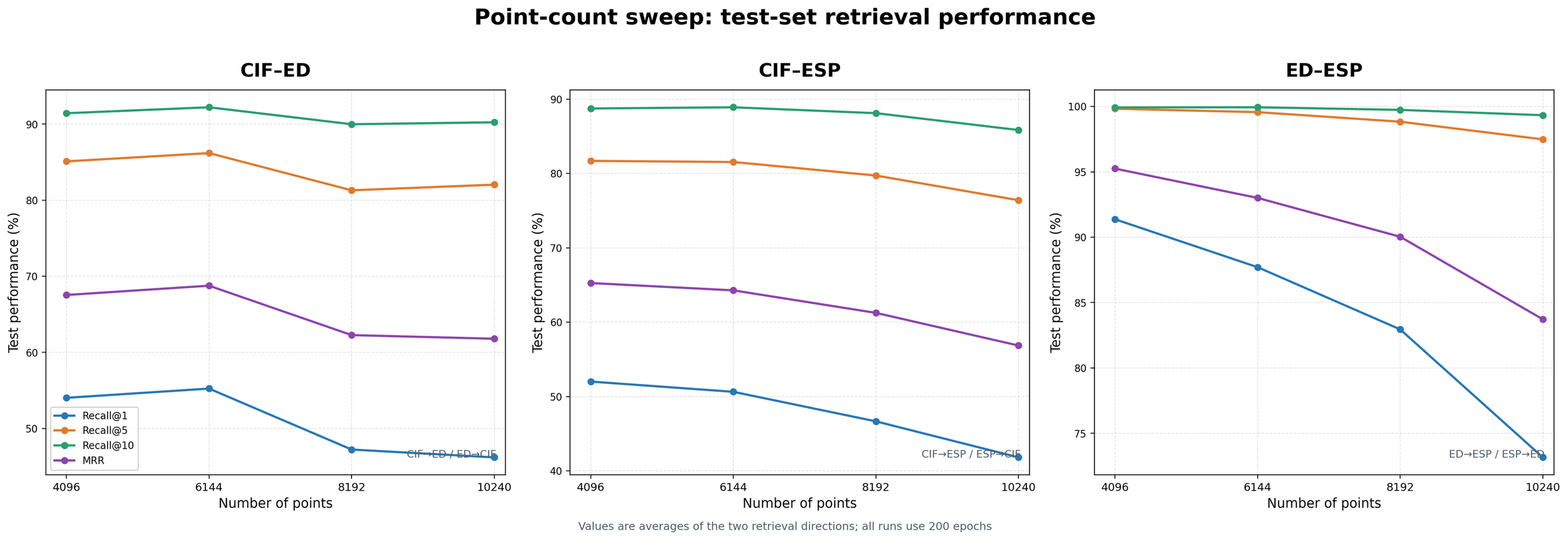}
\caption{
\textbf{Sensitivity of retrieval to point-cloud size.}
Recall@1, Recall@5, Recall@10 and MRR for CIF--ED, CIF--ESP and ED--ESP
using $N=4096$, $6144$, $8192$ or $10240$ points per field.
Metrics are averaged over both directions; all models were trained
for 200 epochs.
}
\label{fig:retrieval_point_sweep}
\end{figure*}

Increasing the point count did not produce a consistent monotonic
improvement across modality pairs
(Supplementary Fig.~\ref{fig:retrieval_point_sweep}). CIF--ED performed
best at an intermediate point count, whereas CIF--ESP and ED--ESP
favoured smaller point clouds under the evaluated protocol.

\begin{figure*}[htbp]
\centering
\includegraphics[width=\textwidth]
{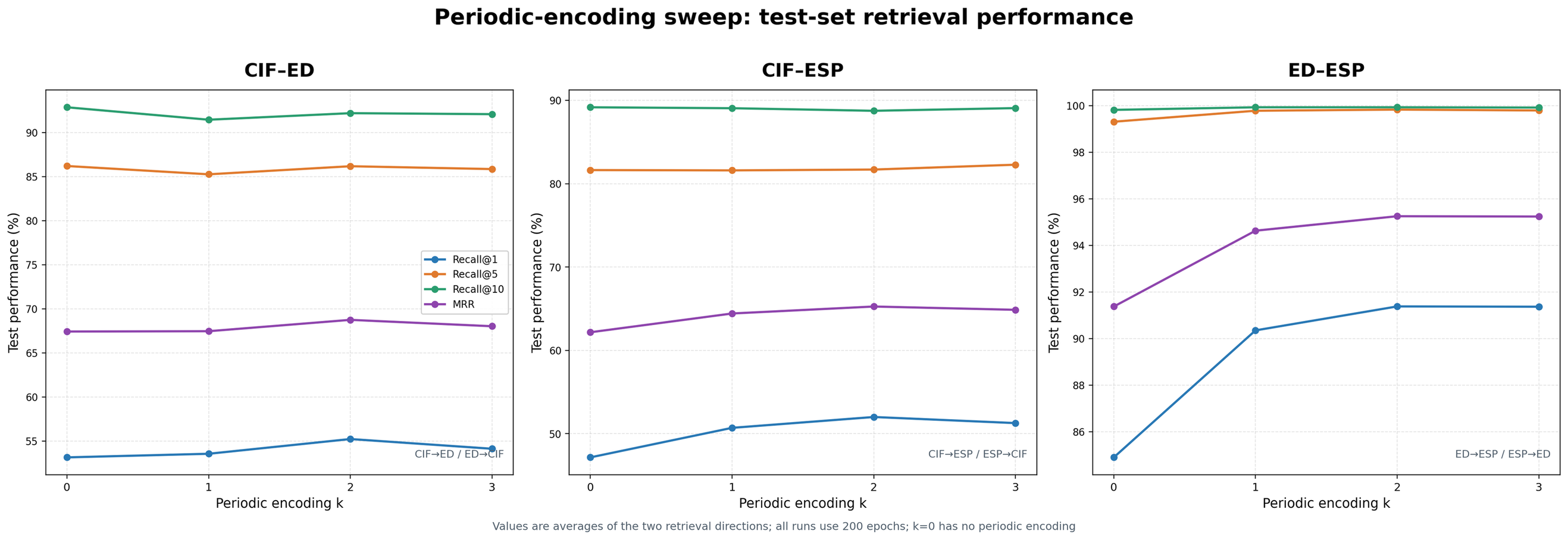}
\caption{
\textbf{Sensitivity of retrieval to periodic positional encoding.}
Recall@1, Recall@5, Recall@10 and MRR as the Fourier-encoding order
varies over $k=0,1,2,3$. The $k=0$ setting omits Fourier features
while retaining Cartesian coordinates. Metrics are averaged over
both directions; all models were trained for 200 epochs.
}
\label{fig:retrieval_periodic_sweep}
\end{figure*}

The Fourier-feature analysis showed a larger effect on CIF--ESP
and ED--ESP than on CIF--ED
(Supplementary Fig.~\ref{fig:retrieval_periodic_sweep}). At $k=0$,
Fourier features of fractional coordinates were omitted while
Cartesian coordinates remained available as geometric inputs.
ED--ESP performance approached a plateau around $k=2$, with limited
additional benefit from the third frequency under the evaluated
settings.

\begin{figure*}[htbp]
\centering
\includegraphics[width=\textwidth]
{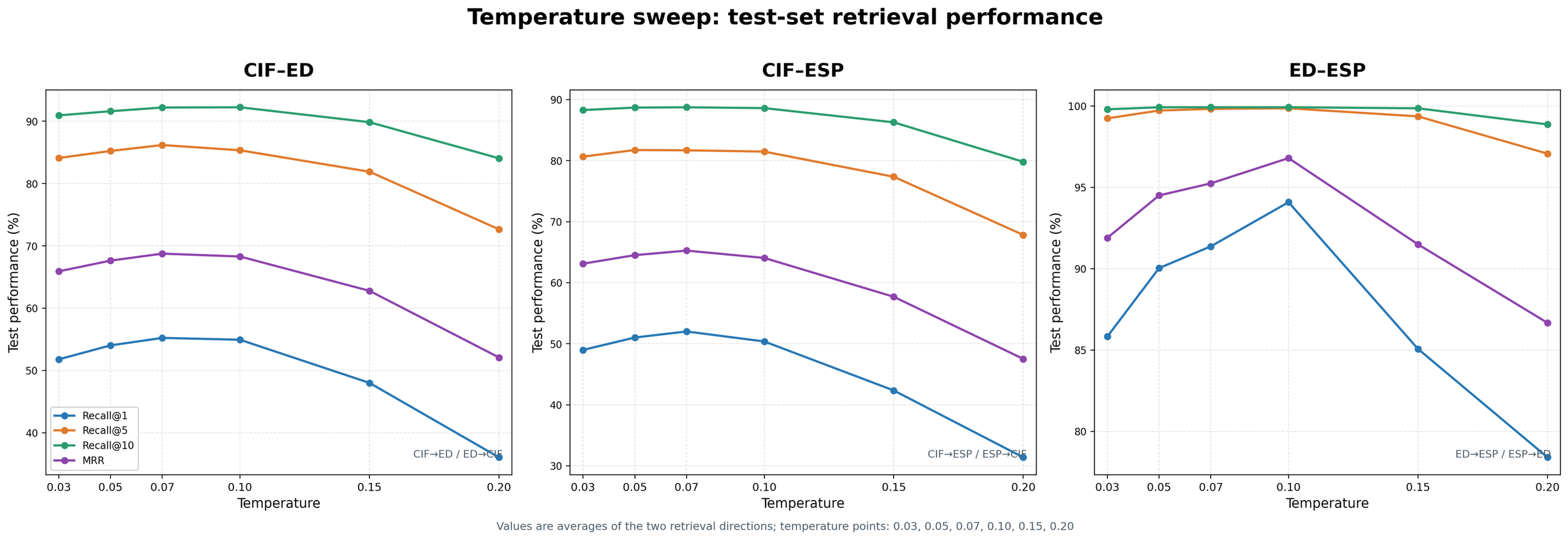}
\caption{
\textbf{Sensitivity of cross-modal retrieval to contrastive temperature.}
Panels show CIF--ED, CIF--ESP and ED--ESP retrieval for
$\tau=0.03$, $0.05$, $0.07$, $0.10$, $0.15$ and $0.20$.
Recall@1, Recall@5, Recall@10 and MRR are averaged over the two
retrieval directions and expressed as percentages.
}
\label{fig:retrieval_temperature_sweep}
\end{figure*}

The contrastive temperature was also varied over
$\tau\in\{0.03,0.05,0.07,0.10,0.15,0.20\}$
(Supplementary Fig.~\ref{fig:retrieval_temperature_sweep}). CIF--ED and
CIF--ESP achieved their highest Recall@1 and MRR near $\tau=0.07$,
whereas ED--ESP reached its highest values for these metrics near
$\tau=0.10$. Increasing the temperature to $0.15$ or $0.20$ reduced
Recall@1 and MRR for all three pairs. Recall@5 and Recall@10 were less
sensitive for ED--ESP and remained close to unity over a broader
temperature range. The preferred temperature therefore depended on
the modality pair and evaluation metric.

\end{appendix}
\end{document}